\documentclass{jfm}

\usepackage{graphicx}
\usepackage{natbib}
\usepackage{verbatim}
\usepackage{tikz}
\usepackage{pgfplots}
\usepackage{pgfplotstable}
\usepackage{amsmath}
\usepackage{caption}
\usepackage{subcaption}
\usepackage{bm}
\usepackage{rotating}
\usepackage{pdflscape}

\ifCUPmtlplainloaded \else
  \checkfont{eurm10}
  \iffontfound
    \IfFileExists{upmath.sty}
      {\typeout{^^JFound AMS Euler Roman fonts on the system,
                   using the 'upmath' package.^^J}%
       \usepackage{upmath}}
      {\typeout{^^JFound AMS Euler Roman fonts on the system, but you
                   dont seem to have the}%
       \typeout{'upmath' package installed. JFM.cls can take advantage
                 of these fonts,^^Jif you use 'upmath' package.^^J}%
      }
  \else
  \fi
\fi

\ifCUPmtlplainloaded \else
  \checkfont{msam10}
  \iffontfound
    \IfFileExists{amssymb.sty}
      {\typeout{^^JFound AMS Symbol fonts on the system, using the
                'amssymb' package.^^J}%
       \usepackage{amssymb}%
       \let\le=\leqslant  
       \let\ge=\geqslant  
      }{}
  \fi
\fi

\ifCUPmtlplainloaded \else
  \IfFileExists{amsbsy.sty}
    {\typeout{^^JFound the 'amsbsy' package on the system, using it.^^J}%
     \usepackage{amsbsy}}
    {}
\fi

\newcommand\be{\begin{equation}}
\newcommand\ee{\end{equation}}
\newcommand\bea{\begin{eqnarray}}
\newcommand\eea{\end{eqnarray}}

\newcommand\base{{\rm base}}

\newcommand{\free}{{\rm outer}}

\makeatletter
\newcommand{\pushright}[1]{\ifmeasuring@#1\else\omit\hfill$\displaystyle#1$\fi\ignorespaces}
\newcommand{\pushleft}[1]{\ifmeasuring@#1\else\omit$\displaystyle#1$\hfill\fi\ignorespaces}
\makeatother

\newcommand\Rey{\mbox{\textit{Re}}}  % Reynolds number
\newsavebox{\astrutbox}
\sbox{\astrutbox}{\rule[-5pt]{0pt}{20pt}}

\newcommand\sumer{{\rm Sumer}}
\newcommand\lap{ {\mathcal{L}}}
\newcommand\dz{ \frac{\partial}{\partial z}}
\newcommand\dzdz{ \frac{\partial^2}{\partial z^2}}

\title[Nonmodal stability analysis of the boundary layer under solitary waves]{
Nonmodal stability analysis of the boundary layer under solitary waves}

\author[J. C. G. Verschaeve {\em et al.}]%
{Joris\ls C.\ls G.\ls Verschaeve$^1$\thanks{Email address for correspondence: joris.verschaeve@gmail.com}
\and \ls
Geir\ls K.\ls Pedersen$^1$ \and \ls Cameron\ls Tropea$^2$}

\affiliation{$^1$University of Oslo, Po.Box 1072 Blindern, 0316 Oslo, Norway\\
$^2$Technische Universit\"at Darmstadt, 64347 Griesheim, Germany}

\pubyear{2010}
\volume{650}
\pagerange{119--126}
\begin{document}

\maketitle

\begin{abstract}
  In the present treatise, a stability analysis
  of the bottom boundary layer under solitary waves
based on energy bounds and nonmodal theory is
performed. The instability mechanism 
of this flow consists of a competition
between streamwise streaks
and two-dimensional perturbations. For
lower Reynolds numbers and early times,
streamwise streaks display larger 
amplification due to their quadratic dependence
on the Reynolds number, whereas two-dimensional
perturbations become dominant for larger 
Reynolds numbers and later times in the deceleration
region of this flow, as the maximum amplification
of two-dimensional perturbations grows exponentially
with the Reynolds number. By means of the present
findings, we can give some indications on the physical
mechanism and on the interpretation of
the results by direct numerical simulation
in \citep{VittoriBlondeaux2008,OzdemirHsuBalachandar2013}
and by experiments in \citep{SumerJensenSorensenFredsoeLiuCarstensen2010}.
In addition, three critical
Reynolds numbers can be defined for which 
the stability properties of the flow change. 
In particular, it is shown that this 
boundary layer changes from
a monotonically stable to a non-monotonically stable
flow at a Reynolds number of $ \Rey_\delta = 18$. 
\end{abstract}

\section{Introduction}

In recent years,
stability and transition processes in the
boundary layer under solitary water waves have received
increased attention in the coastal engineering community,
cf. 
\citep{LiuParkCowen2007,VittoriBlondeaux2008,SumerJensenSorensenFredsoeLiuCarstensen2010,OzdemirHsuBalachandar2013,VerschaevePedersen2014}.
Motivated by the design of
harbors and other coastal installations, this boundary layer is of 
importance for understanding sediment transport phenomena
under water waves and scaling effects in experiments. 

In the present
treatise, the mechanisms leading to instability
and finally to turbulent transition shall be investigated
by means of a nonmodal stability analysis.
%a nonmodal stability analysis of this flow 
%is performed revealing an important ingredient
%of the mechanisms leading to instability
%and finally to turbulent transition. Thereby,
The present boundary layer is not only of interest
for the coastal engineering community, but can also
serve as a useful generic flow for the investigation of
stability and
transition mechanisms of boundary layers displaying
favorable and adverse pressure gradients, such as
the ones developing in front and behind of the
location of maximum thickness of an airplane wing
or turbine blade profile. In addition,
the present flow can be considered a model
for the single stroke of a pulsating flow, such as Stokes'
second problem,
which is of importance for biomedical applications.\\ %[1ex]

Solitary waves, which are either found
as surface or internal waves, are of great interest in the ocean engineering
community for several reasons.
They are nonlinear and dispersive. When frictional effects due
to the boundary layer at the bottom and the top 
are negligible, the shape of solitary waves is preserved during
propagation. Relatively simple approximate analytic solutions exist,
see for instance \citet{Benjamin1966}, \citet{Grimshaw1971}
or \citet{Fenton1972}. In addition,
these waves are relatively easy to reproduce experimentally. 
As such, they are often used in order to investigate the 
effect of a single crest of a train of waves. \\

The first works on the boundary layer under solitary waves
aimed at estimating the dissipative effect on the overall
wave \citep{Shuto1976,Miles1980}. 
The bottom boundary layer has been considered
more relevant than the surface boundary layer for viscous dissipation
\citep{LiuOrfila2004} and
the stability of this boundary layer is also the subject of the
present treatise. \\
\begin{comment}
More detailed experimental and numerical work on 
this boundary layer and its stability properties has
been performed first for the boundary layer under
internal solitary waves, since this boundary layer is important 
for transport of sediments on the bottom of the sea
\citep{BoguckiDickeyRedekopp1997}. 
%As this boundary layer depends on at least four parameters,
%progress on the investigation of its stability properties
%has been slow. 
By means of direct
numerical simulation
\citet{DiamessisRedekopp2005} obtained a relationship for the critical parameters for which transition occurs. However,
experimental results \citep{CarrDavies2006,CarrDavies2010} 
contradict this relationship,
as cases deemed stable displayed
instabilities and cases deemed unstable did not. More recently,
a modified formula for the critical parameters has been advanced in \cite{AghsaeeBoegmanDiamessisLamb2012}
by means of direct numerical simulation.\\
\end{comment}

The earliest experiments on the bottom boundary layer under
solitary waves have been performed
for internal waves by \citep{CarrDavies2006,CarrDavies2010}
and for surface waves by \citet{LiuParkCowen2007}.
The latter showed that an inflection
point develops in the deceleration region behind the crest
of the wave. However, instabilities have not been observed
in the experiments performed by them \citep{LiuParkCowen2007}. 
In 2010, Sumer {\em et al.} used a water tunnel 
to perform experiments on the boundary layer under solitary waves. 
They observed three flow regimes.
By means of a Reynolds number $ \Rey_{\delta} $, defined by
the Stokes length of the boundary layer and
the characteristic particle velocity, as used in \citet{OzdemirHsuBalachandar2013}
and in the present treatise,
these regimes can be characterized as follows. 
For small Reynolds numbers $ \Rey_{\delta} < 630 $($\approx \Rey_\sumer = 2 \cdot 10^5 $, i.e. the Reynolds number defined in
\citet{SumerJensenSorensenFredsoeLiuCarstensen2010}), the
flow does not display any instabilities and is close
to the laminar solution given in \cite{LiuParkCowen2007}. 
For a Reynolds number in the range $ 630  \le \Rey_\delta < 1000 $
($ 2\cdot 10^5 \le \Rey_\sumer < 5 \cdot 10^5 $),
they observed the appearance of regularly spaced vortex rollers in
the deceleration region of the flow. Increasing the Reynolds number
further leads to a transitional flow displaying the emergence of
turbulent spots growing together and causing transition to
turbulence in the boundary layer. This happens at first in the deceleration region. However, the first instance of
spot nucleation moves forward into the acceleration region of the flow
for
increasing Reynolds number. Sumer {\em et al.} did not
control the level of
external disturbances in their experiments
nor did they report
any information on its characteristics, such as length
scale or intensity. \\

Almost parallel to the experiments by Sumer {\em et al.}, 
Vittori and Blondeaux performed direct numerical simulations
of this flow \citep{VittoriBlondeaux2008,VittoriBlondeaux2011}. 
Their results correspond roughly to the findings by
Sumer {\em et al.} in that the flow in their simulations
is first observed to display a laminar regime before displaying
regularly spaced vortex rollers and finally becoming turbulent.
However, the Reynolds numbers at which these regime shifts 
occur
are larger than those in the experiments by Sumer {\em et al.}. 
In particular, Vittori and Blondeaux observed the flow to be laminar until
a Reynolds number somewhat lower than $ \Rey_\delta = 1000 $,
after which the flow in their simulations displays regularly spaced
vortex rollers.
Transition to turbulence has been observed to 
occur for Reynolds numbers somewhat
larger than $ \Rey_\delta = 1000 $. 
They triggered the flow regime changes
by introducing a random disturbance of a specific
magnitude in the computational domain before the arrival of the
wave. \citet{OzdemirHsuBalachandar2013}
performed direct numerical
simulations using the same approach as Vittori and Blondeaux, but
varied the magnitude of the initial disturbance. As a result
they found different flow regimes than what Sumer {\em et al.}
and Vittori and Blondeaux had observed. In the simulations
by {\"O}zdemir {\em et al.} the flow stays laminar until
$ \Rey_\delta = 400 $, then enters a regime
they called 'disturbed laminar' for $  400 < \Rey_\delta < 1500 $, where
instabilities can be observed. For $ \Rey_\delta  > 1500 $ regularly 
spaced vortex rollers appear in the deceleration region
of the flow in their simulations giving rise to a $ K $-type transition before
turbulent break down, if the Reynolds number is large enough. A $ K $-type
transition is characterized by a spanwise instability
giving rise to the development of $ \Lambda $-vortices
arranged in an aligned fashion, cf. \citet{Herbert1988}. 
For very large Reynolds numbers $ \Rey_\sumer > 2400 $,
{\"O}zdemir {\em et al.} reported that the
$ K $-type transition is replaced by a transition which reminded
them of a free stream layer type transition. \\

Next to investigations based on direct numerical simulations
and experiments,
modal stability theories have been employed in the works by 
\citet{BlondeauxPralitsVittori2012}, \citet{VerschaevePedersen2014}
and
\citet{SadekParrasDiamessisLiu2015}. 
Employing
a quasi-static approach for the Orr-Sommerfeld equation, cf. \citep{KerczekDavis1974},
Blondeaux {\em et al.} found that this unsteady flow displayed
unstable regions for all of their Reynolds number considered, even those deemed
stable by direct numerical simulation. 

In order to 
explain the divergences in transitional Reynolds numbers obtained
by direct numerical simulation and experiment, 
\citet{VerschaevePedersen2014} 
performed a stability analysis in the frame of reference
moving with the wave, where the present boundary layer flow is steady. 
For steady flows,
well-established stability methods can be used. By means of the
parabolized stability equation, they showed that
for all Reynolds numbers considered in their analysis, the boundary
layer
displays regions of growth of disturbances. As the flow goes to
zero towards infinity, there exists a point
on the axis of the moving coordinate
where the perturbations reach a maximum amplification
before decaying again for a given Reynolds number. Depending
on the level of initial disturbances in the flow, 
this maximum amount of amplification is sufficient for triggering
secondary instability, such as turbulent spots or $ \Lambda $-vortices,
or not. This explains the diverging
critical Reynolds numbers observed in direct numerical
simulations and experiments for this boundary layer flow.
A particular case in point, mentioned in \cite{VerschaevePedersen2014},
is the experiment on the boundary layer under internal
solitary waves by \cite{CarrDavies2006}.
Although, the amplitudes of the generated internal
solitary waves in these experiments are relatively large
compared to the thickness of the upper layer, the outer
flow on the bottom is relatively well approximated by the
first order solution of \citet{Benjamin1966}, cf. figure 12 in \cite{CarrDavies2006}. 
In these
experiments, the flow displays instabilities for
Reynolds numbers much smaller than in the experiments
by \citet{SumerJensenSorensenFredsoeLiuCarstensen2010} 
or in the direct numerical simulations
by \citet{VittoriBlondeaux2008} 
or \citet{OzdemirHsuBalachandar2013}.
%although these boundary layer flows are very similar in
%the limit of small amplitude solitary waves. 
\citet{VerschaevePedersen2014} proposed, that due
to the characteristic velocity of internal solitary waves
being significantly smaller than that for surface solitary waves,
they are expected to display instabilities much earlier
for comparable levels of background noise.

\citet{SadekParrasDiamessisLiu2015} performed a similar
modal stability analysis as \citet{VerschaevePedersen2014}
by marching Orr-Sommerfeld eigenmodes forward in time using
the linearized and two-dimensional nonlinear Navier-Stokes equations. They
observed that only for Reynolds numbers larger than $ \Rey_\delta = 90 $,
Orr-Sommerfeld eigenmodes display growth and consequently defined this
Reynolds number to be the
critical Reynolds number where the flow changes from a stable to an unstable regime.\\

The modal stability theories employed in \citet{BlondeauxPralitsVittori2012},
\citet{VerschaevePedersen2014} and \citet{SadekParrasDiamessisLiu2015}
capture only parts of the picture. 
In all of these works, only two-dimensional disturbances are considered. In addition,
the amplifications computed in \citet{VerschaevePedersen2014}
and \citet{SadekParrasDiamessisLiu2015}
describe only the so-called exponential growth of the 
most unstable eigenfunction of the Orr-Sommerfeld equation. 
As shown in \citet{ButlerFarrell1992,TrefethenTrefethenReddyDriscoll1993,SchmidHenningson2001,Schmid2007},
perturbations can undergo significant transient growth even
when modal stability theories predict the flow system to
be stable. Nonmodal theory formulates the stability problem
as an optimization problem for the perturbation energy. 
In the present treatise, optimal perturbations
are computed for the unsteady boundary layer flow
under a solitary wave, complementing the modal
analysis performed in
\citep{BlondeauxPralitsVittori2012,VerschaevePedersen2014,SadekParrasDiamessisLiu2015}.
In particular, we shall investigate the following
questions. 

In \citet{SadekParrasDiamessisLiu2015}, a critical
Reynolds number is found based on a modal analysis.
However, as perturbations can display
growth even for cases where modal analysis predicts
stability, this question needs to be treated
in the framework of energy methods \citep{Joseph1966}. 
Using
an energy bound derived in \citep{DavisKerczek1973},
we shall show that a critical Reynolds number
$ \Rey_A > 0 $ can be found, 
such that for all Reynolds numbers smaller than $ \Rey_A $,
the flow is monotonically stable, meaning 
that all perturbations are damped for all times.

\citet{OzdemirHsuBalachandar2013} supposed that a by-pass transition starts to develop 
in their simulations for some cases, but could not explain why 
then suddenly two-dimensional perturbations emerge
producing a $ K $-type transition typical for growing
Tollmien-Schlichting waves.
In the present treatise, we shall show that nonmodal theory is
able to describe this competition between streaks and
two-dimensional perturbations (i.e. nonmodal Tollmien-Schlichting waves),
which allows us to predict the onset of growth of streaks and
two-dimensional perturbations, their maximum amplification
and the point in time when this maximum is reached. 
Furthermore, the dependence on the Reynolds number
of the maximum amplification shall be investigated. 
The results obtained in the present treatise indicate
why in
the direct numerical simulations
by \citet{VittoriBlondeaux2008,VittoriBlondeaux2011}
and \citet{OzdemirHsuBalachandar2013},
in all cases investigated, two dimensional perturbations lead to turbulent
break-down, although one would expect, at least
for some cases, turbulent break-down via three dimensional structures
for a purely random seeding. On the other hand
\citet{SumerJensenSorensenFredsoeLiuCarstensen2010}
observed the growth of two-dimensional
structures only for a certain range of Reynolds numbers,
before the appearance of turbulent spots. A $K$-type
transition has not been observed in their experiments. 
Turbulent spots are in general attributed to 
the secondary instability of streamwise streaks,
see for example \citep{AnderssonBrandtBottaroHenningson2001,
BrandtSchlatterHenningson2004}.
Though,
the random break-down of Tollmien-Schlichting
waves is also thought to produce turbulent spots,
cf. \citep{ShaikhGaster1994,Gaster2016}.
The present analysis is limited to the primary
instability of streamwise streaks and nonmodal Tollmien-Schlichting
waves. It gives, however, indications for a possible
secondary instability mechanism of competing streaks
and Tollmien-Schlichting waves. \\

The present treatise is organized as follows. In the following
section, section \ref{sec:problem}, we describe the flow system
and present equations for energy bounds and the nonmodal governing
equations. The solutions of these equations applied to 
the present flow are presented and discussed in section
\ref{sec:results}. In section \ref{sec:otherworks}, we shall relate the
current findings to results obtained previously in the literature. 
The present treatise is concluded in section \ref{sec:conclusions}. 

\section{Description of the problem} \label{sec:problem}

\subsection{Specification of base flow} \label{sec:base}

The outer flow 
of the present boundary layer is given by the celebrated first order solution for the inviscid horizontal velocity for solitary waves
\citep{Benjamin1966,Fenton1972}. For a given point at the bottom,
the outer flow can thus be written as in \citet{SumerJensenSorensenFredsoeLiuCarstensen2010}:
\be
U_\free^*(t^*) = U_0 {\rm sech}^2 \left( \omega_0 t^* \right). \label{eq:solitonFormula}
\ee
In the limit of vanishing amplitude of the solitary wave, not
only the nonlinearities in the inviscid solution become negligible,
but they can also be neglected in the boundary layer equations. Following
\citet{LiuOrfila2004}, the horizontal component
in the boundary layer $ U_\base $ can be written as 
\be
U_\base = U_\free + u_{bl}, \label{eq:base}
\ee
where $ u_{bl} $ contains the rotational part of the velocity
and ensures that the no-slip boundary condition is satisfied. 
Neglecting the nonlinearities, we obtain the following
boundary layer equations for $ u_{bl} $
\citep{LiuParkCowen2007,ParkVerschaevePedersenLiu2014}:
\bea
\frac{\partial}{\partial t} u_{bl} &=& \frac{1}{2} \frac{\partial^2}{\partial z^2} u_{bl} \label{eq:bc1} \\
u_{bl}(0,t) &=& - U_\free(t)  \label{eq:bc2}\\
u_{bl}(\infty,t) &=& 0 \label{eq:bc3} \\
u_{bl}(z,-\infty) &=&0 \label{eq:bc4}
\eea
Equation (\ref{eq:bc1}) is the linearized momentum equation. Equations (\ref{eq:bc2}) and
(\ref{eq:bc3}) are the boundary conditions of the problem, with equation (\ref{eq:bc2}) representing
the no-slip boundary condition and equation (\ref{eq:bc3}) representing the outer flow
boundary condition. Equation (\ref{eq:bc4}) is the initial condition, which is advanced in
time from $ - \infty $. The resulting base flow $ U_\base $, equation (\ref{eq:base}),
is valid on the entire time axis $ t \in (-\infty,\infty) $.
The scaling used in equations (\ref{eq:bc1}-\ref{eq:bc4}) is given by $ \omega_0 $ for the time,
\be
t = \omega_0 t^*,
\ee
by $ U_0 $ for the velocity,
\be
U_\free = \frac{1}{U_0} U_\free^*,
\ee
and by the Stokes boundary layer thickness $ \delta $ for the wall normal variable $ z $:
\be
z = \frac{z^*}{\delta},
\ee
where
\be
\delta = \sqrt{ \frac{2\nu}{\omega_0} }.
\ee
For the solution of equations (\ref{eq:bc1}-\ref{eq:bc4}),
a Shen-Chebyshev discretization in wall normal direction
is chosen, whereas the resulting system is integrated
in time by means of a Runge-Kutta integrator, cf.
reference \citep{Shen1995} and appendix \ref{sec:implementation}
for details.
Summing up, we
consider solitary waves of small amplitudes for which formula (\ref{eq:solitonFormula})
is a good approximation of the outer flow, such as the solitary wave
experiments in \citet{CarrDavies2006,CarrDavies2010,LiuParkCowen2007} or
the water channel experiments in \citet{SumerJensenSorensenFredsoeLiuCarstensen2010} and \citet{TanakaWinartaSuntoyoYamaji2011}. As shown in \cite{VerschaevePedersen2014}, for larger amplitude solitary
waves the nonlinear effects are not negligible anymore and significant qualitative
differences arise, making the present nonmodal approach not applicable anymore. 

\begin{figure}
    \includegraphics[width=\textwidth]{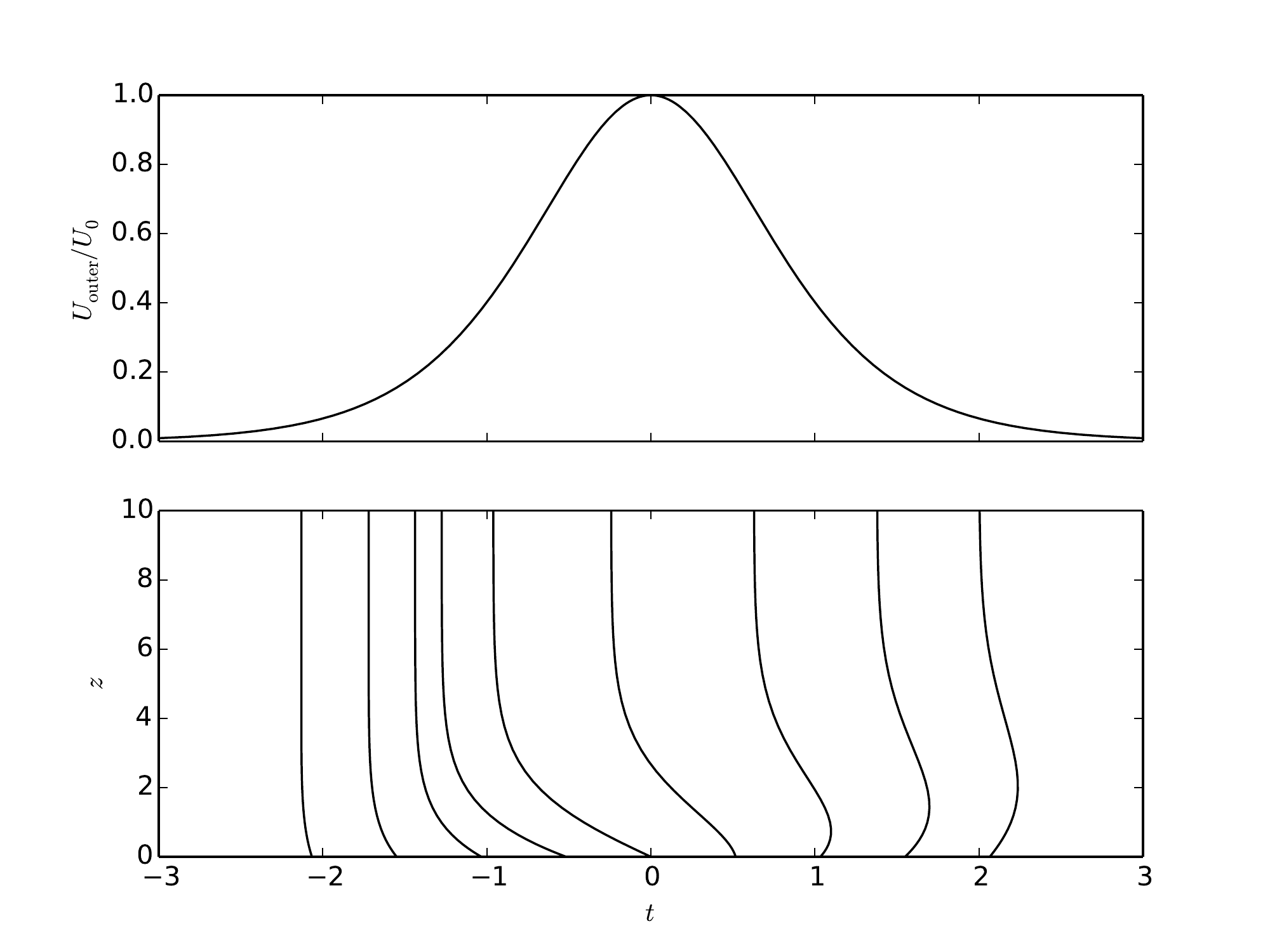}
    \caption{Inviscid outer flow $ U_\free $ at the bottom
and profiles of the horizontal velocity component in the boundary layer under a solitary wave moving from right to left. 
%The dimensionless wave height is $ \epsilon = 0.1 $.
The profiles 
have been multiplied by 40.  
The value at $ z = 0 $ of the profiles shown corresponds to the
point in time $ t $, at which the profile has been taken.
The horizontal velocity
vanishes at $ z = 0 $ in order to satisfy the no-slip boundary condition.}
\label{fig:profilesAndSolitaryWave}
\end{figure}

\subsection{Stability analysis by means of an energy bound}

In the present treatise,
we use the same definition for 
the Reynolds number as in \citet{OzdemirHsuBalachandar2013}.
This Reynolds number $ \Rey_\delta $
is based on the Stokes length $ \delta $ 
and the characteristic velocity $ U_0 $:
\be
\Rey_{\delta} = \frac{U_0 \delta}{\nu} = U_0 \sqrt{ \frac{ 2}{ \nu \omega_0} },
\ee
where $ \nu $ is the kinematic viscosity of the fluid. 
The Reynolds number $ \Rey_\sumer $ used in \citet{SumerJensenSorensenFredsoeLiuCarstensen2010} is related to $ \Rey_\delta $ by the following formula:
\be
\Rey_\delta = \sqrt{ 2 \Rey_\sumer }. 
\ee
%The time $ \tau $ in the resulting Navier-Stokes equations is then scaled the
%following way:
%\be
%\tau = \frac{U_0}{\delta} t^* = \frac{\Rey_\delta}{2} t.
%\ee
We introduce a perturbation velocity $ \mathbf{u}' = (u',v',w' )$ in the 
streamwise, spanwise and wall normal direction, defined by:
\be
\mathbf{u}' = \left(u',v',w'\right) = \left(u_{ns},v_{ns},w_{ns} \right) 
- \left(U_\base \left(z,t \right),0,0\right),
\ee
where $ (u_{ns},v_{ns},w_{ns} ) $ satisfies the Navier-Stokes equations. The
energy of the perturbation is given by:
\be
E_p = \frac{1}{2} \int \limits_V u'^2 + v'^2 + w'^2 \, dV,
\ee
which is integrated over $ V = \{ (x,y,z) \, | \, z > 0 \} $. For time
dependent flows in infinite domains, \citet{DavisKerczek1973} derived
a bound for the perturbation energy of the nonlinear Navier-Stokes
equations:
\be
\frac{E_p(t)}{E_{p,0}} \le \exp \frac{\Rey_\delta}{2} \int \limits_{t_0}^t
\mu(t') \, dt',
\ee
where $ \mu $ is the largest eigenvalue of the following linear system:
\bea
\frac{1}{\Rey_\delta} \Delta \mathbf{u}' - \mathbf{S}_\base(t) \cdot
\mathbf{u}' - \nabla p &=& \frac{1}{2} \mu \mathbf{u}' \label{eq:davis1} \\
\nabla \cdot \mathbf{u}' &=& 0, \label{eq:davis2}
\eea
where the tensor $ \mathbf{S}_\base $ is the rate of strain tensor
given by the base flow, equation (\ref{eq:base}).
%\be
%\mathbf{S}_\base = \frac{1}{2} \left( \nabla \mathbf{U}_\base
%+ \left( \nabla \mathbf{U}_\base \right)^T \right).
%\ee
We remark
that \citet{DavisKerczek1973} appear to have overlooked a sign and a factor two in their
equations. As the rate of strain tensor depends on time,
the eigenvalue $ \mu $ is a function of $ t $.
If $ \mu < 0 $ for all times, then the flow is
monotonically stable for this Reynolds number, meaning that
all perturbations will decay for all times. This allows us
to investigate, if there exists a Reynolds number $ \Rey_A $,
at which $ \mu $ switches sign from negative to positive at some point in time.
As the base flow is independent of $ x $ and $ y $, we consider a single Fourier
component of $ \mathbf{u}' $:
\be
(u',v',w')(x,y,z,t) = (u,v,w)(z,t) \exp {\rm i} \left( \alpha x + \beta y \right).
\ee
This allows us to eliminate $ p $ from the equations (\ref{eq:davis1}-\ref{eq:davis2}), resulting into 
\bea
\frac{1}{\Rey_\delta} \lap^2 w + \frac{ {\rm i } \alpha}{2}
\left\{ \frac{\partial^2}{\partial z^2} U_\base w + 2 \dz U_\base \dz w \right\} + \frac{{\rm i} \beta}{2}  \dz U_\base \zeta  &=& \frac{1}{2}\mu \lap w,
\label{eq:davis1a}\\
-\frac{1}{\Rey_\delta} \lap \zeta - \frac{ {\rm i} \beta}{2} \dz U_\base w &=& \frac{1}{2}\mu (-\zeta) \label{eq:davis2a}
\eea
where $ \lap $ is the Laplacian defined by:
\be
\lap = -k^2 + \dzdz,
\ee
where $ k^2 = \alpha^2 + \beta^2 $. 
The system of four equations (\ref{eq:davis1}-\ref{eq:davis2}),
has been reduced to two, by means of the normal vorticity component $ \zeta $:
\be
\zeta 
= {\rm i} \left( \alpha v - \beta u \right).
\ee
A Galerkin formulation for the system (\ref{eq:davis1a}-\ref{eq:davis2a}) is 
chosen based on Shen-Legendre polynomials for the biharmonic equation
for the normal component $ w $ and Shen-Legendre polynomials 
for the Poisson equation
for the normal vorticity $ \zeta $, cf. reference \citep{Shen1994}. Thereby,
the Hermitian property of the system (\ref{eq:davis1a}-\ref{eq:davis2a}) 
is conserved in the discrete setting, guaranteeing
purely real eigenvalues. Details of
the implementation are given in appendix \ref{sec:implementation}. 

\subsection{The nonmodal stability equations}

The nonmodal stability analysis is based on the linearized Navier-Stokes
equations, which can be written in the present setting as follows,
\bea
\left( \frac{2}{\Rey_\delta}\frac{\partial}{\partial t} + {\rm i} \alpha U_\base - \frac{1}{\rm Re_\delta} \mathcal{L}
\right) \mathcal{L} w
- {\rm i} \alpha w \dzdz U_\base  &=& 0, \label{eq:sys1} \\
\left( \frac{2}{\Rey_\delta} \frac{\partial}{\partial t} + {\rm i} \alpha U_\base  - \frac{1}{\rm Re_\delta} \mathcal{L}
\right) \zeta - {\rm i} \beta w \dz U_\base 
&=& 0. \label{eq:sys2}
\eea
We refer to \citet{SchmidHenningson2001,Schmid2007} for a thorough derivation
of equations (\ref{eq:sys1}) and (\ref{eq:sys2}).
%We remark that $ U_\base $ varies on the convective time scale
%in equations (\ref{eq:sys1}) and (\ref{eq:sys2}):
%\be
%U_\base = U_\base \left(z,\tau \frac{2}{\Rey_\delta}\right).
%\label{eq:scalingU}
%\ee
Given an initial
perturbation $ (w_0,\zeta_0 ) $ at time $ t_0 $, equations 
(\ref{eq:sys1}) and (\ref{eq:sys2}) can be integrated to
obtain the temporal evolution of $ (w,\zeta)$ for $ t > t_0 $.
Nonmodal theory formulates the stability problem as finding
the initial condition $(w_0,\zeta_0 ) $ maximizing the perturbation
energy $ E(t) $ of $ (w,\zeta) $ at time $ t > t_0 $. 
This perturbation energy $ E $ is the sum
of two contributions, one from the wall normal component $ w $ 
and one from the normal vorticity component $ \zeta$:
\be
E(t) = E_w(t) + E_\zeta(t) = 
\frac{1}{2} \int \limits_0^\infty \frac{1}{k^2} \left| \dz w \right|^2 + \left| w \right|^2 \, dz 
+ \frac{1}{2} \int \limits_0^\infty \frac{1}{k^2} \left| \zeta \right|^2 \, dz.
\label{eq:energyNonmodal}
\ee
The optimization problem can then be formulated by maximizing $ E$ for
a perturbation $ (w,\zeta) $ satisfying (\ref{eq:sys1}) and (\ref{eq:sys2})
and having an initial energy $ E_0 $. One way of solving this
optimization problem is 
by means of the adjoint equation as in \citet{LuchiniBottaro2014}. Another
approach for finding the optimal perturbation, which is employed in
the present treatise, consists in formulating
the discrete problem first and computing the evolution matrix $ \mathbf{X}(t,t_0) $ of the
system of ODEs, cf. references \citet{TrefethenTrefethenReddyDriscoll1993,SchmidHenningson2001,Schmid2007} for details. The energy $ E $ is then
given in terms of $ \mathbf{X} $ and the initial condition. Details of
the implementation are given in appendix \ref{sec:implementation}. 
By computing $ E(t) $ one way or the other, we can compute
the amplification $ G $ from time $ t_0 $ to $ t $ 
of the optimal perturbation for wave numbers $ \alpha $ and $ \beta$:
\be
G(\alpha,\beta,t_0,t,\Rey_\delta) = \max_{ (w_0,\zeta_0) } \frac{ E(t) }{E(t_0) }. \label{eq:amplification}
\ee
We remark that the initial condition $ (w_0,\zeta_0 ) $
from which the optimal perturbation starts,
might be different for each point in time $ t $,
when tracing $ G $ as a function of $ t $, cf. section \ref{sec:results}.
The maximum amplification $ G_{\max}(\Rey_\delta)$, which can be reached
for a given Reynolds number $ \Rey_\delta $, is obtained by
maximizing $ G$ over time, initial time and wavenumbers:
\be
G_{\max} = \max_{\alpha,\beta,t_0,t} G. \label{eq:maximumA}
\ee
In the following, we shall distinguish between three types of perturbations:
\vspace{2mm}
\begin{itemize}
\item streamwise streaks.\\
These are perturbations independent of the streamwise coordinate $ x$. They
can be computed by setting $ \alpha = 0 $. 
\item Two-dimensional perturbations.\\
These perturbations are independent of the spanwise coordinate $ y $ and
can be computed by setting $ \beta = 0 $. In this case,
equations (\ref{eq:sys1}) and (\ref{eq:sys2}) are decoupled. These
two-dimensional perturbations can be considered nonmodal
Tollmien-Schlichting waves resulting from an optimization of the
initial conditions of (\ref{eq:sys1}) and (\ref{eq:sys2}). Therefore,
they display larger growth than modal Tollmien-Schlichting waves
resulting from the Orr-Sommerfeld equation. This shall be presented more
in detail in section \ref{sec:otherworks}. 
\item Oblique perturbations.\\
These are all remaining perturbations with $ \alpha \neq 0 $ and
$ \beta \neq 0 $. 
\end{itemize}
\vspace{2mm}
%and $ \exists $ 

\section{Results and discussion} \label{sec:results}

\subsection{Monotonic stability} \label{sec:absoluteStability}

\begin{figure}
  \centering
    \includegraphics[width=\textwidth]{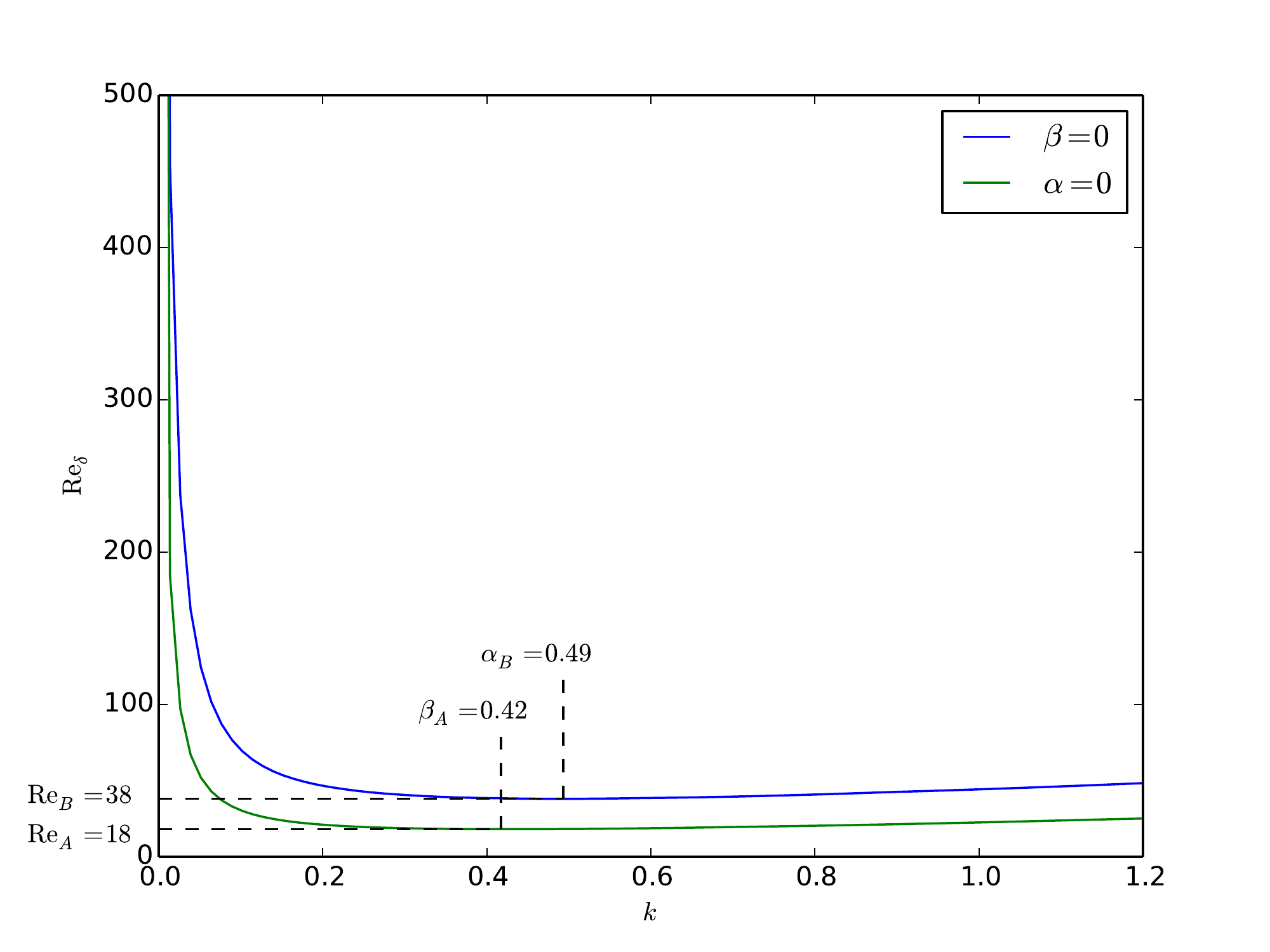}
    \caption{Isolines of $ \mu = 0 $ for the energy bound of \citet{DavisKerczek1973}, equations (\ref{eq:davis1a}) and (\ref{eq:davis2a}), as a function of
      the wave number $ k^2 = \alpha^2 + \beta^2 $ and the Reynolds number $ \Rey_\delta $. The blue and green lines correspond to the cases $ \beta = 0 $ and
$ \alpha = 0 $, respectively. All other cases have their critical Reynolds
number in the space between these lines.  }
    \label{fig:davisKerczek}
\end{figure}

In this section, 
we shall determine the critical Reynolds number $ \Rey_A $ 
behind which perturbations display growth. To this aim,
the energy criterion in \citet{DavisKerczek1973}
shall be used. We solve 
equations (\ref{eq:davis1a}) and (\ref{eq:davis2a}) for a given
pair of wave numbers $ (\alpha,\beta) $ and note the Reynolds number
$ \Rey_\delta $ for which the largest eigenvalue $ \mu $ changes from
minus to plus. At first, we compute the curves  of critical Reynolds numbers
$ \Rey_\delta(\alpha) $ and $ \Rey_\delta(\beta) $ by setting
$ \beta = 0 $ and $ \alpha = 0 $, respectively. These
curves are plotted in figure \ref{fig:davisKerczek}. 
As it turns out, all other cases, i.e. $ \alpha \neq 0 $ and $ \beta \neq 0 $,
have their critical Reynolds number lying in the region
between these two curves. From figure \ref{fig:davisKerczek}, we
can infer that the flow is monotonically stable for all Reynolds numbers
$ \Rey_\delta $ smaller than $ \Rey_A = 18 $. The physical significance
of this critical Reynolds number is, however, limited. For
example, the water depth of a surface solitary wave with
amplitude ratio $ \epsilon = 0.1 $ would be approximately $ 1 \, {\rm cm} $ for this case. For these small water depths, other physical effects,
such as capillary effects and not least the dissipative effect of the
boundary layers on the solitary wave, are not negligible anymore.
The solitary wave solution would thus not be valid in the first place.
%However,
%the present flow is an illustrative example 
%when considering the lower range of
%validity of Rayeleigh's inflexion point theorem or Fj\o rtoft's theorem
%for inviscid flows \citep{DrazinReid1981}. 
%Another example of such a flow is the related case of Stokes second problem,
%for which \citet{DavisKerczek1973} computed the corresponding
%critical Reynolds number.
From figure 
\ref{fig:davisKerczek}, we observe that streamwise streaks
will grow first. Two-dimensional perturbations, on the other hand,
can only grow for flows with a Reynolds number larger than $ \Rey_B = 38 $. 

\subsection{Optimal perturbation}

\subsubsection{Theoretical considerations}

Before turning to the computation of the amplification $ G $,
equation (\ref{eq:amplification}), we shall first consider
a scaling argument, as in \citet{Gustavsson1991,SchmidHenningson2001}. 
For streamwise streaks ($ \alpha = 0 $), equations (\ref{eq:sys1})
and (\ref{eq:sys2}) 
can be written as:
\bea
\left( \frac{\partial}{\partial t}  - \frac{1}{2} \mathcal{L}
\right) \mathcal{L} {w}
 &=& 0, \label{eq:sys1streak2} \\
\left( \frac{\partial}{\partial t} - \frac{1}{2} \mathcal{L}
\right) \tilde{\zeta} - {\rm i} \beta {w} \dz U_\base 
&=& 0, \label{eq:sys2streak2}
\eea
where 
$ \tilde{\zeta} $ is scaled by $ \Rey_\delta /2$:
\be
\tilde{\zeta} = \frac{2}{\Rey_\delta} \zeta(z,t) .
\ee
Equation (\ref{eq:sys1streak2}) corresponds to slow
viscous damping of $ w $, as also the homogeneous part of
equation (\ref{eq:sys2streak2}) for $ \tilde{\zeta} $. On the
other hand the second term in (\ref{eq:sys2streak2}) represents
a forcing term which varies on the temporal scale of
the outer flow. Therefore, streamwise streaks display
temporal variations on the time scale of the outer flow.
%From equations (\ref{eq:sys1streak2}) and (\ref{eq:sys2streak2}),
%we can infer that for streamwise streaks $ w $ and $ \zeta $ display
%temporal variations on a convective time scale. 
As for steady flows \citep{Gustavsson1991,SchmidHenningson2001},
the energy $ E_\zeta $ is proportional to the square of the
Reynolds number for the present unsteady flow:
\be
E_\zeta \propto \Rey_\delta^2. 
\ee
%This motivates the choice of $ \Rey_\delta $ by 
%\citet{OzdemirHsuBalachandar2013} in the present analysis. 
For
large Reynolds numbers $ E_\zeta $ will dominate. Therefore,
the maximum amplification $ G $ for streamwise streaks is expected to
behave as 
\be
\max_{\beta,t_0,t} G(\alpha=0,\beta,t_0,t,\Rey_\delta) \approx \Rey^2_\delta \quad \Rey_\delta >> 1. \label{eq:streakGrowth}
\ee
This quadratic growth of streamwise streaks can be contrasted to the
exponential growth of $ E_w $ for perturbations with $ \alpha > 0 $, as we shall see in the following. To
this aim, we use a decomposition (or integrating factor)
as in the parabolized stabiltiy equation \citep{BertolottiHerbertSpalart1992} for the normal velocity component $ w $:
\be
w = \tilde{w}(z,t) \exp \int \limits_{t_0}^t \omega(t') \, dt', \label{eq:pseDecomposition}
\ee
where the imaginary part of $ \omega $ accounts for
the oscillatory character of $ w $ and the real part
of $ \omega $ is the growth rate of the perturbation.
In order to define the shape function $ \tilde{w} $ univocally, all growth is restricted to $ \omega $. Somewhat different to \citep{BertolottiHerbertSpalart1992},
we define the normalization condition on the entire kinetic energy $ \tilde{E} $ of the shape function $ \tilde{w} $ :
\be
\tilde{E} = \frac{1}{2} \int \limits_0^\infty \frac{1}{k^2} | D\tilde{w} |^2 + | \tilde{w} |^2 \, dz,
\ee
where we have write $ D = \partial/ \partial z $. 
Thus, the normalization constraint on $ \tilde{w} $ is given by the following two conditions:
\be
 \int \limits_0^\infty \frac{\partial \tilde{w}^\dagger }{\partial t}  \mathcal{L} \tilde{w} \, dz
 = \int \limits_0^\infty  \tilde{w}^\dagger \mathcal{L}  \frac{\partial \tilde{w} }{\partial t} \, dz
 = 0 
\ee
From this, it follows, that we can define the energy of the shape function to be unity
for all times:
\be
\frac{\partial }{\partial t} \int \limits_0^\infty \tilde{w}^\dagger  \mathcal{L} \tilde{w} \, dz
=0
\quad \mbox{or} \quad \tilde{E} = 
-\frac{1}{2k^2}\int \limits_0^\infty \tilde{w}^\dagger \mathcal{L} \tilde{w} \, dz =  1.
\ee
Equation (\ref{eq:sys1}) becomes then:
\be
\partial_t \mathcal{L} \tilde{w} + \omega \mathcal{L} \tilde{w}
= \frac{1}{2} \mathcal{L}^2 \tilde{w} + {\rm i} \alpha \frac{1}{2} \Rey_\delta
\left( D^2 U_0 - U_0 \mathcal{L} \right) \tilde{w}
\ee
Multiplying by $ \tilde{w}^\dagger $ and integrating in $ z $, leads to
a formula for $ \omega $:
\be
\omega = -\frac{1}{4 k^2} \int \limits_0^\infty \tilde{w}^\dagger \mathcal{L}^2 \tilde{w} \, dz
- \frac{{\rm i} \alpha }{4 k^2} \Rey_\delta 
\int \limits_0^\infty \tilde{w}^\dagger D^2 U_\base \tilde{w} - \tilde{w}^\dagger U_\base \mathcal{L} \tilde{w} \, dz \label{eq:omegaPSE}
\ee
The growth rate, ie. the real part of $ \omega $, is given by:
\be
\omega_r = - \frac{1}{4 k^2}
\int \limits_0^\infty \mathcal{L} \tilde{w}^\dagger \mathcal{L} \tilde{w}
\, dz  + \Rey_\delta  \frac{\alpha}{4 k^2}
\int \limits_0^\infty D U_\base \left\{ \tilde{w}_r D \tilde{w}_i - \tilde{w}_i D \tilde{w}_r \right\} \, dz \label{eq:growth}
\ee
The first term on the right hand side represents viscous dissipation and is always negative. The
second term, however, can, depending on $ U_\base $ and $ \tilde{w} $, be positive or negative.
Only when this term is positive and in magnitude larger than the viscous dissipation,
growth of $ E_w $ can be observed. We observe that this term is multiplied
by $ \alpha/(\alpha^2 + \beta^2 ) $, which for a given $ \alpha $ is maximal for
$ \beta = 0 $. This indicates that the possible growth rate for two-dimensional
perturbations is larger than that for oblique perturbations when considering exponential growth in $ E_w $ and neglecting quadratic growth in $ E_\zeta $.
We shall return to this point, when discussing the numerical results. 
For the decomposition in equation (\ref{eq:pseDecomposition}), the continuity
equation can be written as:
\be
{\rm i} \alpha \tilde{u} + {\rm i} \beta \tilde{v} = - D \tilde{w},
\ee
where we have normalized the horizontal velocities:
\be
\tilde{u} = u \exp - \int \limits_{t_0}^{t} \omega \, dt',
\quad \tilde{v} = v \exp - \int \limits_{t_0}^{t} \omega \, dt'.
\ee
Then the growth rate $ \omega_r $,
equation (\ref{eq:growth}), can be written as:
\bea
\omega_r& =& - \frac{1}{4 k^2}
\int \limits_0^\infty | \mathcal{L} \tilde{w}|^2
\, dz  - \frac{ \Rey_\delta}{4}
\int \limits_0^\infty  \left( \tilde{u}^\dagger_k , \tilde{w}^\dagger \right)
\mathbf{S}_k
 \left( \begin{array}{c} \tilde{u}_k \\ \tilde{w} \end{array} \right),
\eea
where $ \tilde{u}_k $ is the projection of the horizontal velocity vector onto the
wavenumber vector $ \mathbf{k} = ( \alpha , \beta ) $,
\be
\tilde{u}_k = \frac{1}{k} \left( \alpha \tilde{u} + \beta \tilde{v} \right),
\ee
and $ \mathbf{S}_k $, the two dimensional rate of strain
tensor of the projection of the base flow on the wavenumber vector
$ \mathbf{k} $:
\be
\mathbf{S}_k = \frac{1}{2} \left( \begin{array}{cc} 0 &  D U_k \\ D U_k  & 0
\end{array} \right), \quad
U_k = \frac{1}{k} \alpha U_\base.
\ee
When considering two-dimensional perturbations ($ \beta = 0 $), the growth rate $ \omega_r $ simplifies to
\bea
\omega_r& =& - \frac{1}{4 \alpha^2}
\int \limits_0^\infty | \mathcal{L} \tilde{w}|^2
\, dz  - \frac{ \Rey_\delta}{4}
\int \limits_0^\infty  \left( \tilde{u}^\dagger , \tilde{w}^\dagger \right)
\mathbf{S}_{2D}
 \left( \begin{array}{c} \tilde{u} \\ \tilde{w} \end{array} \right), \label{eq:tilting}
\eea
where the $ \mathbf{S}_{2D} $ is the two-dimensional rate of strain tensor
of the base flow:
\be
\mathbf{S}_{2D} = \frac{1}{2} \left( \begin{array}{cc} 0 & D U_\base \\ D U_\base & 0 \end{array} \right).
\ee
In this case (ie. $\beta = 0$), equations (\ref{eq:sys1}) and (\ref{eq:sys2}) are decoupled. As can
be seen from equation (\ref{eq:sys2}),
the normal vorticity $ \zeta $ experiences only dampening. Growth
can, therefore, only arise in the energy $ E_w $
associated to the normal velocity component $ w $, equation (\ref{eq:energyNonmodal}). As mentioned above, the first term on the right hand side in equation (\ref{eq:tilting}) is always negative and
represents the viscous dissipation stabilizing the flow. As the eigenvalues
of $ \mathbf{S}_{2D} $ are given by $ D U_\base/2 $ and $ - D U_\base/2 $, the second term on
the right hand side in equation (\ref{eq:tilting}) can, depending on $ \tilde{w} $, be positive
or negative. All possible growth of two-dimensional perturbations is thus due
to the second term where
the velocity vector $ \left( \tilde{u} , \tilde{w} \right)^T $ is being tilted by the
rate of strain tensor $ \mathbf{S}_{2D} $. Equation (\ref{eq:tilting}) is
an illustrative formula for the Orr-mechanism. The growth mechanism itself is thus always inviscid. This
holds for any two-dimensional perturbation, also those being the eigenfunctions
of the Orr-Sommerfeld equation, the modal Tollmien-Schlichting waves, which
are commonly thought of as slow viscous instabilities, cf. for example
\citep{Jimenez2013} and \citep{BrandtSchlatterHenningson2004}.
Whether growth of two-dimensional perturbations is fast or slow is, as
formula (\ref{eq:tilting}) suggests, primarily
a property of the base flow profile $ U_\base $. As we shall see below, 
velocity profiles having an inflection point allow for larger growth rates
than profiles without.\\

As the Reynolds number multiplies the second term in
equation (\ref{eq:tilting}), we can conclude that
for large $ \Rey_\delta $, the maximum amplification
of two-dimensional perturbations roughly behaves like:
\be
\max_{\alpha,t_0,t} G(\alpha,\beta=0,t_0,t,\Rey_\delta) \approx
e^{c \Rey_\delta}, \quad \Rey_\delta >> 1 \label{eq:TSgrowth}
\ee
where $ c $ is some constant. This exponential growth of the maximum amplification
with the Reynolds number has also been observed for other flows displaying an adverse
pressure gradient. For example, \citet{Biau2016} observed that the maximum amplification
of two-dimensional perturbations for Stokes' second problem grows exponentially
with the Reynolds number.

In the following, we
shall see that the
competition of
the maximum amplification
between the quadratic growth in $ \Rey_\delta $ 
of streamwise streaks, equation (\ref{eq:streakGrowth}), and
the exponential growth in $ \Rey_\delta $ of two-dimensional structures, 
equation (\ref{eq:TSgrowth}), composes the essential primary instability 
mechanism of this flow. \\

\subsubsection{Numerical results} \label{sec:numericalResults}

\begin{sidewaysfigure}%[ht]
  \vspace*{13cm}
  \centering
%  \begin{minipage}{\textwidth}
  \includegraphics[height=0.6\textheight]{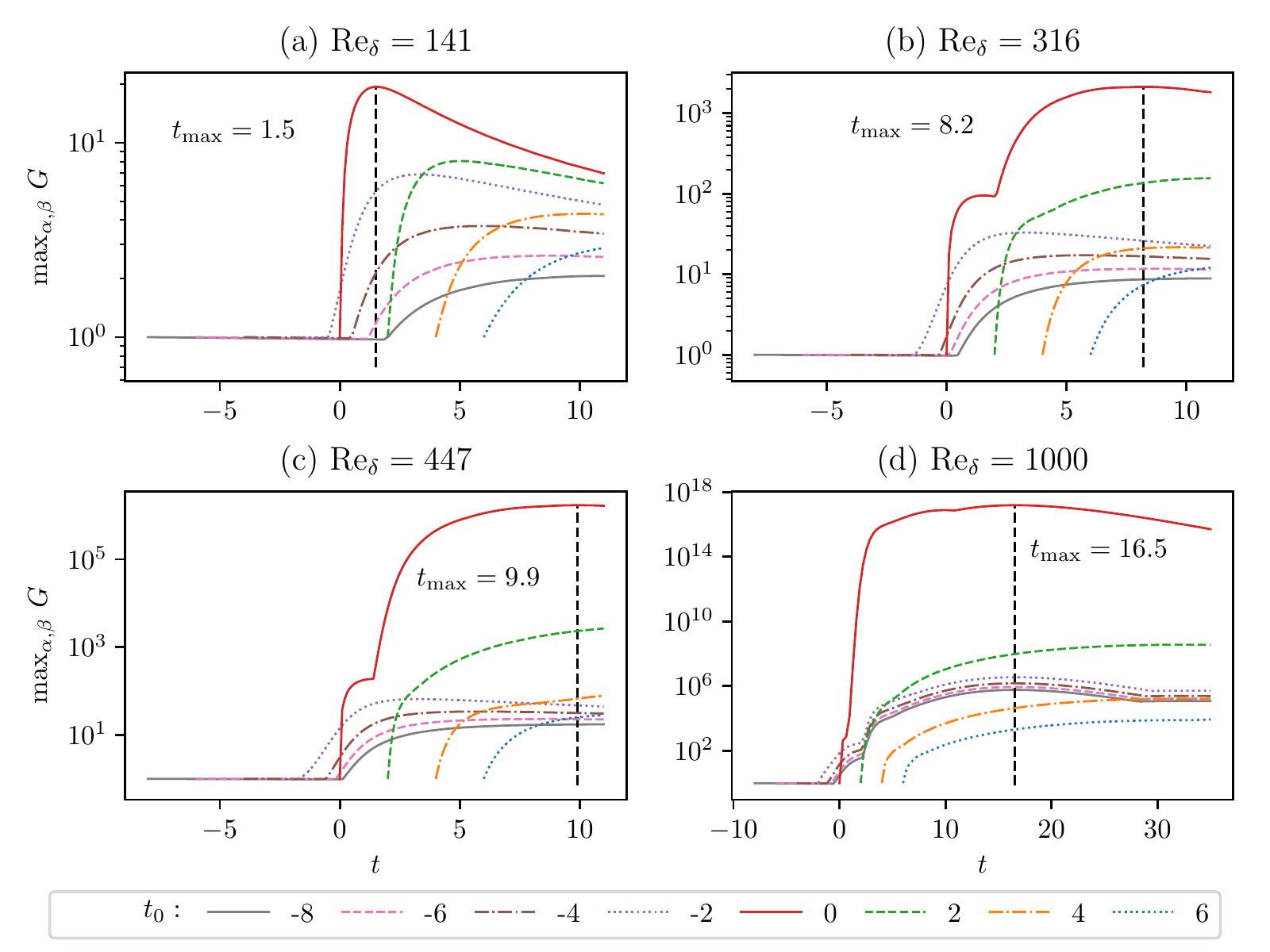}\\
  \caption{Temporal evolution of the amplification $ G $ maximized 
    over the wavenumbers $ \alpha $ and $ \beta $ for different
    Reynolds numbers $ \Rey_\delta $ and initial times $ t_0 $.}
  \label{fig:temporalEvolution}
%  \end{minipage}
\end{sidewaysfigure}

The amplification $ G $, equation (\ref{eq:amplification}),
for the present flow problem depends on five parameters,
the wavenumbers $ \alpha $ and $ \beta $, the initial time $ t_0 $,
the time $ t $ and the Reynolds number $ \Rey_\delta $. 
We start our numerical analysis by tracing the evolution of $ \max_{\alpha,\beta} G $
for a given Reynolds number $ \Rey_\delta $ and a given initial time $ t_0 $. 
In figure \ref{fig:temporalEvolution}, we plot the temporal
evolution of $ \max_{\alpha,\beta} G $ for the Reynolds numbers $ \Rey_\delta = 141, 316 , 447$ and $ 1000 $
($ \Rey_\sumer = 10^4 , 5\cdot 10^4 , 10^5 , 5\cdot 10^5 $)
and initial times $ t_0 = -8, -6, \ldots, 6 $. For the 
case $\Rey_\delta =141 $, cf. figure \ref{fig:temporalEvolution}a, 
we observe
that growth of perturbations is mainly restricted to the deceleration region
of the flow, i.e. where $ t > 0 $. Only the optimal perturbation starting
at $ t_0 = -2 $ displays some growth before the arrival
of the crest of the solitary wave. 
Among the initial conditions $ t_0 $ chosen, the optimal perturbation
with $ t_0 =0 $ displays the maximum amplification at $ t_{\max} = 1.5 $
with $ G \approx 20 $. 
This is due to the acceleration region of the flow ($ t < 0 $) having 
a damping effect on the perturbations starting before $ t = 0 $. On the
other hand the perturbations starting at later times $ t_0 \ge 2 $
already miss out a great deal of the destabilizing effect of the adverse
pressure gradient. All curves display a maximum at some time.
For some cases, this maximum lies outside of the plotting domain. 
For a slightly larger Reynolds number, cf. figure \ref{fig:temporalEvolution}b
with $ \Rey_\delta = 316 $, we observe a qualitatively similar behavior
for the perturbations starting at $ t_0 < 0 $ with the 
difference that growth of these perturbations
sets in somewhat earlier in time than in the $ \Rey_\delta = 141 $ case and leads
also to higher amplifications. However, the optimal perturbation
starting at $ t_0 = 0 $ behaves differently than the corresponding
one for the $ \Rey_\delta = 141 $ case. At early times, i.e. for $ t \lesssim 2 $, 
the evolution
of this perturbation is similar to the $ \Rey_\delta = 141 $
case. The perturbation grows to a maximum $ G \approx 100 $ at
$ t \approx 1.5 $,
before decaying again, but, at time $ t \approx 2 $, the 
amplification curve displays 
a kink and a sudden growth to $ G \approx 2000 $ at time $ t_{\max} = 8.2 $. 
A similar, however, less expressive kink is also visible in the curve for
$ t_0 = 2 $. Increasing the Reynolds number to $ \Rey_\delta = 447 $,
cf. figure \ref{fig:temporalEvolution}c, does not change the picture
qualitatively. However, the maximum amplification of 
the optimal perturbation starting
at $ t_0 = 0 $ has increased
by a factor of approximately thousand
compared to the $ \Rey_\delta = 316 $ case. In comparison,
the maximum of
the optimal perturbation starting at $ t_0 = -2 $ has
only increase by a factor of approximately 1.25 when going
from $ \Rey_\delta = 316 $ to $ \Rey_\delta = 447 $. 
This violent growth for the optimal perturbation starting at $ t_0 $ is
also visible for the $ \Rey_\delta = 1000 $ case, cf. figure 
\ref{fig:temporalEvolution}d. However, for this case,
even the curves of
the perturbations
starting at earlier times display a similar kink and sudden growth
in the deceleration region. 

\begin{sidewaysfigure}
  \vspace*{13cm}
  \newcommand{\scl}{0.42}
  \centering
  \includegraphics[width=\scl\textwidth]{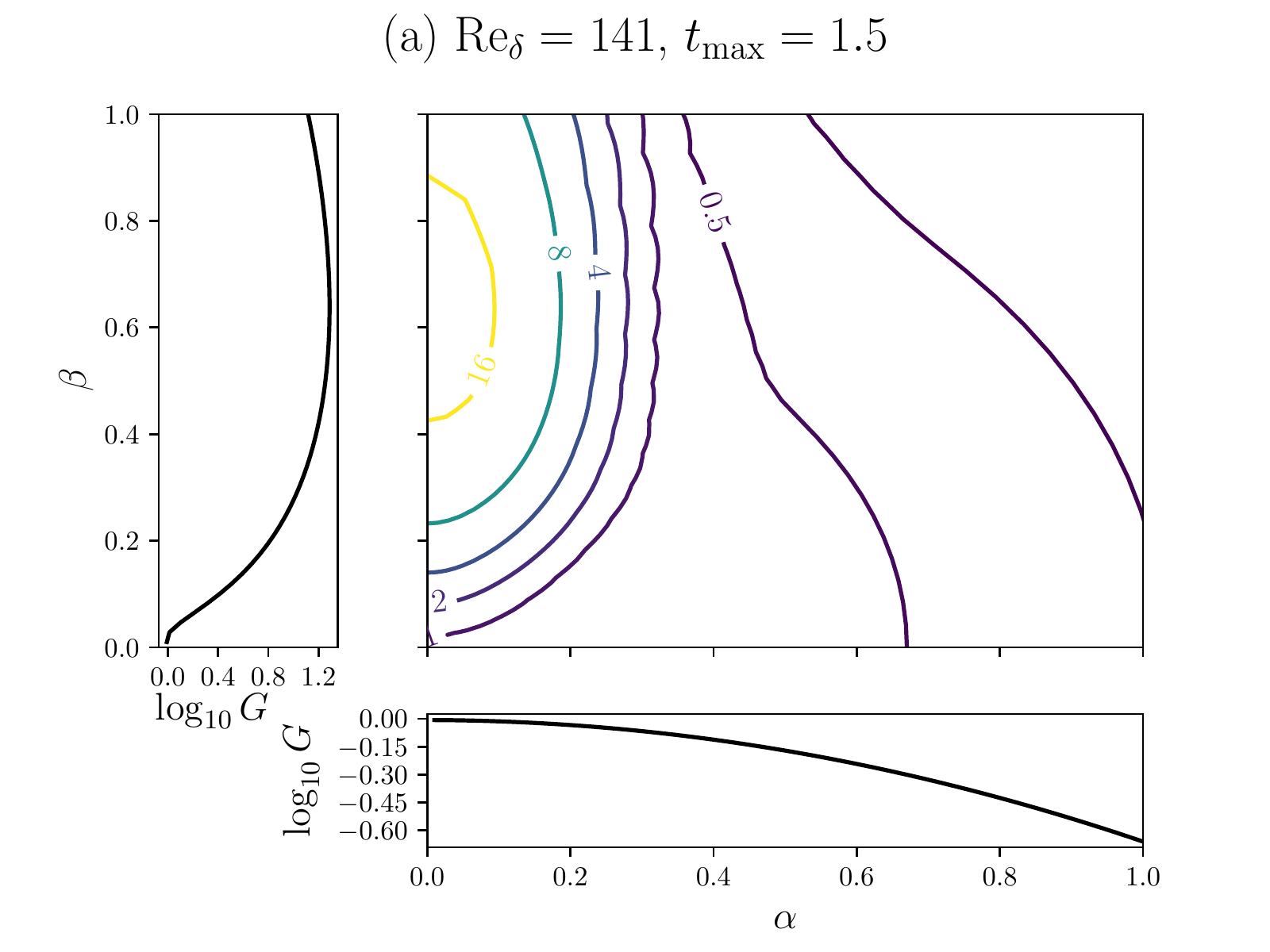}
  \includegraphics[width=\scl\textwidth]{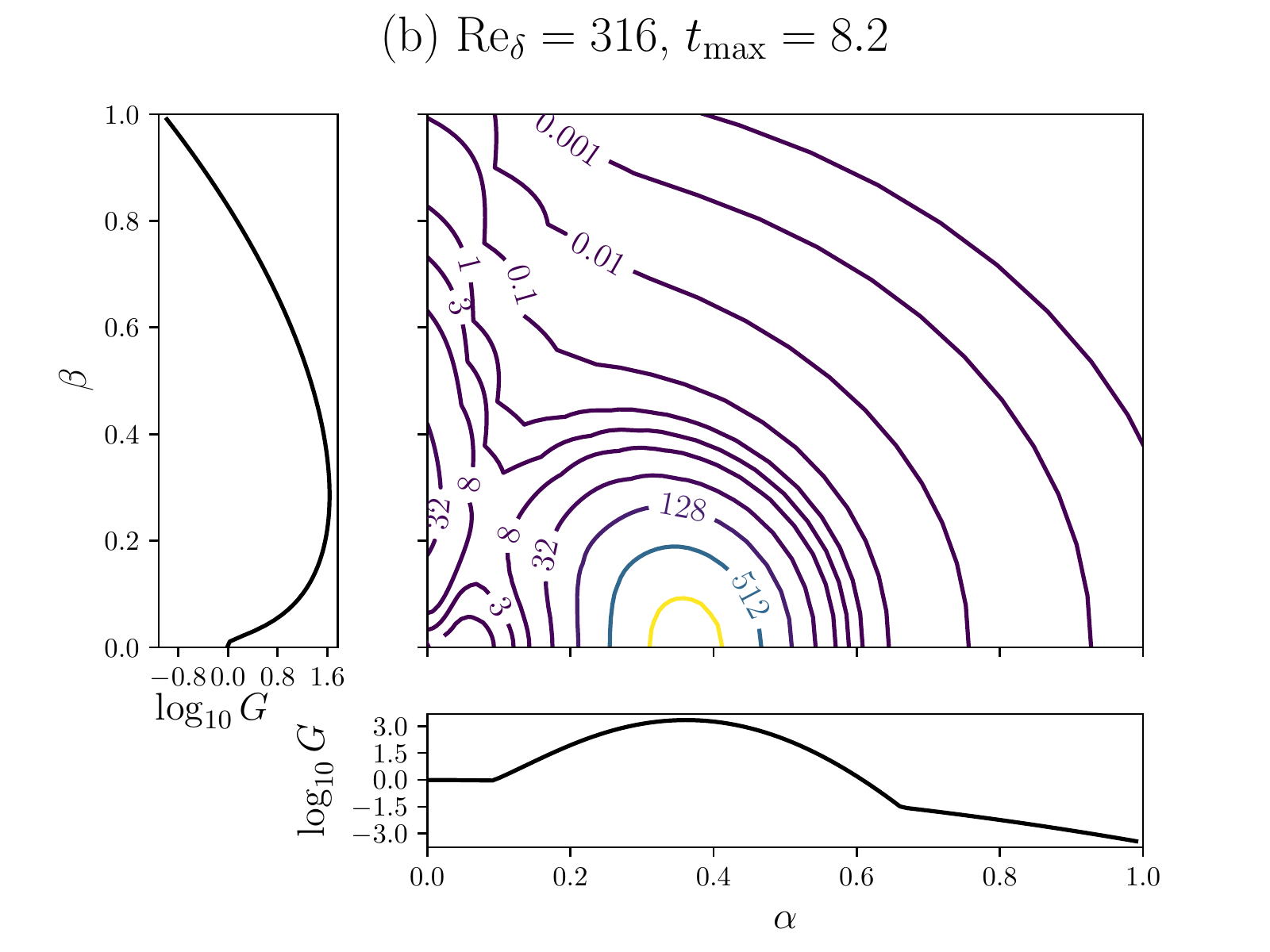}\\
  \includegraphics[width=\scl\textwidth]{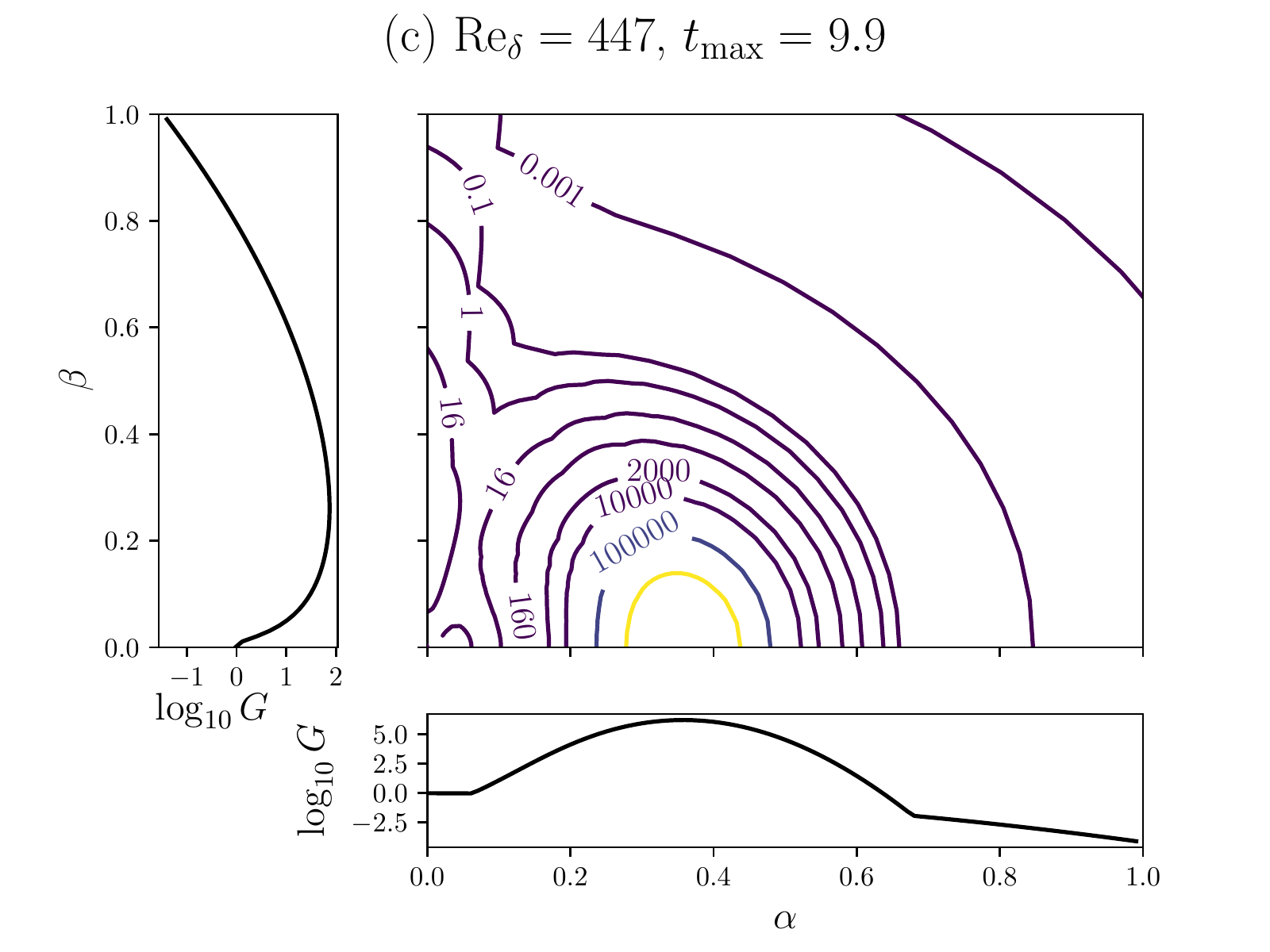}
  \includegraphics[width=\scl\textwidth]{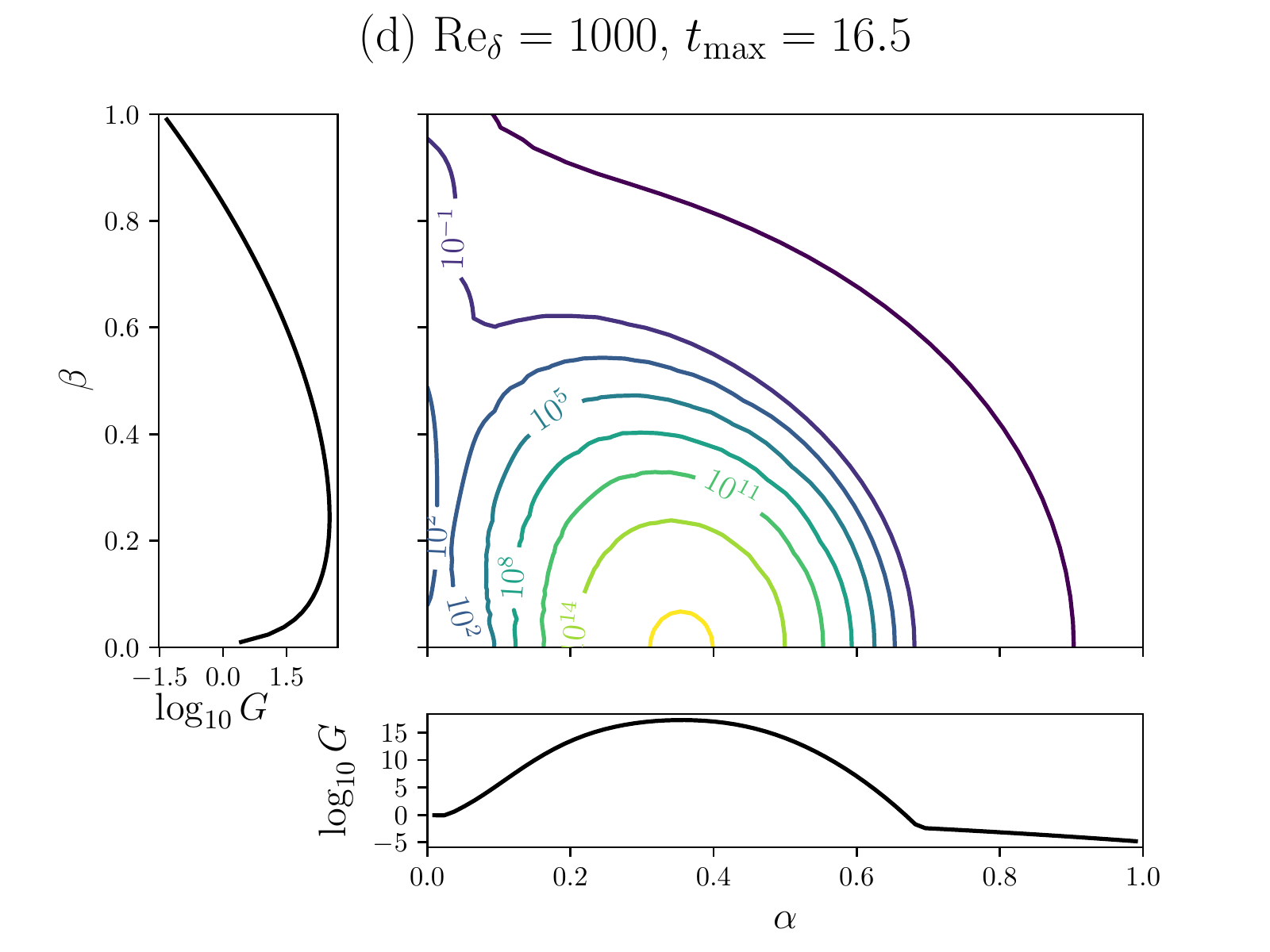}
  \caption{Contour plots of the amplification 
    $ G(\alpha,\beta,t_0 = 0,t_{\max} , \Rey_\delta )$ at $ t_{\max} = 1.5, 8.2 , 9.9 , 16.5 $
    for the cases $ \Rey_\delta = 141, 316, 447 , 1000 $, respectively. The plots to the left and below the contour plot show a slice along the $ \beta $- and
$ \alpha $-axes, respectively.}
\label{fig:contour}
\end{sidewaysfigure}

In figure \ref{fig:contour}, we show
contour plots of the amplification $ G(\alpha,\beta,t_0 = 0,t_{\max} , \Rey_\delta )$ at $ t_{\max} = 1.5, 8.2 , 9.9 , 16.5 $
for the cases $ \Rey_\delta = 141, 316, 447 , 1000 $, respectively. 
For the case $ \Rey_\delta = 141 $, cf. figure \ref{fig:contour}a, 
we find a single
maximum lying on the $ \beta $-axis. On the other hand,
the $ \Rey_\delta = 316 $ case
is different, cf. figure \ref{fig:contour}b. Whereas
all two-dimensional perturbations display decay at $ t_{\max} = 1.5 $ for
the $ \Rey_\delta = 141 $ case, the
amplification of two-dimensional perturbations displays a peak at around
$ \alpha = 0.35 $ for the $ \Rey_\delta = 316 $ case. A second peak,
lying on the $ \beta $ axis, is significantly smaller than the 
peak of two-dimensional perturbations on the $ \alpha $-axis. Increasing
the Reynolds number, cf. figures \ref{fig:contour}c and \ref{fig:contour}d, 
increases the magnitude of the peaks, with the peak on the $\alpha$-axis
growing faster with $ \Rey_\delta $ than the peak on the $ \beta$-axis. 
This competition between streamwise streaks and two-dimensional structures
is characteristic for flows with adverse pressure gradients
and has also been observed for steady flows. 
The Falkner-Skan boundary layer with adverse pressure
gradient displays contour levels similar
to the present ones, cf. for example \citet[figure 10d]{LevinHenningson2003}
or \citet{CorbettBottaro2000}. Another example is the flow of
three dimensional swept boundary layers investigated in 
\citet{CorbettBottaro2001}. 

\begin{figure}
  \centering
    \includegraphics[width=\textwidth]{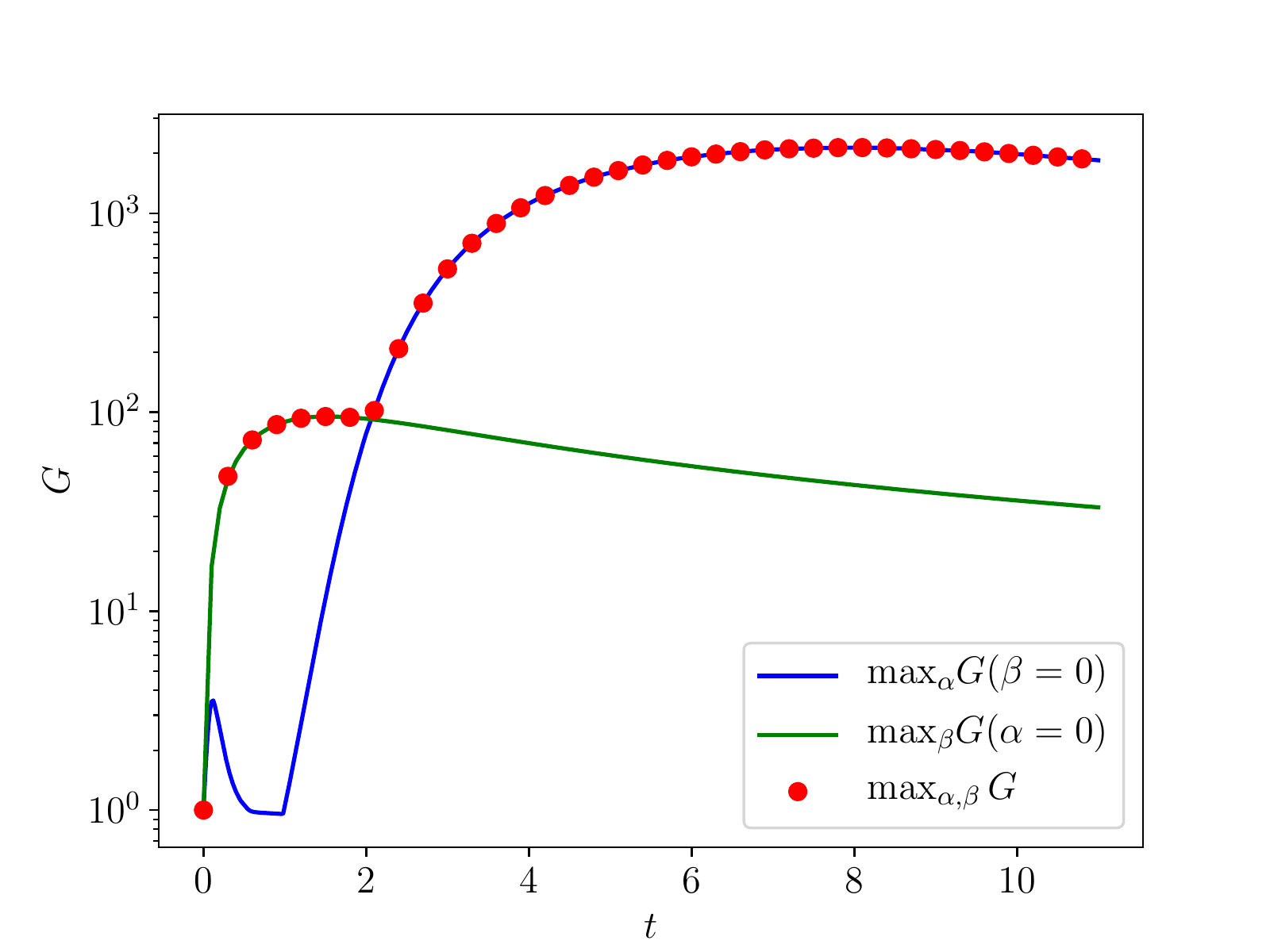}
    \caption{ Temporal evolution
of $ \max_\beta G(\alpha=0,\beta,t_0 = 0 , t , \Rey_\delta = 316) $,
 $ \max_\alpha G(\alpha,\beta=0,t_0 = 0 , t , \Rey_\delta = 316) $
and  $ \max_{\alpha,\beta} G(\alpha,\beta,t_0 = 0 , t , \Rey_\delta = 316) $. }
    \label{fig:temporalEvolutionStreakVSTS}
\end{figure}

\begin{figure}
  \centering
    \includegraphics[width=\textwidth]{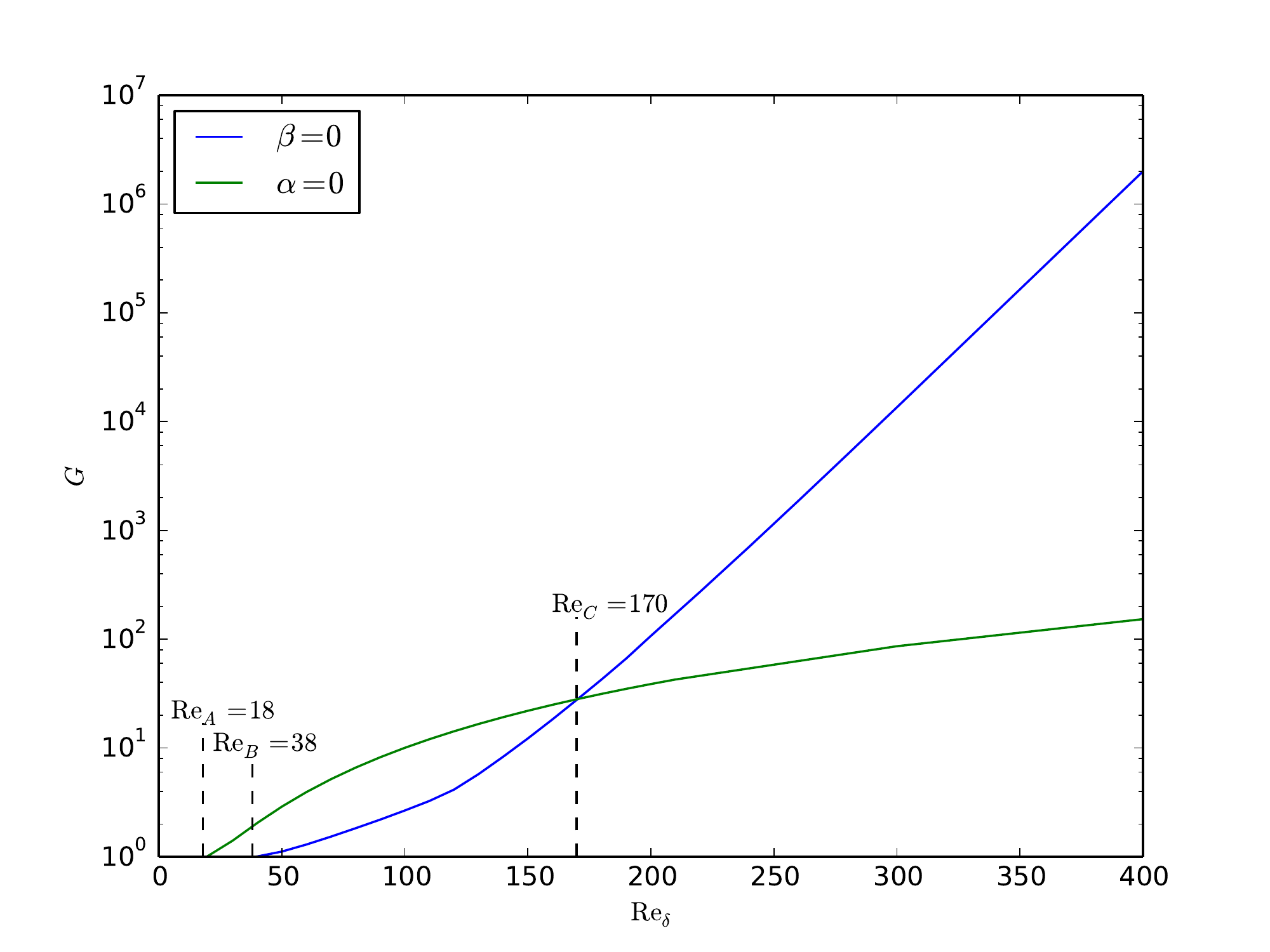}
    \caption{Maximum amplification of streamwise streaks $ \max_{\beta,t_0,t} G(\alpha=0,\beta,t_0,t,\Rey_\delta) $ and two-dimensional perturbations 
$ \max_{\alpha,t_0,t} G(\alpha,\beta=0,t_0,t,\Rey_\delta) $.}
    \label{fig:maxAmplificationStreaksVSTS}
\end{figure}

The competition between streamwise streaks
and two-dimensional perturbations can also be observed in the 
temporal evolution of the amplification of the optimal perturbation. 
In figure \ref{fig:temporalEvolutionStreakVSTS}, we
compare the temporal evolution
of $ \max_\beta G(\alpha=0,\beta,t_0 = 0 , t , \Rey_\delta = 316) $,
 $ \max_\alpha G(\alpha,\beta=0,t_0 = 0 , t , \Rey_\delta = 316) $
and  $ \max_{\alpha,\beta} G(\alpha,\beta,t_0 = 0 , t , \Rey_\delta = 316) $. 
For early times ($ 0 < t \lesssim 2 $) 
the streamwise streaks display a larger amplification
than the two-dimensional perturbations, but at time $ t \approx 2 $, 
the two-dimensional perturbations overtake the streaks. Maximizing over $ \alpha $ and $ \beta $,
chooses either perturbation displaying maximum amplification. 
%It is remarkable that 
The amplification of oblique perturbations seems
to be most often smaller than that of streamwise streaks or two-dimensional
perturbations. This allows us to trace the maximum amplification $ G_{\max} $,
equation (\ref{eq:maximumA}), by considering
only the amplification of the cases $ (\alpha = 0,\beta) $ and
 $ (\alpha,\beta = 0) $ 
instead of maximizing over all possible wave numbers $ (\alpha,\beta)$. 
Growth of streamwise streaks is associated to
the lift-up effect \citep{EllingsenPalm1975}, whereas the growth of
two-dimensional perturbations is associated to the Orr-mechanism
\citep{Jimenez2013}. We remark that
other growth mechanisms exists, such as the Reynolds stress mechanism,
cf. \citet{ButlerFarrell1992}, which can lead to the maximum amplification
of streaks not being exactly on the $ \beta $ axis, 
but having a non-zero $ \alpha $-component. However, as also shown for
other flows \citep{ButlerFarrell1992}, this $ \alpha $-component 
is negligibly small and, therefore, not considered in the present
treatise. 
In figure \ref{fig:maxAmplificationStreaksVSTS}, the amplification
of streamwise streaks and two-dimensional perturbations maximized over
the initial time $ t_0 $ and time $ t $ is plotted against the Reynolds number.
As predicted in section \ref{sec:absoluteStability}
by the energy bound of \citet{DavisKerczek1973}, streamwise streaks start
to grow for Reynolds numbers larger than $ \Rey_A = 18 $,
whereas two-dimensional perturbations start growing
for $ \Rey_B > 38 $. We can define a third critical Reynolds number
$ \Rey_C = 170 $ for this flow, which stands for
the value when the maximum amplification
of two-dimensional perturbations overtakes the maximum amplification
of streamwise streaks. This happens for rather low levels of
amplification, the maximum amplification being $ G_{\max} = 28 $ for
$ \Rey_\delta = 170 $. As in \citet{Biau2016} for Stokes second problem,
the amplification of two-dimensional perturbations is observed
to be exponential. For flows with a Reynolds number
larger than $ \Rey_C $, which are most relevant cases, the dominant
perturbations are therefore likely to be two-dimensional (up to secondary instability). 
This supports the observation by \citet{VittoriBlondeaux2008} and
\citet{OzdemirHsuBalachandar2013} of a transition process
via the development of two-dimensional vortex rollers. However,
when starting early, i.e. for initial times $ t_0 < -1 $,
streamwise streaks start growing before
two-dimensional structures, as can be seen in figure \ref{fig:temporalEvolution}d.
The competition between streamwise streaks and two-dimensional
structures to first reach secondary instability, might therefore
not only be determined by the maximum amplification reached,
but also by
the point in time, when the amplification of the perturbation
is sufficient to trigger secondary instability, be it streaks
or two-dimensional perturbations. We shall discuss this point
further in section \ref{sec:otherworks}. 

\begin{figure}
  \centering
    \includegraphics[width=\textwidth]{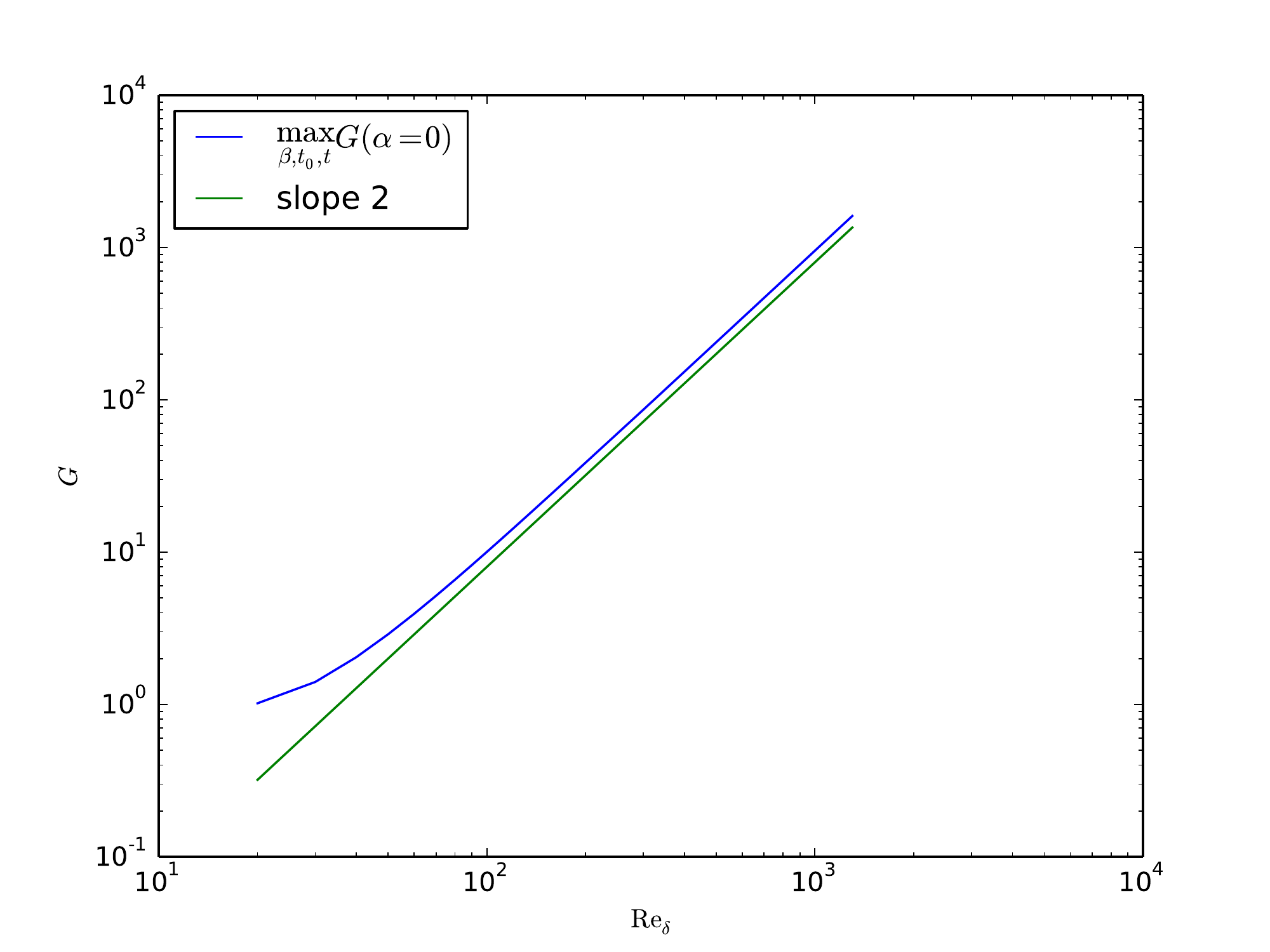}
    \caption{Maximum amplification of streamwise streaks, $ \max_{\beta,t_0,t} G(\alpha=0,\beta,t_0,t,\Rey_\delta) $, versus Reynolds number.}
    \label{fig:loglogPlot}
\end{figure}

\begin{figure}
  \centering
  \begin{subfigure}[b]{0.5\textwidth}
    \includegraphics[width=\textwidth]{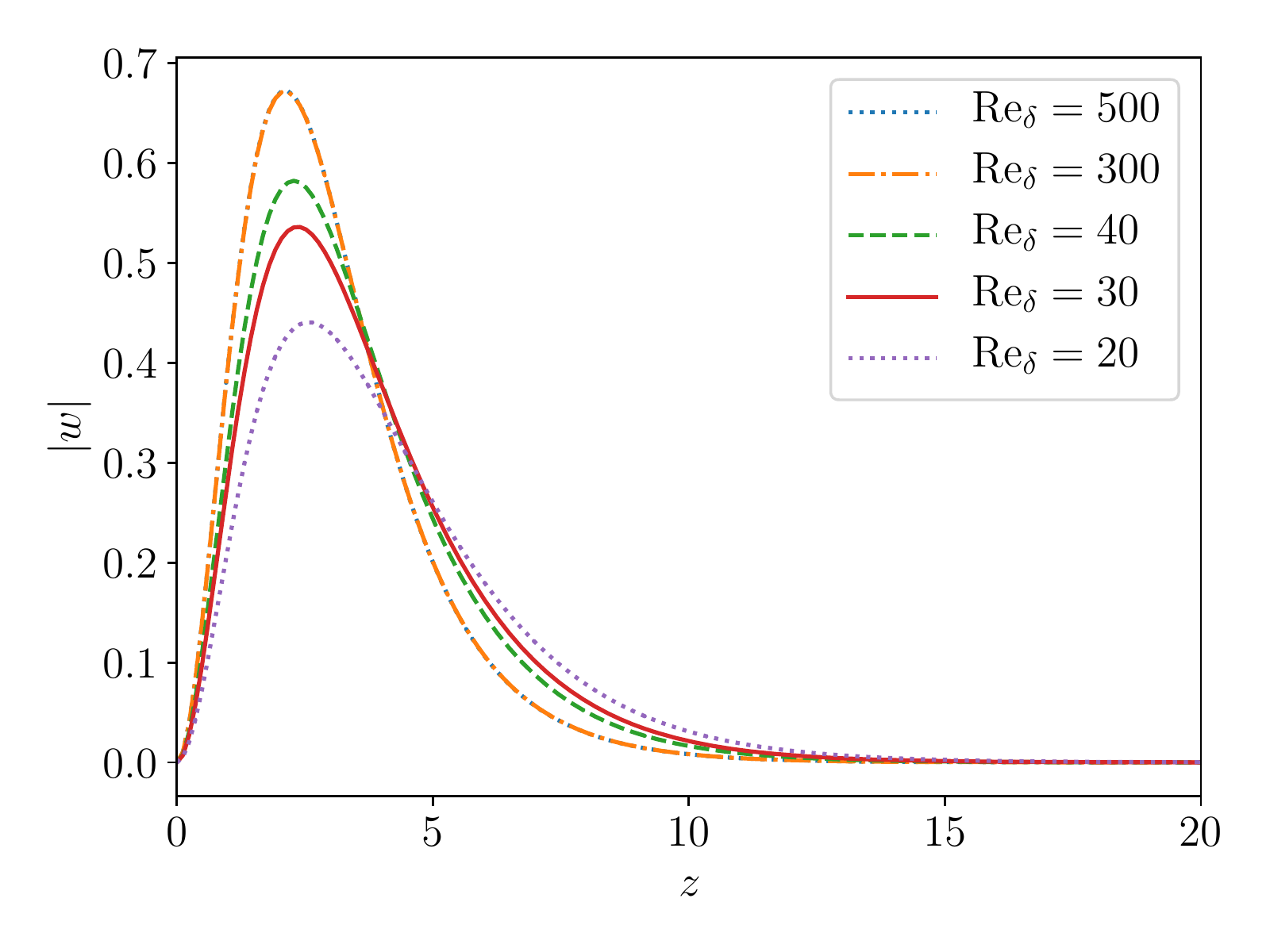}
    \caption{ $ w $ }
    \label{fig:contourB}
  \end{subfigure} \nolinebreak
  \begin{subfigure}[b]{0.5\textwidth}
    \includegraphics[width=\textwidth]{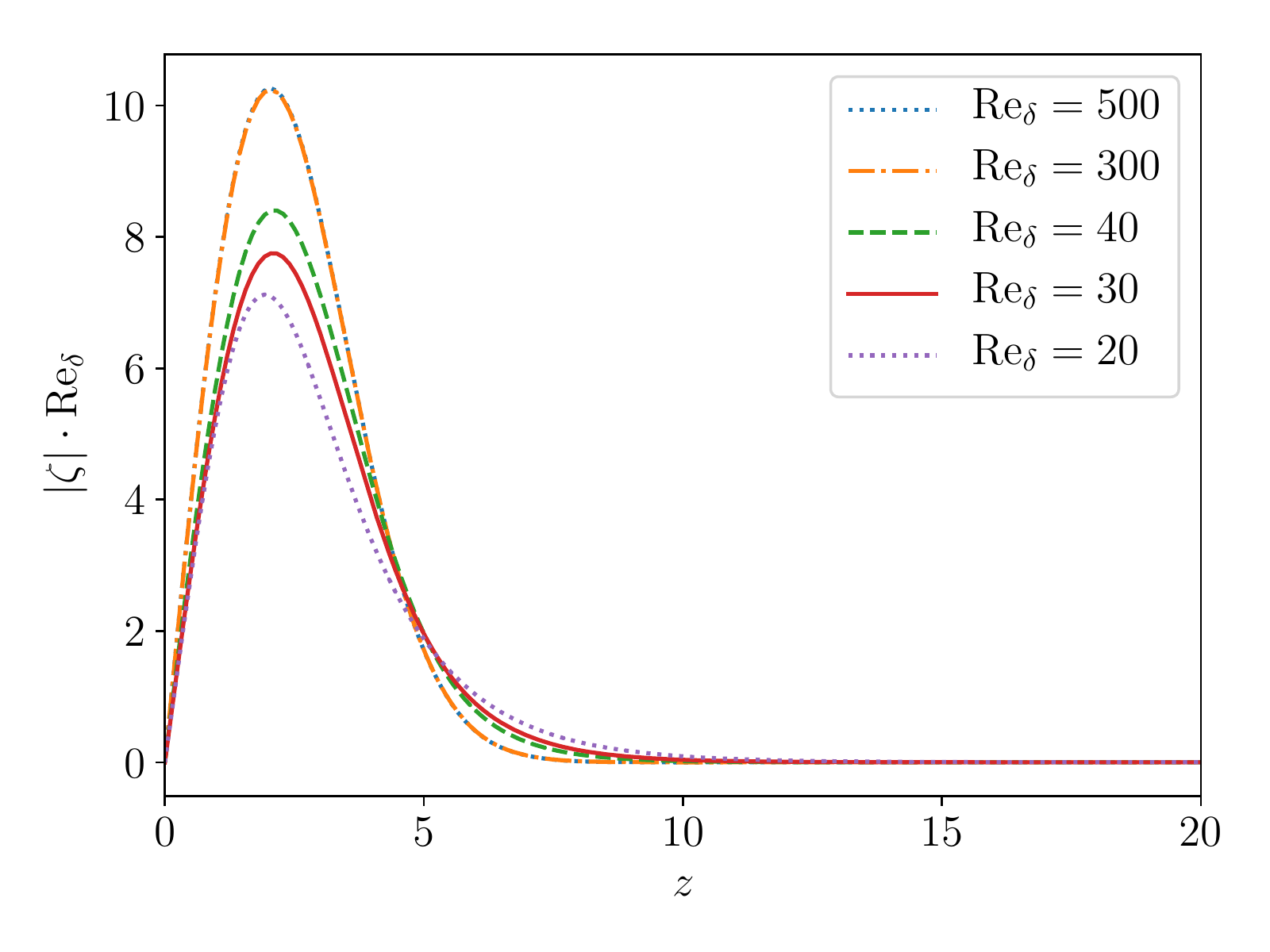}
    \caption{ $ \zeta $ }
    \label{fig:contourC}
  \end{subfigure} 
  \caption{Initial condition for the streamwise streak with maximum amplification,
    $ \max_{\beta,t_0,t} G(\alpha=0,\beta,t_0,t,\Rey_\delta) $, for
    different Reynolds numbers.}
  \label{fig:initialConditionStreak}
\end{figure}

\begin{figure}
  \centering
  \begin{subfigure}[b]{0.5\textwidth}
    \includegraphics[width=\textwidth]{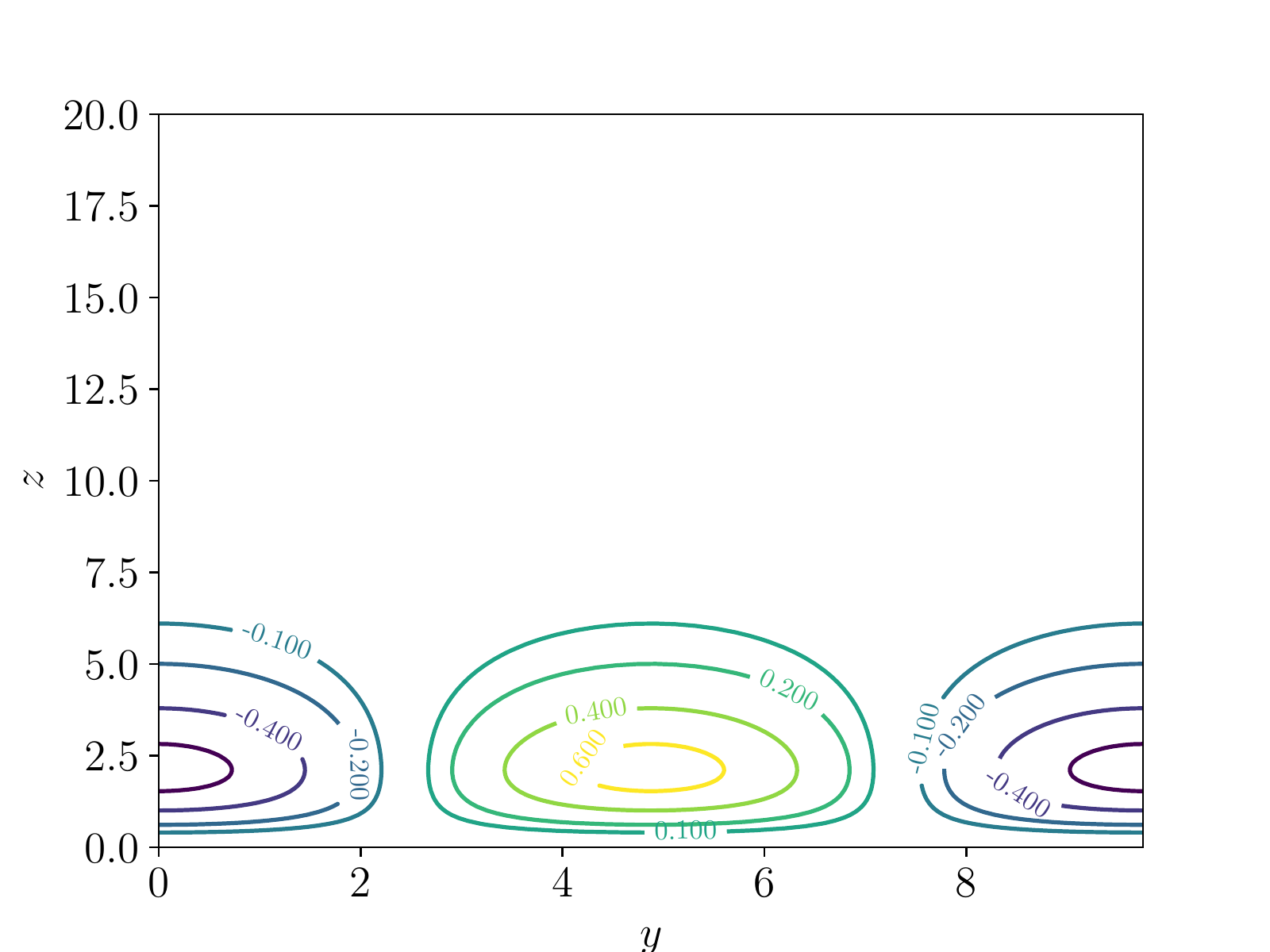}
    \caption{ $ w(z,t_0) \cdot \exp {\rm i} \beta y $ }
    \label{fig:contourB}
  \end{subfigure} \nolinebreak
  \begin{subfigure}[b]{0.5\textwidth}
    \includegraphics[width=\textwidth]{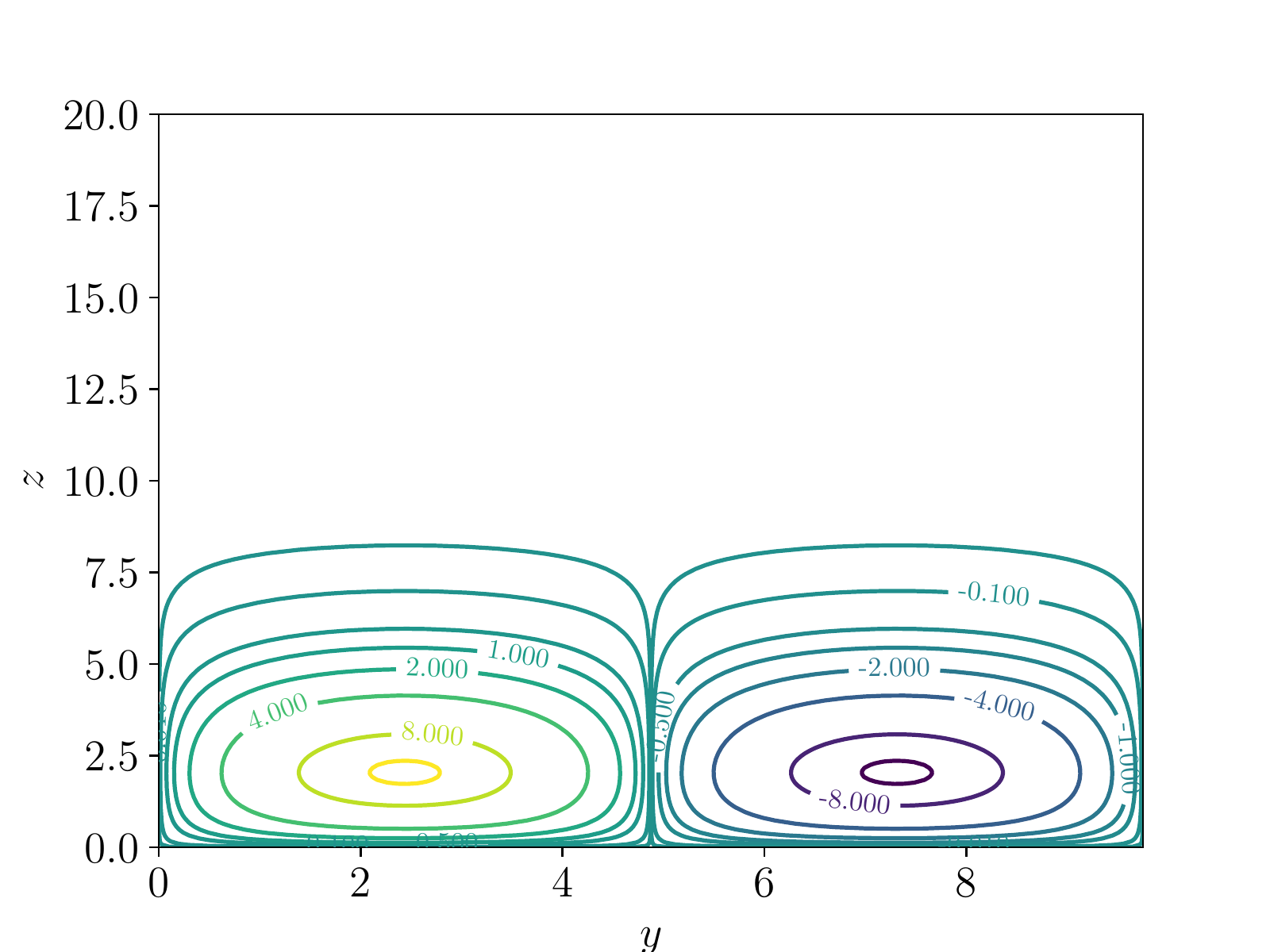}
    \caption{ $ \zeta(z,t_0) \cdot \exp {\rm i} \beta y \cdot \Rey $ }
    \label{fig:contourC}
  \end{subfigure} 
  \caption{Contour plots of the real part of $ w(z,t_0) \cdot \exp {\rm i} \beta y $
    and the real part of $ \zeta(z,t_0) \cdot \exp {\rm i} \beta y \cdot \Rey $,
    which are the initial condition at $ t_0 $
    for the optimal perturbation for the case $ \Rey = 500 $, $ \beta_{\max} = 0.64 $, $ t_0 = 0.11 $, $ t = 1.53 $.}
  \label{fig:initialConditionStreakContour}
\end{figure}

\begin{figure}
  \centering
  \begin{subfigure}[b]{0.5\textwidth}
    \includegraphics[width=\textwidth]{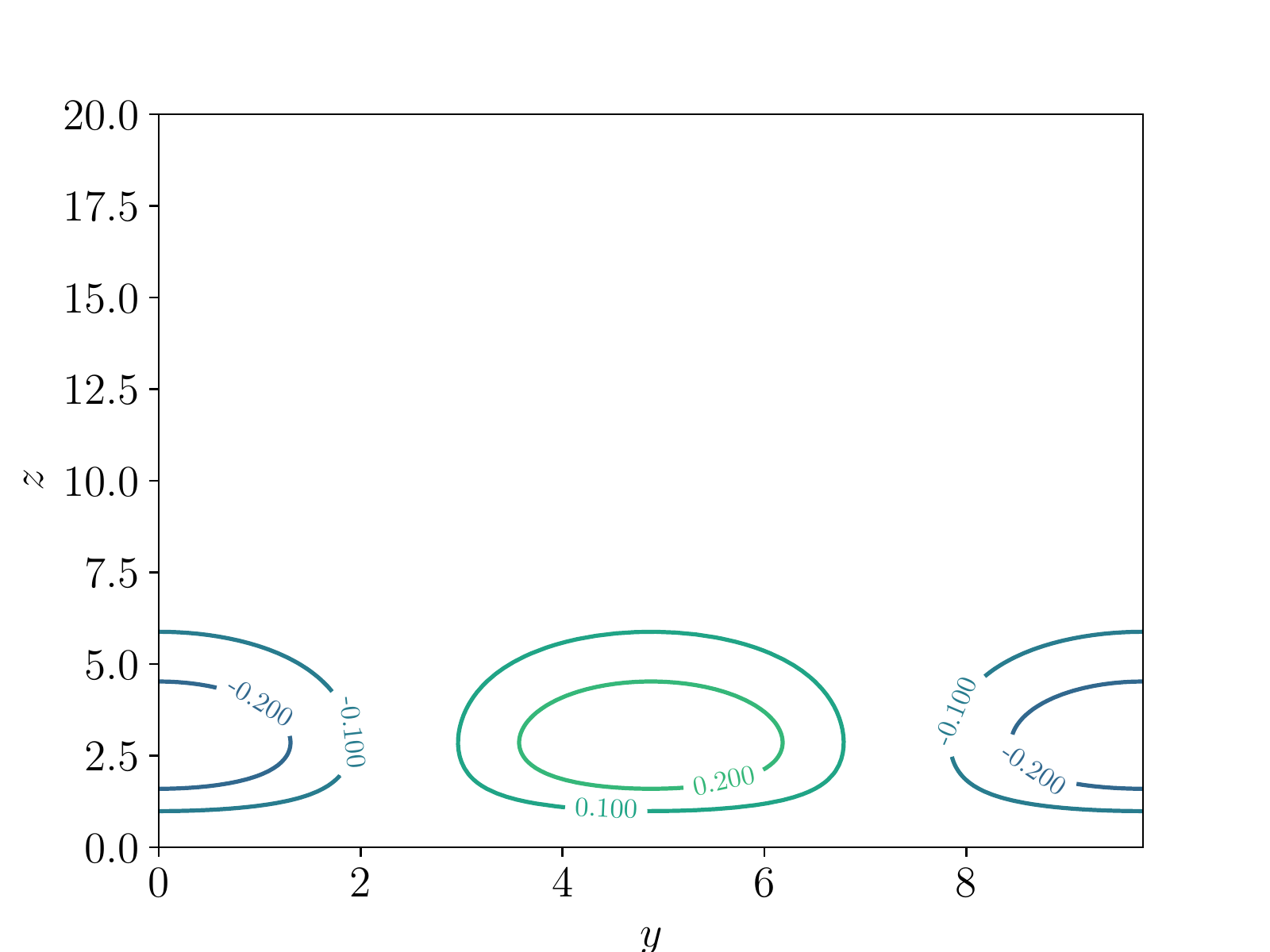}
    \caption{ $ w(z,t) \cdot \exp {\rm i} \beta y $ }
    \label{fig:contourB}
  \end{subfigure} \nolinebreak
  \begin{subfigure}[b]{0.5\textwidth}
    \includegraphics[width=\textwidth]{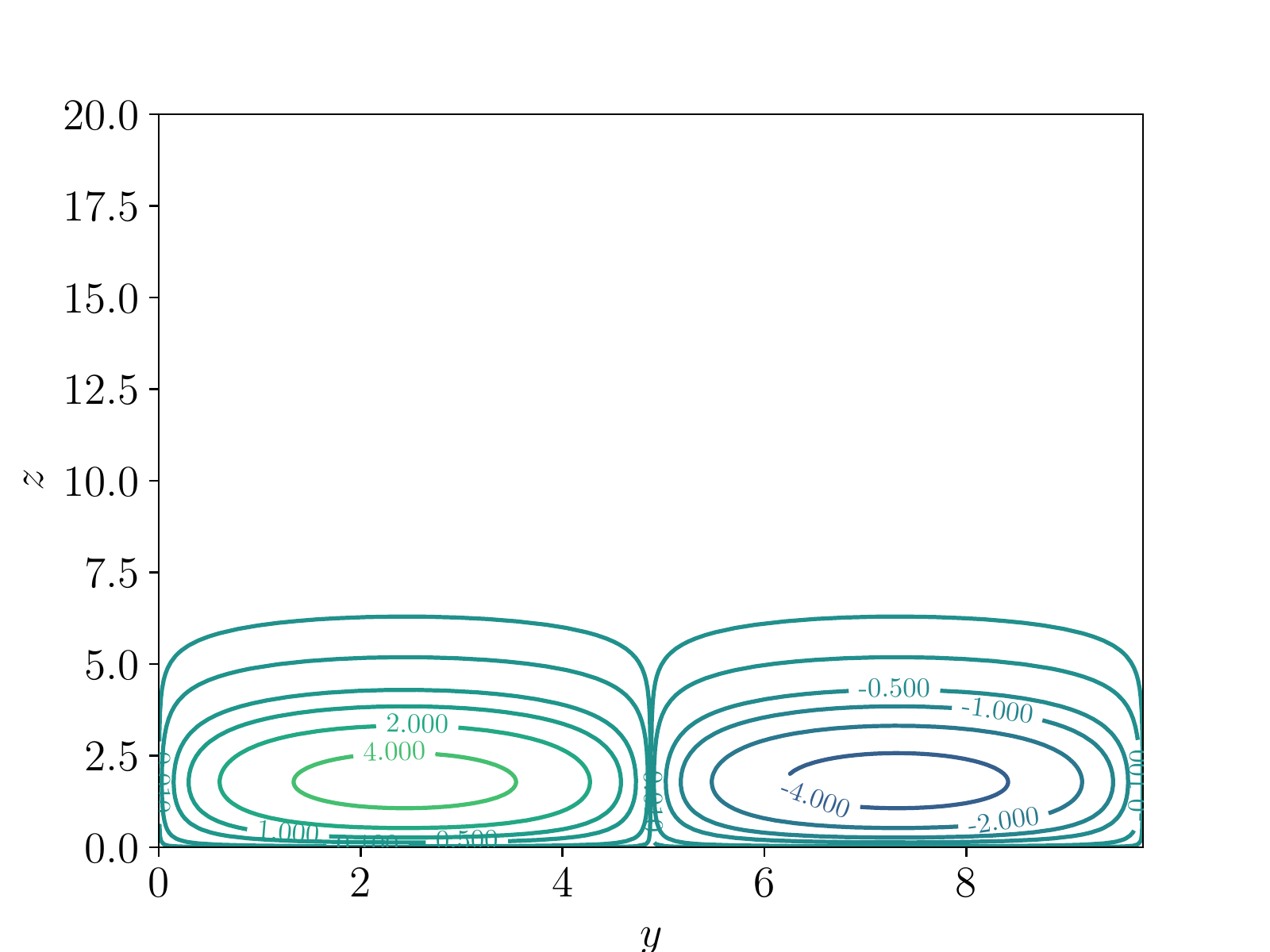}
    \caption{ $ \zeta(z,t) \cdot \exp {\rm i} \beta y  \cdot \Rey_\delta \cdot 10^{-3}$ }
    \label{fig:contourC}
  \end{subfigure} 
  \caption{Contour plots of the real part of $ w(z,t) \cdot \exp {\rm i} \beta y $
    and the real part of $ \zeta(z,t)  \cdot \exp {\rm i} \beta y  \cdot \Rey \cdot 10^{-3} $,
    which are obtained by advancing the initial condition in figure \ref{fig:initialConditionStreakContour} to time $ t  = 1.53 $
    for the optimal perturbation for the case $ \Rey = 500 $, $ \beta_{\max} = 0.64 $, $ t_0 = 0.11 $.}
  \label{fig:initialConditionStreakContourFinal}
\end{figure}

When plotting the maximum amplification of streamwise streaks in 
a log-log plot, cf. figure \ref{fig:loglogPlot}, we find the
expected quadratic behavior of the maximum amplification.
In line with this quadratic growth in $ \Rey_\delta $, a straightforward calculation,
cf. appendix \ref{sec:initialCondition},
shows that when normalizing the energy $ E =  E_w + E_\zeta $, equation (\ref{eq:energyNonmodal}) of the initial condition of the
optimal streamwise streak to one, the amplitude of the initial normal vorticity
scales inversely with the Reynolds number,
whereas the amplitude of the normal velocity converges to a constant in the
asymptotic limit:
\be
\max_z \zeta(z,t_0) \propto \frac{1}{\Rey_\delta}, \quad
\max_z w(z,t_0) \propto {\rm const} \quad \mbox{for} \quad \Rey_\delta \rightarrow \infty. \label{fig:scalingStreaksInfty}
\ee
This can also be observed in
figure \ref{fig:initialConditionStreak}, where we show that
for larger Reynolds numbers, the graphs of $ | \zeta | \cdot \Rey_\delta $ 
and $ | w | $ collapse.
In order to visualize the
spatial structure of the optimal streamwise streak,
we consider the case $ \Rey_\delta = 500 $ with
a maximum amplification of: 
\be
\max_{\beta,t_0,t} G(\alpha=0,\beta,t_0,t,\Rey_\delta = 500 ) = 238.6,
\ee
where the parameters at maximum are given by:
\be
\beta = 0.64, \quad t_0 = 0.11, \quad t = 1.53. 
\ee
In figure \ref{fig:initialConditionStreakContour}, contour plots of
the real part of the
initial condition at $ t_0 = 0.11 $ of the optimal perturbation
in the $ (y,z) $-plane is shown.
When advancing this initial condition to $ t = 1.53 $, where
the energy of the streamwise streak is maximum, cf.
figure \ref{fig:initialConditionStreakContourFinal}, we observe
that the amplitude of the normal velocity component $ w $ has
decreased by approximately a factor of two, whereas the amplitude
of the normal vorticity $ \zeta $ increased by approximately
a factor of five hundred. \\

\begin{figure}
  \centering
%  \begin{subfigure}[b]{0.5\textwidth}
%    \includegraphics[width=\textwidth]{nonModalInitialCondition2DW.pdf}
%    \caption{ $ w $ }
%    \label{fig:contourB}
%  \end{subfigure} \nolinebreak
%  \begin{subfigure}[b]{0.5\textwidth}
%    \includegraphics[width=\textwidth]{nonModalInitialConditionScaled2DU.pdf}
%    \caption{ $ u $ }
%    \label{fig:contourC}
%  \end{subfigure} 
  \includegraphics[width=0.7\textwidth]{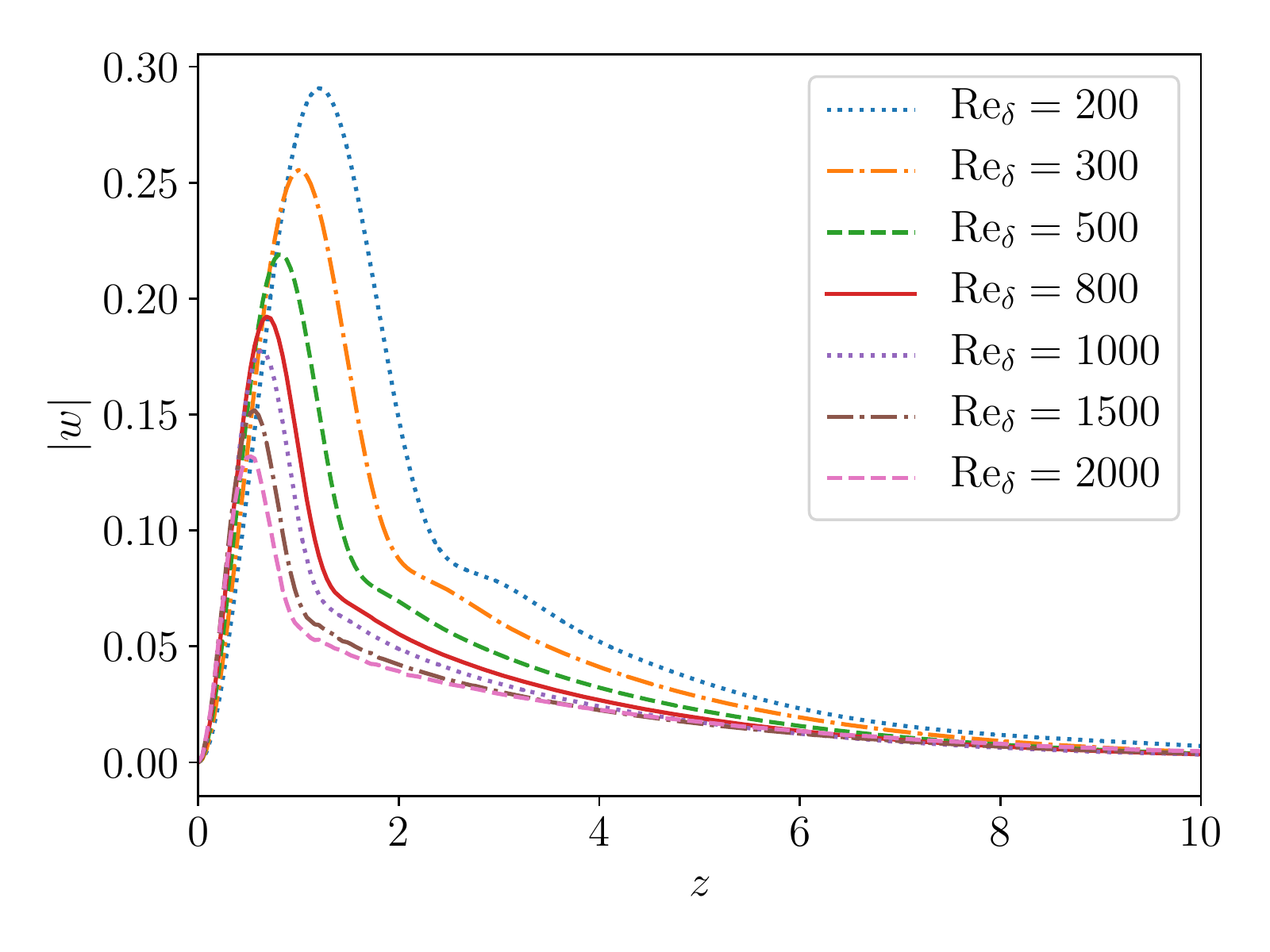}
  \caption{Initial condition $ w $ %and its derivative $ u = {\rm i} Dw/\alpha $
    for the two-dimensional perturbations with maximum amplification,
    $ \max_{\alpha,t_0,t} G(\alpha,\beta=0,t_0,t,\Rey_\delta) $, for
    different Reynolds numbers.}
  \label{fig:initialCondition2D}
\end{figure}

\begin{figure}
  \centering
%  \begin{subfigure}[b]{0.5\textwidth}
%    \includegraphics[width=\textwidth]{nonModalInitialCondition2DW.pdf}
%    \caption{ $ w $ }
%    \label{fig:contourB}
%  \end{subfigure} \nolinebreak
%  \begin{subfigure}[b]{0.5\textwidth}
%    \includegraphics[width=\textwidth]{nonModalInitialConditionScaled2DU.pdf}
%    \caption{ $ u $ }
%    \label{fig:contourC}
%  \end{subfigure} 
  \includegraphics[width=0.7\textwidth]{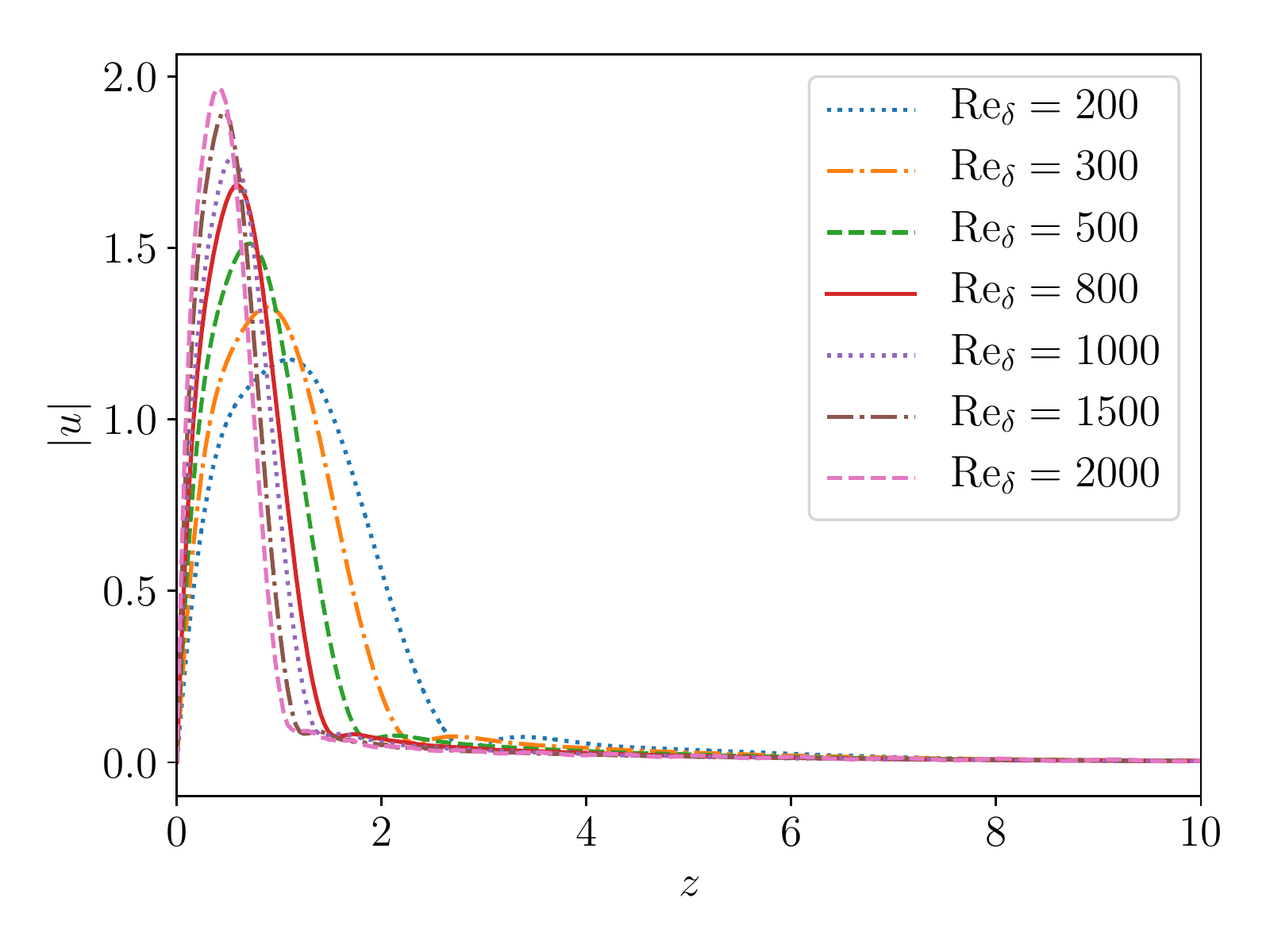}
  \caption{The horizontal component $ u = {\rm i} Dw/\alpha $ of the initial condition 
    for two-dimensional perturbations with maximum amplification,
    $ \max_{\alpha,t_0,t} G(\alpha,\beta=0,t_0,t,\Rey_\delta) $, for
    different Reynolds numbers.}
  \label{fig:initialCondition2DU}
\end{figure}

\begin{figure}
  \centering
  \begin{subfigure}[b]{0.5\textwidth}
    \includegraphics[width=\textwidth]{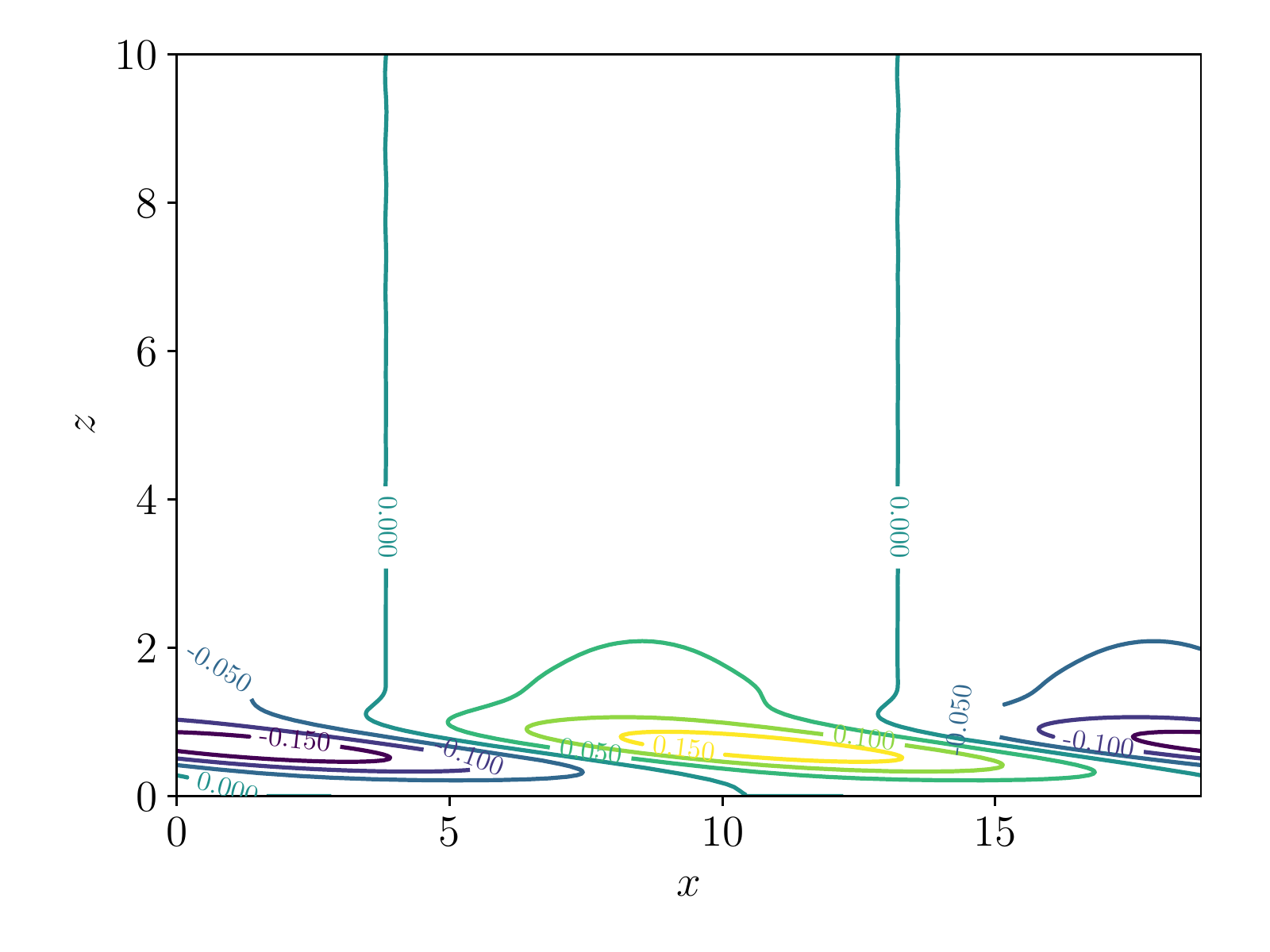}
    \caption{$ t_0 = 0.26 $}
    \label{fig:contourB}
  \end{subfigure} \nolinebreak
  \begin{subfigure}[b]{0.5\textwidth}
    \includegraphics[width=\textwidth]{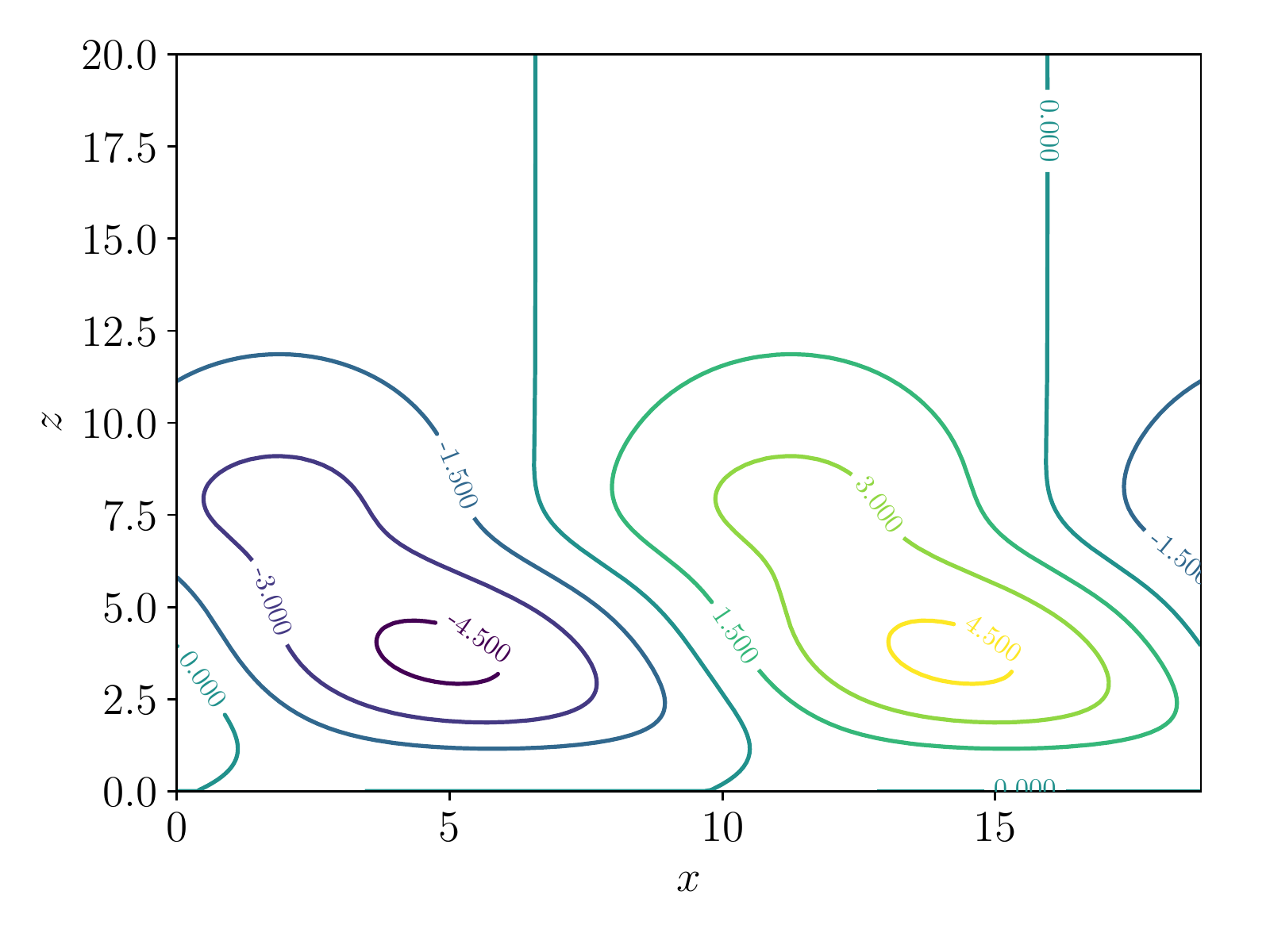}
    \caption{$ t = 14.2 $}
    \label{fig:contourC}
  \end{subfigure} 
  \caption{Contour plots of the real part of
    $ w \cdot \exp {\rm i} \alpha x $,
    at initial time $ t_0 = 0.26 $ and at $ t = 14.2 $ ($ w $ multiplied by
    $ 10^{-8} $),
    when it reaches its maximum amplification,
    for the optimal perturbation for the case $ \Rey = 1000 $ with
    $ \alpha_{\max} = 0.33 $.}
  \label{fig:initialCondition2DContour}
\end{figure}

For two-dimensional perturbations, on the other hand, the energy
is distributed between the normal component
$ w $ and the horizontal component $ u = {\rm i} D w/\alpha $.
As can be observed from figure \ref{fig:initialCondition2D},
for increasing Reynolds number the amplitude of $ w $
decreases. Following, its share of the initial energy
goes down as well. Since the initial energy is normalized to one, this
implies that the energy contribution associated to $ u $
must increase. Corresponding to this energy increase, we observe
that the amplitude of $ u $ increases for increasing Reynolds number,
cf. figure \ref{fig:initialCondition2DU}.
We choose the case $ \Rey_\delta = 1000 $ in order to visualize
the spatial structure of the optimal two-dimensional perturbation.
For this case the maximum amplification is given by:
\be
\max_{\alpha,t_0,t} G(\alpha,\beta=0,t_0,t,\Rey_\delta = 1000 ) = 1.34 \cdot 10^{18},
\ee
where the parameters at maximum are given by:
\be
\alpha = 0.33, \quad t_0 = 0.26, \quad t = 14.2. 
\ee
In figure \ref{fig:initialCondition2DContour}, contour plots in
the $(x,z)$-plane of the real part of
$ w \cdot \exp {\rm i} \alpha x $ at initial time $ t_0 $ and
at time $ t $ when it reaches maximal amplification are plotted.
Initially, the perturbation is confined to a thin layer inside the boundary
layer. While reaching its maximum amplification its spatial structure
grows in wall normal direction. 

\section{Relation to previous results in the literature}
\label{sec:otherworks}

\begin{figure}
  \centering
  \includegraphics[width=0.7\textwidth]{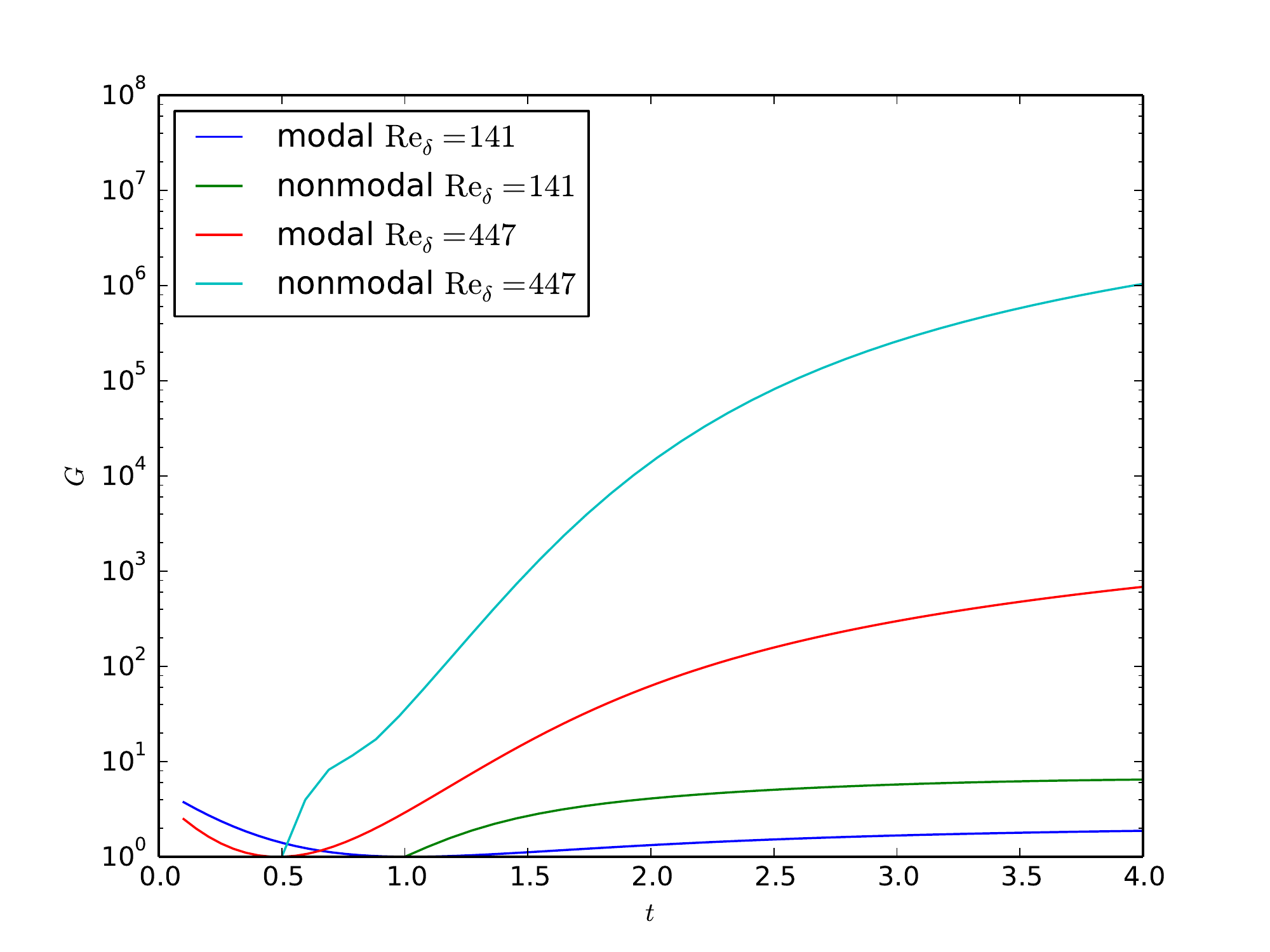}
  \caption{Amplification $ G(\alpha=0.35,\beta=0,t_0,t,\Rey_\delta) $
    of the nonmodal two-dimensional perturbation versus corresponding
    amplification of the
    modal Tollmien-Schlichting wave with $ \alpha = 0.35 $
    computed by means of the Orr-Sommerfeld
    equation, for $ \Rey_\delta = 141, 447 $. The initial time
    $ t_0 $ is taken from the minimum of the modal Tollmien-Schlichting
    waves.}
  \label{fig:modalVSnonmodal}
\end{figure}

A question which suggests itself immediately, is the relation
between the present nonmodal stability analysis and
the modal stability analyses performed previously in 
\citet{BlondeauxPralitsVittori2012}, \citet{VerschaevePedersen2014}
and \citet{SadekParrasDiamessisLiu2015}.
Naturally,
the amplifications of the optimal perturbations are expected to
be larger
than the corresponding ones of the modal Tollmien-Schlichting waves.
This can be seen in figure \ref{fig:modalVSnonmodal}, where
we have solved the Orr-Sommerfeld equation
for the present problem in a quasi-static fashion for the 
wave number $ \alpha = 0.35 $ and Reynolds numbers $ \Rey_\delta = 141 $ and
$ \Rey_\delta = 447$. The amplification of the optimal perturbation can be several orders
of magnitude larger than that of the corresponding modal Tollmien-Schlichting
wave. On the other hand the main conclusions
by \citet{VerschaevePedersen2014} are still supported by the 
present analysis. Although attempted by several experimental and
direct numerical studies \citep{VittoriBlondeaux2008,SumerJensenSorensenFredsoeLiuCarstensen2010,OzdemirHsuBalachandar2013}%,DiamessisRedekopp2005,
%CarrDavies2006,CarrDavies2010,AghsaeeBoegmanDiamessisLamb2012}
, a
well defined transitional Reynolds number cannot be given for
this flow. As also pointed out in the present analysis, depending
on the characteristics of the external perturbations, such as length scale and
intensity, the flow might 
transition to turbulence for different Reynolds numbers. Without
control of the external perturbations, any experiment on
the stability properties of this flow will hardly be repeatable. 
On the other hand, as we have shown above, 
a critical Reynolds number $ \Rey_A $
can be defined for which the present flow switches from a
monotonically stable to a non-monotonically stable flow. This
critical Reynolds number has, however, little practical bearing.\\

Concerning the direct numerical simulations by 
\citet{VittoriBlondeaux2008,VittoriBlondeaux2011}
and \citet{OzdemirHsuBalachandar2013}, the present study
gives an indication for
the transition process happening via two-dimensional
vortex rollers observed in their direct numerical simulations. In addition,
we are able to answer the question raised by 
\citet{OzdemirHsuBalachandar2013} about the possible mechanism of 
a by-pass transition. However, quantitative
differences between the direct numerical
results by \citet{OzdemirHsuBalachandar2013} and the present ones
exist. \citet{OzdemirHsuBalachandar2013} introduced
a random disturbance at $ t_0 = - \pi $ with
different amplitudes in their simulations
and monitored the evolution of the amplitude of these disturbances, 
cf. figure 10 in
\citet{OzdemirHsuBalachandar2013}. From this figure, we see the
characteristic kink of two-dimensional perturbations overtaking
streamwise streaks appearing in their simulations
only for $ \Rey_\delta = 2000$
and higher. If we compare this to the optimal perturbations
with initial times $ t_0 = -4 $ and $ t_0 = -2 $ in figure
\ref{fig:temporalEvolution}, we see this kink developing 
already for a much lower Reynolds number,
namely $ \Rey_\delta = 1000 $, cf. figure \ref{fig:temporalEvolution}d.
The reasons for this discrepancy are unclear.
Although \citet{OzdemirHsuBalachandar2013} employed
perturbation amplitudes with values up to 20 \% of the base flow, which
might trigger nonlinear effects, the acceleration region of
the flow has a strong damping effect, such that the initial perturbation
growth starting in the deceleration region is most likely governed by linear effects. 
We might, however,
point out that, in order for a Navier-Stokes solver to
capture the growth of two-dimensional
perturbations correctly an extremely fine resolution in space and
time is needed,
as can be seen in \citet[Appendix A]{VerschaevePedersen2014}
for modal Tollmien-Schlichting waves. In particular, when the
resolution requirements are not met, these perturbations tend to
be damped instead of amplified. In this respect, it is interesting
to note, that \citet{VittoriBlondeaux2008,VittoriBlondeaux2011}
found that regular vortex tubes appeared in their simulation
for a Reynolds number around $ \Rey_\delta = 1000$ ($\Rey_\sumer = 5 \cdot 10^5$),
which corresponds relatively well with the present findings. However,
it cannot be excluded that this is for the wrong reason, as
a larger level of background noise resulting from, for example the
numerical approximation error by their low order solver, might be
present in their simulations.

The Reynolds number in the experiments by \citet{LiuParkCowen2007}
lies in the range $ \Rey_\delta = 72-143 $ which is
larger than $ \Rey_A = 18 $. However, as can be seen from figure
\ref{fig:temporalEvolution}, the maximum amplification for
these cases is around a factor of $ 30 $. Therefore,
without any induced disturbance, growth of streamwise streaks from
background noise is probably not observable and has not been observed
in \citet{LiuParkCowen2007}.
On the other hand, in the experiments by \citet{SumerJensenSorensenFredsoeLiuCarstensen2010} 
vortex rollers
appeared in the range $ 630  \le \Rey_\delta < 1000 $. Assuming that
the initial level of external perturbations in the experiments
is higher than in the direct numerical simulations, the
observation by Sumer {\em et al.} fits the present picture. 
However, for $ \Rey_\delta > 1000 $, they observed the
development of turbulent spots in the deceleration
region of the flow. This is in contrast to the
results by \citet{OzdemirHsuBalachandar2013} of a $ K$-type transition.
The present analysis
supports the finding of a transition process via the growth of
two-dimensional perturbations. However, whether these
nonmodal Tollmien-Schlichting waves break down via a $ K $-type
transition as in \citet{OzdemirHsuBalachandar2013} or whether
they break up randomly producing turbulent spots 
\citep{ShaikhGaster1994,Gaster2016} is difficult
to say from this primary instability analysis. In addition,
more information on the initial disturbances in the
experiments is needed to make any conclusions. 
Whereas random noise is applied in 
\citet{VittoriBlondeaux2008,VittoriBlondeaux2011}
and
\citet{OzdemirHsuBalachandar2013},
the initial disturbance in 
\citet{SumerJensenSorensenFredsoeLiuCarstensen2010} 
might stem from residual motion in their facility,
exhibiting probably certain
characteristics. Depending on these characteristics,
other perturbations than the one showing optimal amplification, might
induce secondary instability. In addition,
it cannot be
excluded that a completely different instability mechanism
is at work in the experiments of \citet{SumerJensenSorensenFredsoeLiuCarstensen2010}. The focus in the present analysis is on the response to
initial conditions and does not take into account
any response to external forcing, which would be modeled
by adding a source term to the equations (\ref{eq:sys1})
and (\ref{eq:sys2}). It is possible that the present flow
system displays some sensitivity to certain frequencies of
vibrations present in the experimental set-up altering
the behavior of the system for larger Reynolds numbers. In particular,
different perturbations, such as streamwise streaks, might be favored,
leaving the possibility open that the turbulent spots, nevertheless,
result from the break-down of streamwise streaks
\citep{AnderssonBrandtBottaroHenningson2001,BrandtSchlatterHenningson2004}.

\section{Conclusions} \label{sec:conclusions}

In the present treatise, a nonmodal stability analysis 
of the bottom boundary layer flow under solitary waves
is performed. Two
competing mechanism can be identified:
Growing streamwise streaks and growing
two-dimensional perturbations (nonmodal Tollmien-Schlichting waves). 
By means of an energy bound, 
it is shown that the present flow is monotonically stable
for Reynolds numbers below $ \Rey_\delta = 18 $ after which it
turns non-monotonically stable, with streamwise streaks growing
first. Two-dimensional perturbations display growth only for
Reynolds numbers larger than $ \Rey_\delta = 38 $. However,
their maximum amplification overtakes
that of streamwise streaks at  $ \Rey_\delta = 170 $.
As for steady flows, the maximum amplification of 
streamwise streaks displays quadratic growth with $ \Rey_\delta $ 
for the present unsteady flow. On the other hand,
the maximum amplification
of two-dimensional perturbations shows a near exponential
growth with the Reynolds number in the deceleration region of the flow. Therefore,
during primary instability,
the dominant perturbations in the deceleration region
of this flow are to be expected two-dimensional. This
corresponds to the findings in the direct numerical simulations
by \citet{VittoriBlondeaux2008} and \citet{OzdemirHsuBalachandar2013}
and in the experiments by \citet{SumerJensenSorensenFredsoeLiuCarstensen2010}
of growing two-dimensional vortex rollers in the deceleration region
of the flow. However, further investigation of the
secondary instability mechanism and of receptivity to external (statistical)
forcing is needed in order to explain the subsequent break-down to 
turbulence in the boundary layer.

The boundary layer under solitary waves is a relatively simple
model for a boundary layer flow with a favorable and an adverse pressure gradient. But just for this reason it allows to analyze stability mechanisms being
otherwise shrouded in more complicated flows.

The implementation of the numerical method has been done 
using the open source libraries Armadillo \citep{armadillo},
FFTW \citep{fftw} and GSL \citep{gsl}. At this occasion, the
first author would like to thank Caroline Lie for pointing
out a mistake in \citet{VerschaevePedersen2014}. In figures
20,22,24 and 26 in \citet{VerschaevePedersen2014}, the frequency
$ \omega $ is incorrectly scaled. However, this
does not affect any of the conclusions of the article. The
first author apologizes for any inconvenience this might represent.

\begin{appendix} 

\section{Numerical implementation} \label{sec:implementation}

\subsection{Numerical implementation for the energy bound}

We expand $ \zeta $ and $ w $
in equations (\ref{eq:davis1a}-\ref{eq:davis2a})
on the Shen-Legendre polynomials $ \phi_j $ and $ \psi_j $ for
the Poisson and biharmonic operator, respectively,
cf. \citep{Shen1994}:
\be
\zeta = \sum \limits_{j=0}^{N-2} \zeta_j \phi_j(z) \quad
w = \sum \limits_{j=0}^{N-4} w_j \psi_j(z), \label{eq:expansionZetaW}
\ee
where $ N $ is the number of Legendre polynomials.
The semi infinite domain $ [0,\infty) $ is truncated at $ h $,
where $ h $ is chosen large enough by numerical inspection.
The basis functions $ \phi_j $ and $ \psi_j $ are linear
combinations of Legendre polynomials, such that
a total number of $ N $ Legendre polynomials is used
for each expansion in (\ref{eq:expansionZetaW}). 
The basis functions $ \phi_j $ satisfy
the homogeneous Dirichlet conditions, whereas
$ \psi_j $ honors the clamped boundary conditions.
A Galerkin formulation is then chosen for the
discrete system:
\be
\left( \begin{array}{cc} \mathbf{A} & \mathbf{B} \\
\mathbf{B}^T & \mathbf{D} \end{array} \right) \left( \begin{array}{c} 
\mathbf{w} \\ \bm{\zeta} \end{array} \right) = \mu
\left( \begin{array}{cc} \mathbf{E} & \mathbf{0} \\
\mathbf{0} & \mathbf{H} \end{array} \right)  \left( \begin{array}{c} 
\mathbf{w} \\ \bm{\zeta} \end{array} \right). \label{eq:sysNonModal}
\ee
The elements of the matrices are given by:
\bea
A_{ij} &=& \frac{1}{\Rey} \left\{ \int \limits_0^h D^2 \psi_i D^2 \psi_j \, dz
+ 2 \left( \alpha^2 + \beta^2 \right) \int \limits_0^h D \psi_i D \psi_j \, dz
+ \left( \alpha^2 + \beta^2 \right)^2  \int \limits_0^h \psi_i \psi_j \, dz
\right\}  \nonumber \\
& & + \frac{{\rm i} \alpha}{2} \left\{ \int \limits_0^h \psi_i \partial_z^2 U_\base  \psi_j \, dz + 2 \int \limits_0^h \psi_i \partial_z U_\base \partial_z \psi_j \, d z \right\} 
\eea
\bea
B_{ij} &=& \frac{{\rm i} \beta}{2} \int \limits_0^h \psi_i
\partial_z U_\base \phi_j \, dz 
\eea
\bea
D_{ij} &=&  \frac{ 1}{\Rey} 
\left\{  \int \limits_0^h D \phi_i D \phi_j \, dz 
+ \left( \alpha^2 + \beta^2 \right) 
\int \limits_0^h \phi_i \phi_j \, dz \right\} 
\eea
\bea
2E_{ij} &=&   - \int \limits_0^h D \psi_i D \psi_j \, dz 
- \left( \alpha^2 + \beta^2 \right) \int \limits_0^h \psi_i \psi_j \, dz
\eea
\bea
2H_{ij} & = & - \int \limits_0^h \phi_i \phi_j \, dz
\eea
For the verification and validation of the method, manufactured solutions
have been used. In addition, the Reynolds numbers $ \Rey_A $ and $ \Rey_B $
for Stokes' second problem have been computed, resulting into
$  \Rey_A = 18.986 $ and $ \Rey_B = 38.951 $, corresponding well
with the numbers $ 19.0 $ and $ 38.9 $ obtained by \citet[table 1]{DavisKerczek1973}.

\subsection{Numerical implementation for the nonmodal analysis}

The basis functions $ \psi_j $ and $ \phi_j $ for
$ w $ and $ \zeta $ are in this case
given by the 
Shen-Chebyshev polynomials, cf. \citet{Shen1995}, instead of
the Shen-Legendre polynomials as before. 
This allows us to use the 
fast Fourier transform for computing derivatives. The
equations (\ref{eq:sys1}-\ref{eq:sys2})
are written in discrete form as:
\be
\frac{2}{\Rey_\delta} \left( \begin{array}{cc} \mathbf{L}^\psi & 0 \\
0 & \mathbf{M}^\phi \end{array} \right) 
\frac{d}{d t} \left( \begin{array}{c} 
\mathbf{w} \\ \bm{\zeta} \end{array} \right)
= \left( \begin{array}{cc} \mathbf{L}^{OSE} & \mathbf{0} \\
\mathbf{L}^C & \mathbf{L}^{SC} \end{array} \right)  \left( \begin{array}{c} 
  \mathbf{w} \\ \bm{\zeta} \end{array} \right), \label{eq:discreteNonmodal}
\ee
where the elements of the matrices are given by:
\bea
M_{ij}^\psi &=& \int \limits_0^h \psi_i \psi_j \, dz \label{eq:MatrixMw}\\
G_{ij}^\psi &=& \int \limits_0^h \frac{d}{dz} \psi_i \frac{d}{dz} \psi_j \, dz \label{eq:MatrixG}\\
A^\psi_{ij}&=& \int \limits_0^h \frac{d^2}{dz^2} \psi_i \frac{d^2}{dz^2} \psi_j \, dz \\
M_{ij}^\phi &=& \int \limits_0^h \phi_i \phi_j \, dz \label{eq:MatrixMphi}\\
G^{\phi}_{ij} &=& \int \limits_0^h \frac{d}{dz} \phi_i \frac{d}{dz} \phi_j \, dz \\
P^1_{ij} &=& \int \limits_0^h \partial_z^2 U_\base \psi_i \psi_j \, dz \\
P^2_{ij} &=& \int \limits_0^h  U_\base \psi_i \left( D^2 - (\alpha^2 + \beta^2) \right) \psi_j \, dz \\
P^3_{ij} &=& \int \limits_0^h  U_\base \phi_i \phi_j \, dz \\
L^\psi_{ij} &=& - G_{ij}^\psi - (\alpha^2+\beta^2) M_{ij}^\psi \\
{L}_{ij}^{OSE} &=& {\rm i} \alpha P^1_{ij} - {\rm i} \alpha P^2_{ij} 
+ \frac{1}{\rm Re} \left( A_{ij}^\psi + 2 \left( \alpha^2 + \beta^2 \right) G_{ij}^\psi 
+ \left( \alpha^2 + \beta^2 \right)^2 M_{ij}^\psi \right)\\
{L}^C_{ik} &=& {\rm i} \beta \int \limits_0^h \partial_z U_0 \phi_i \psi_k \, dz \\
{L}^{SC}_{ij} &=& - {\rm i} \alpha P^3_{ij} + \frac{1}{\rm Re} \left( -G_{ij}^\phi - 
(\alpha^2+\beta^2) M_{ij}^\phi \right) 
\eea
For the Shen-Chebyshev polynomials, $ \mathbf{L}^\psi $ and $ \mathbf{M}^\phi $
are sparse banded matrices. Therefore, the system (\ref{eq:discreteNonmodal})
can be efficiently advanced in time, allowing us to compute the 
evolution matrix $ \mathbf{X}(t,t_0) $ for a wide range of parameters.
The amplification $ G $, equation (\ref{eq:amplification}), for the discrete
case can then be computed as suggested in \citet{TrefethenTrefethenReddyDriscoll1993,SchmidHenningson2001,Schmid2007}. We write
\be
\mathbf{q} = \left( \begin{array}{c} 
\mathbf{w} \\ \bm{\zeta} \end{array} \right),
\ee
and note that the energy $ E $, equation (\ref{eq:energyNonmodal}), in the discrete case is
given by:
\be
E = \mathbf{q}^* \mathbf{W} \mathbf{q},
\ee
where
\be
\mathbf{W} = \frac{1}{2}\left(
\begin{array}{cc}
  \frac{1}{k^2} \mathbf{G}^\psi + \mathbf{M}^\psi & 0 \\
  0 & \frac{1}{k^2} \mathbf{M}^\phi
\end{array}
\right).
\ee
Matrices $ \mathbf{G}^\psi $, $ \mathbf{M}^\psi $ and $ \mathbf{M}^\phi $ are defined in equations (\ref{eq:MatrixG}),
(\ref{eq:MatrixMw}) and (\ref{eq:MatrixMphi}), respectively. The Cholesky factorization of $ \mathbf{W}$ is given by:
\be
\mathbf{F}^T\mathbf{F} = \mathbf{W}. 
\ee
The coefficients $ \mathbf{q}(t) $ at time $ t $ can be obtained by means of the
evolution matrix $ \mathbf{X} $:
\be
\mathbf{q}(t) = \mathbf{X}(t,t_0) \mathbf{q}_0,
\ee
where $ \mathbf{q}_0 $ is the initial condition at $ t_0 $. From this it follows
that $ \mathbf{X}(t_0,t_0) $ reduces to the identity matrix. 
The amplification $ G $ can then be computed by
\bea
G(\alpha,\beta,t_0,t,\Rey_\delta) &=&\max_{ \mathbf{q}_0 } \frac{ \mathbf{q}(t)^\dagger \mathbf{W} 
  \mathbf{q}(t) } { \mathbf{q}_0^\dagger \mathbf{W} \mathbf{q}_0 } \\
&=& \max_{ \mathbf{q}_0 } \frac{ \mathbf{q}_0^\dagger \mathbf{X}^\dagger \mathbf{W} \mathbf{X}
  \mathbf{q}_0 } { \mathbf{q}_0^\dagger \mathbf{W} \mathbf{q}_0 } \\
&=& \max_{ \mathbf{b} } \frac{ \mathbf{b}^\dagger \mathbf{F}^{-T} \mathbf{X}^\dagger
  \mathbf{W} \mathbf{X}\mathbf{F}^{-1} \mathbf{b} } { \mathbf{b}^\dagger \mathbf{b} } \\
&=& \left| \left| \mathbf{F} \mathbf{X}\mathbf{F}^{-1} \right| \right|^2 ,
\eea
where the matrix norm $ \left| \left| \mathbf{F} \mathbf{X}\mathbf{F}^{-1} \right| \right| $
is given by the maximum singular value of $ \mathbf{F} \mathbf{X}\mathbf{F}^{-1}  $,
cf. \citet{TrefethenTrefethenReddyDriscoll1993,SchmidHenningson2001,Schmid2007}.\\

\begin{figure}
  \centering
  \includegraphics[width=0.7\textwidth]{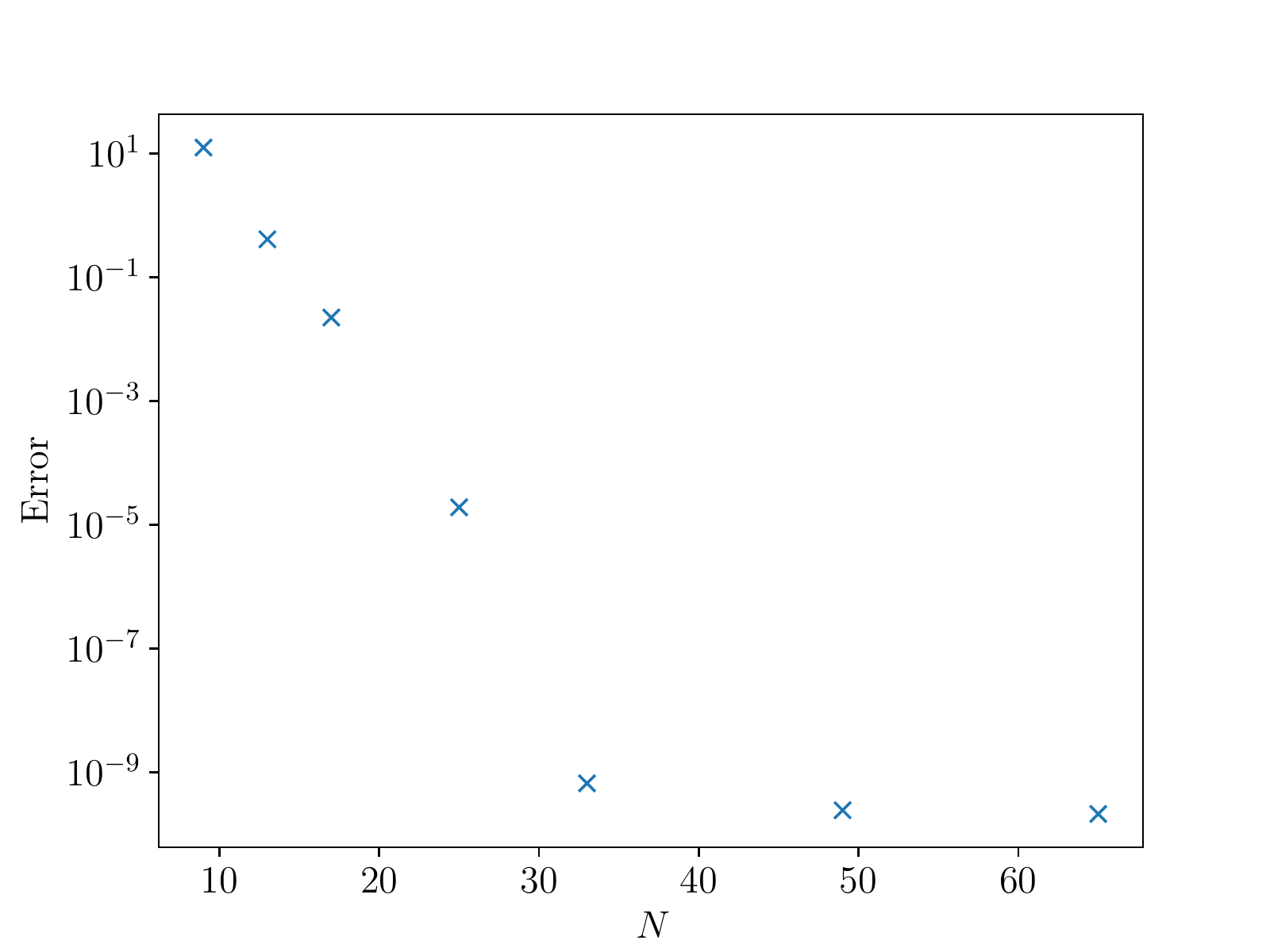}
  \caption{Error convergence of the manufactured problem given by
    equation \ref{eq:toyProblem}. }
  \label{fig:toyError}
\end{figure}

The present method consists of two steps. First, the evolution matrix $ \mathbf{X} $
needs to be computed by solving equation (\ref{eq:discreteNonmodal}) with the
identity matrix as initial condition at time $ t_0 $. Then the amplification $ G $ can
be computed using $ \mathbf{X} $. In order to verify the well functioning of the
present time integration, the following manufactured solution has been used:
\be
w = \cos (\omega_1 t) \sin^2 ( 5 \pi z) \quad
\zeta = \cos (\omega_2 t) \sin ( 3 \pi z) \quad
U_\base = \cos (\omega_3 t) \left( 1 - \exp \left( -2z \right) \right). \label{eq:toyProblem}
\ee
A forcing term is defined by the resulting term, when injecting the above solution into equations (\ref{eq:sys1}) and (\ref{eq:sys2}). Equations (\ref{eq:discreteNonmodal}) are advanced by means
of the adaptive {\tt Runge-Kutta-Cash-Karp-54} time integrator
included in the {\tt boost} library. The absolute and relative error of the time
integration are set to $ 10^{-10} $. For verification, we use the above manufactured solution
with the following parameter values:
\be
\Rey_\delta = 123 \quad \alpha = 0.3 \quad \beta = 0.234 \quad h = 1 \quad \omega_1 = 1.234 \quad \omega_2 = 1.123 \quad \omega_3 = 0.4567 \quad t_0 = 0,
\ee
and compare reference and numerical solution by computing a mean error on the Chebyshev knots.
The behavior of the error for increasing $ N $ is displayed in figure \ref{fig:toyError}. We observe
that the error displays exponential convergence until approximately $ 10^{-9} $, when the
error contribution due to the time integration becomes dominant. In addition, the analytic
solution of the energy of this problem can be used to verify parts of the amplification
computation (results not shown). \\

\begin{figure}
  \centering
  \includegraphics[width=0.7\textwidth]{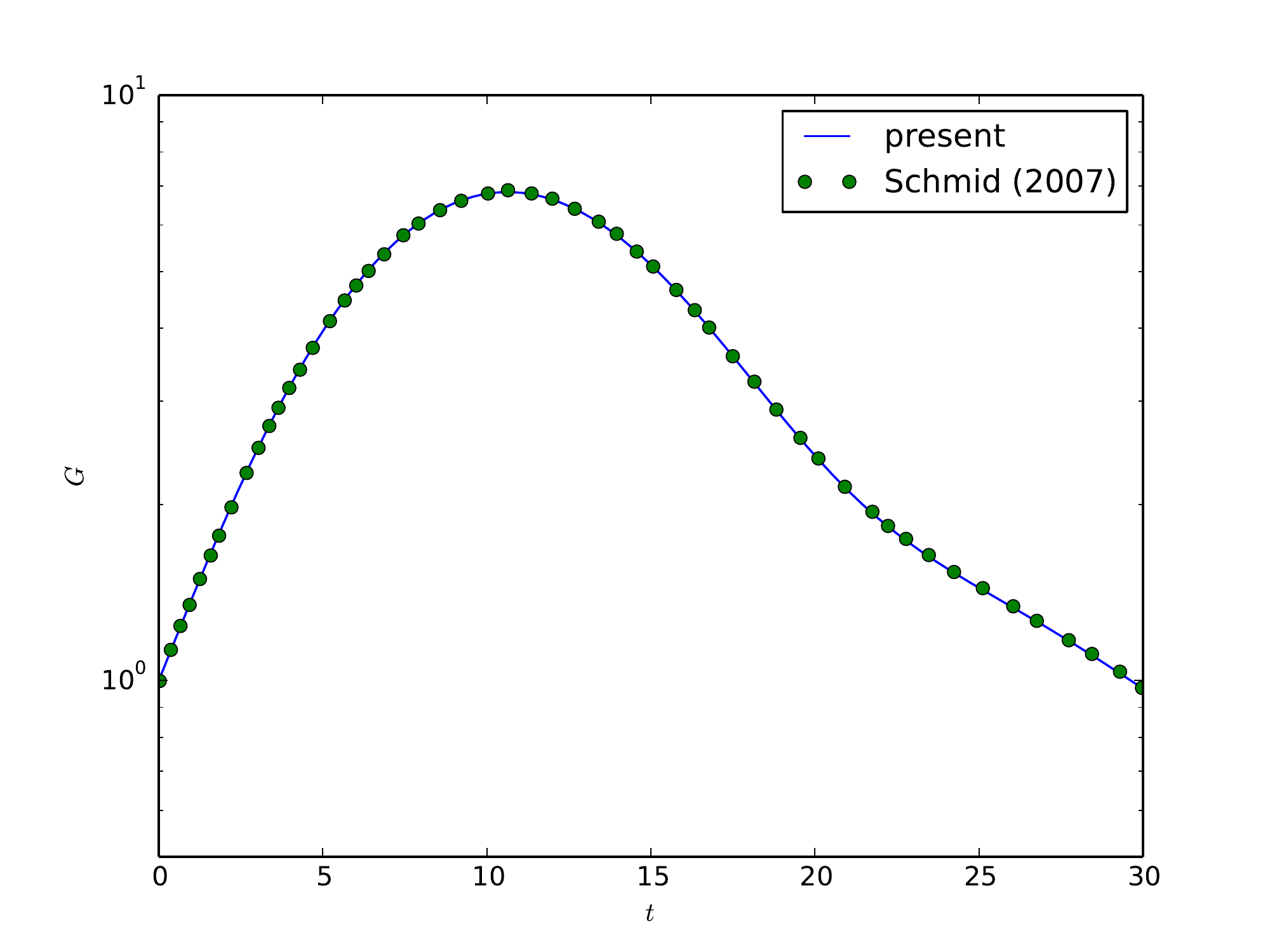}
  \caption{Amplification $ G(\alpha=1.,\beta=0,t_0 = 0.,t,\Rey = 1000.) $
    of the nonmodal perturbation for Poiseuille flow. 
    The present results collapse onto
    the data from figure 3 in \cite{Schmid2007}.}
  \label{fig:verification2}
\end{figure}

For validation purposes, the case of
transient growth for
Poiseuille flow with a Reynolds number $ \Rey = 1000 $ and $ \alpha = 1 $
in \cite{Schmid2007} has been computed by means of the present method for $ N = 65 $.
As can be seen from figure \ref{fig:verification2}, the
results by the present method correspond well to
the data digitized from figure 3 in \cite{Schmid2007}. \\

\begin{figure}
  \centering
  \includegraphics[width=0.7\textwidth]{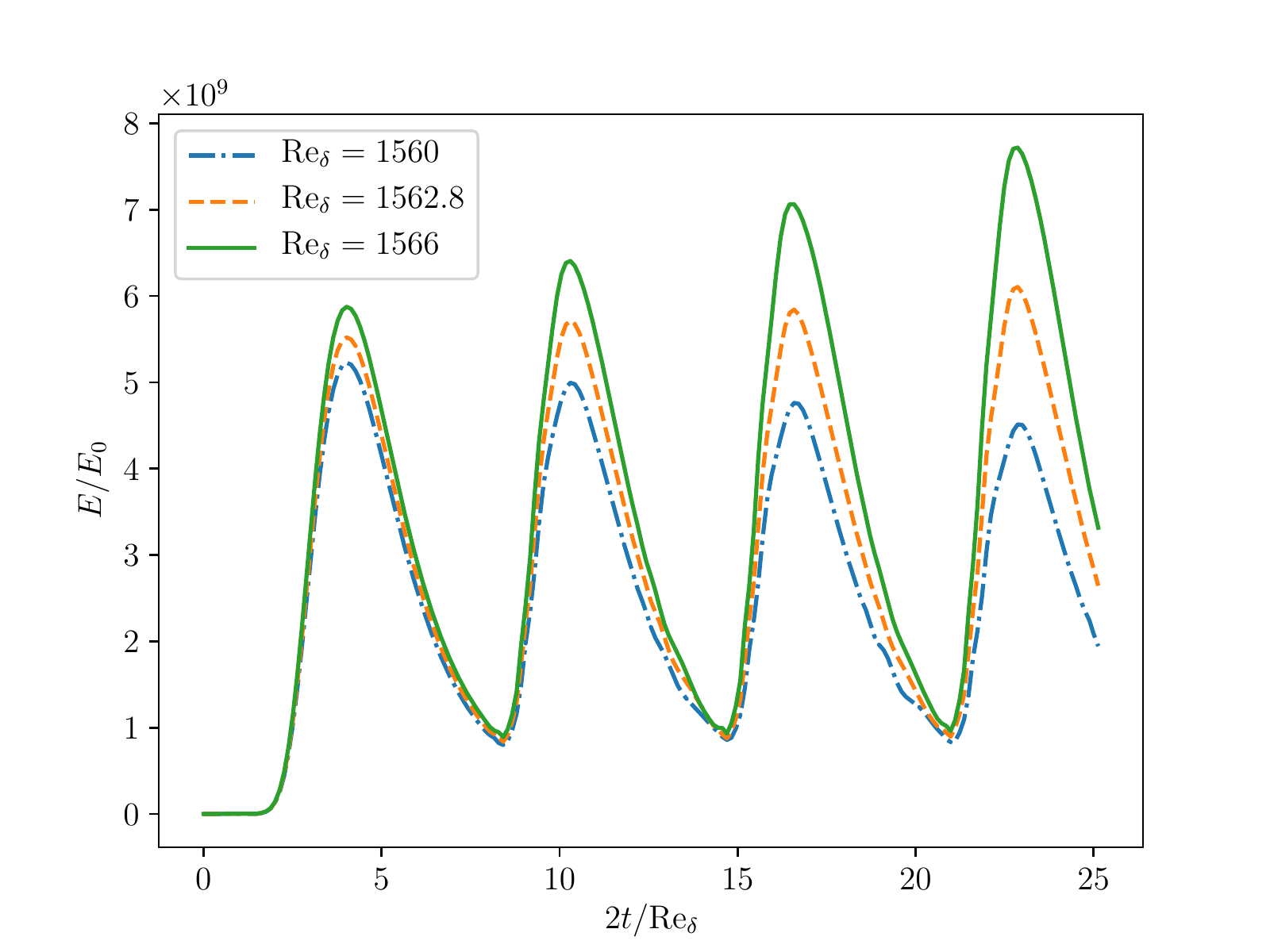}
  \caption{Temporal evolution of the amplitude $ E/E_0 $ when advancing
    the Orr-Sommerfeld eigenmode at time $ t_0 = 0 $ forward in time
    with the present method. }
  \label{fig:stokesAmplitude}
\end{figure}

\begin{figure}
  \centering
  \includegraphics[width=0.7\textwidth]{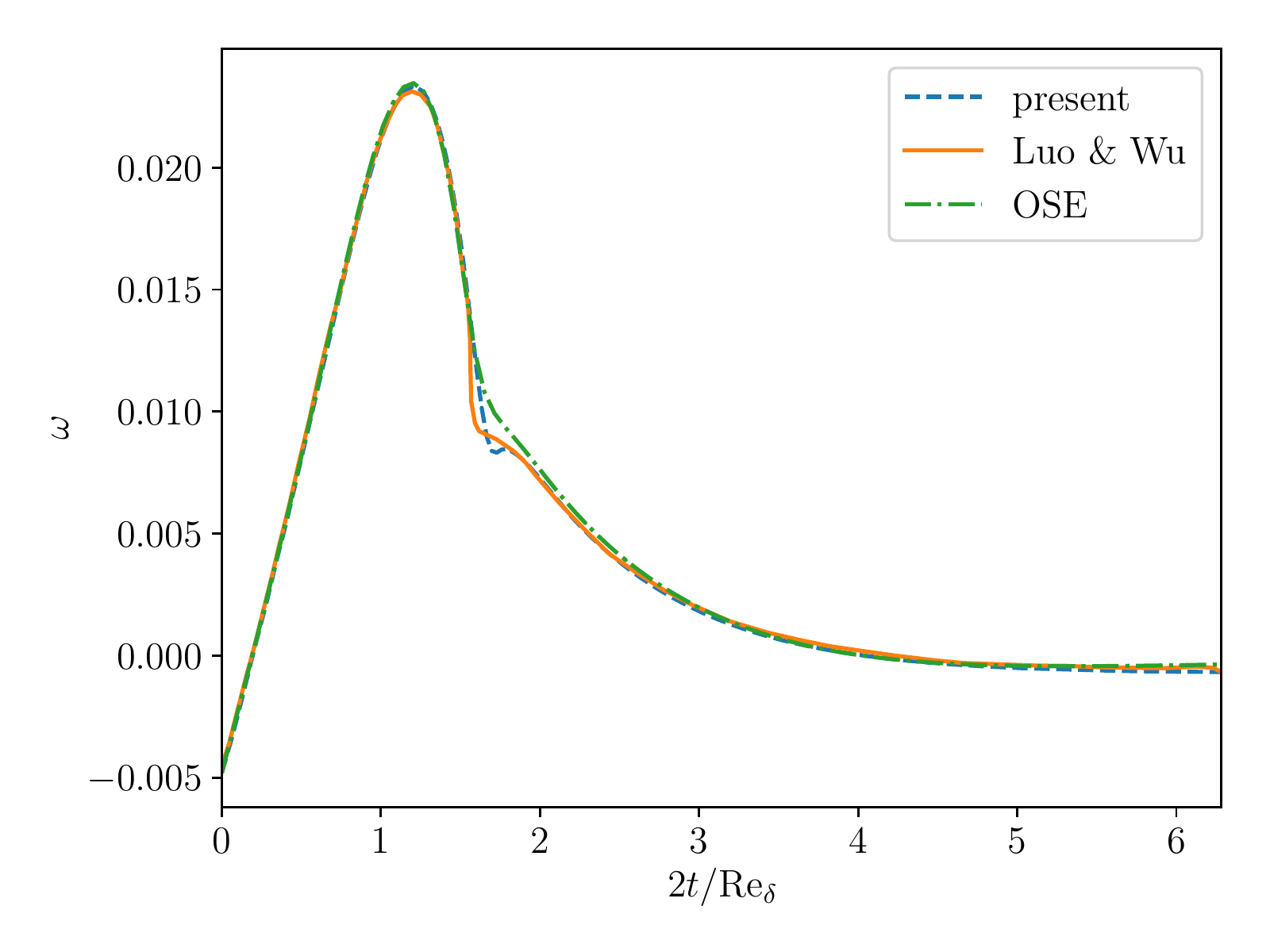}
  \caption{Growth rate of the perturbation when advancing
    the Orr-Sommerfeld eigenmode at time $ t_0 = 0 $ forward in time
    with the present method. }
  \label{fig:stokesOmega}
\end{figure}

Furthermore, the validation with an unsteady base
flow is performed by means of Stokes second problem whose base flow is given by
\be
U_\base = \exp(-z) \cos \left( \frac{2}{\Rey_\delta} t - z \right).
\ee
The results in \citet{LuoWu2010} define a test case for the present method. In
\citet{LuoWu2010}, the temporal evolution of eigenmodes of the Orr-Sommerfeld
equation for $ t_0 = 0 $ is investigated. They consider three
cases defined by $ \Rey_\delta = 1560 $, $ 1562.8 $ and $ 1566 $
and $ \alpha = 0.3 $ and $ \beta = 0 $. As initial condition, the
eigenmodes corresponding to the following eigenvalues $ \omega_{\rm OSE} $ for
each $ \Rey_\delta $ are used:
\[
\begin{array}{l|r}
  \Rey_\delta & \omega_{\rm OSE} \\
  \hline
  1560 & -0.004847-0.196045 {\rm i} \\
  1562.8 & -0.00482994-0.196076{\rm i} \\
  1566 & -0.00481052-0.196111{\rm i} 
  \end{array}
\]
As a main result from the investigation in \cite{LuoWu2010}, the maximum amplitude
of the perturbation for $ \Rey_\delta = 1560 $ decreases from cycle to cycle,
whereas for $ \Rey_\delta = 1562.8 $ the maximum amplitude displays almost no growth from cycle to cycle. However,
for $ \Rey_\delta = 1566 $, the maximum amplitude increases from cycle to
cycle. This can also be observed when using the present method, cf. figure \ref{fig:stokesAmplitude},
where we have used $ N = 97 $.
The amplitude is in our case defined by the ratio between the perturbation
energy at time $ t $ and at time $ t_0 = 0 $. \cite{LuoWu2010} defined the amplitude
differently, namely
by the first coefficient of the expansion of the perturbation on all Orr-Sommerfeld modes.
Therefore, the exact numerical values in figure \ref{fig:stokesAmplitude} and in figure 7 in
\cite{LuoWu2010} are not comparable. When comparing the growth rate $ \omega $ of
the present perturbation, given by:
\be
\omega = \frac{1}{E} \frac{ dE}{dt} 
\ee
with the growth rate given by the real part of the eigenvalue resulting from the Orr-Sommerfeld
equation for the case $ \Rey_\delta = 1566 $, we confirm the observation by \cite[figure 10]{LuoWu2010} that during one cycle the growth rate is relatively well approximated by the Orr-Sommerfeld
solution. In addition, the growth rate taken from figure 10 in \cite{LuoWu2010} by
digitization follows closely the present one, even if the definition of the
amplitude is a different one, cf. figure \ref{fig:stokesOmega}. \\

Returning to the present flow, we shall consider the case
\be
\Rey_\delta = 1000 \quad \alpha = 0.6 \quad \beta = 0.14 \quad h = 30 \quad t_0 = 0
\quad t = 6,  \label{eq:exampleProb}
\ee
for determining the discretization parameters. Before solving the nonmodal equations
(\ref{eq:discreteNonmodal}), the base flow solution needs to be generated.
This is done by numerically solving the boundary layer equations
(\ref{eq:bc1}-\ref{eq:bc4}), applying the same discretization techniques as for
the nonmodal equations (\ref{eq:sys1}-\ref{eq:sys2}). The
present boundary layer solver has been verified by comparison to the solution
obtained by means of the integral formula in \cite{LiuParkCowen2007}. 
An important ingredient in the numerical solution of the boundary layer
equations (\ref{eq:bc1}-\ref{eq:bc4}) is the choice
of a finite value $ t_{-\infty} $ for imposing
the boundary condition (\ref{eq:bc3}). As the outer flow
dies off exponentially towards $ t \rightarrow \pm \infty $, we choose $ t_{-\infty} = -8 $
and $ t_{-\infty} = -12 $ as starting point. For these values the magnitude of the outer flow
amounts to $ U_\free(t_{-\infty} = -8) = 4.50141 \cdot 10^{-7} $ and
$ U_\free(t_{-\infty} = -12) = 1.51005 \cdot^{-10} $, respectively. Choosing $ N = 129 $, we
solve the above nonmodal example problem, equation (\ref{eq:exampleProb}), for $ U_\base $ computed
with $ t_{-\infty} = -8 $ and $ t_{-\infty} = -12 $. The resulting amplification $ G $ is
given by:
\bea
G(0.6,0.14,0,6,1000) = 1.11855 \cdot 10^{9} & & \mbox{for} \quad t_{-\infty} = -8 \\
G(0.6,0.14,0,6,1000) = 1.11869 \cdot 10^{9} & & \mbox{for} \quad t_{-\infty} = -12.
\eea
Choosing $ t_{-\infty} = -12 $ and varying the number of Chebyshev polynomials $ N $,
we observe the following values for G:
\[
\begin{array}{r|l}
  N & G(0.6,0.14,0,6,1000)\\
  \hline 
  33 & 2.22803 \cdot 10^{13} \\
  49 & 3.51768 \cdot 10^8 \\
  65 & 1.13902 \cdot 10^{9} \\
  97 & 1.11865 \cdot 10^{9} \\
  129 & 1.11869 \cdot 10^{9}
  \end{array}
\]
For the simulations in section \ref{sec:results}, computations with $ N = 97 $ and $ N =129 $
have been performed to ensure that the results are accurate.

\section{Scaling of the initial condition for streamwise streaks} \label{sec:initialCondition}

For streamwise streaks ($ \alpha = 0 $), we have the governing equations
given by equations (\ref{eq:sys1streak2}) and (\ref{eq:sys2streak2}).
We shall first find the general solution of $ \tilde{\zeta} $. 

The sine transform of $ \tilde{\zeta} $ is defined as:
\be
\Theta (\gamma,t) = \int \limits_0^\infty \tilde{\zeta} \sin (\gamma z) \, dz
\ee
Taking the sine transform of equation (\ref{eq:sys2streak2}), gives us:
\be
\frac{\partial}{\partial t} \Theta + \frac{1}{2} \left( \gamma^2 + \beta^2 \right) \Theta - F = 0, \label{eq:sineTransformed}
\ee
where
\be
F(\gamma,t) = {\rm i} \beta \int \limits_0^\infty w D U_\base \sin( \gamma z) \, dz.
\ee
Solving equation (\ref{eq:sineTransformed}) gives us for $ \Theta $:
\be
\Theta(\gamma,t) = \left( \Theta(\gamma,0)
+ \int \limits_0^t \frac{F(\gamma,\tau)}{ e^{-\frac{1}{2} \left( \beta^2 + \gamma^2 \right) \tau } }\, d\tau \right) e^{ -\frac{1}{2} \left( \beta^2 + \gamma^2 \right) t}.
\ee
The general solution of $ \tilde{\zeta} $ can thus be written as:
\bea
\tilde{\zeta} &=& \frac{2}{\pi} \int \limits_0^\infty \Theta(\gamma,0)
e^{ -\frac{1}{2} \left( \beta^2 + \gamma^2 \right) t} \sin(\gamma z) \, d\gamma \nonumber \\
 & & + \frac{2}{\pi} \int \limits_0^\infty 
e^{ -\frac{1}{2} \left( \beta^2 + \gamma^2 \right) t} 
\int \limits_{t_0}^t \frac{F(\gamma,\tau)}{ e^{-\frac{1}{2} \left( \beta^2 +\gamma^2 \right) \tau }} \, d\tau  \sin ( \gamma z ) \, d\gamma.
\label{eq:generalSolutionZeta}
\eea
Motivated by the findings in section (\ref{sec:numericalResults}), we
shall assume that in the asymptotic limit $ \Rey_\delta \rightarrow \infty $,
the initial condition of $ w $ and $ \tilde{\zeta} $
can approximately be written as:
\be
w = w_m (\Rey_\delta) \hat{w}(z, t_0) \quad
\tilde{\zeta} = \zeta_m (\Rey_\delta) \hat{\zeta}(z,t_0),
\ee
where only the coefficients $ w_m $ and $ \zeta_m $ depend on $ \Rey_\delta$.
Subsequently, using equation (\ref{eq:generalSolutionZeta}),
we can write $ w $ and $ \tilde{\zeta} $ as:
\bea
\tilde{\zeta} &=& \zeta_m a(z,t) + w_m b(z,t), \\
w &=& w_m c(z,t),
\eea
where $ a $, $ b $ and $ c $ are some functions of $ z $ and $ t$,
with $ b(z,t_0) = 0 $. The energy $ E = E_w + E_\zeta $,
equation (\ref{eq:energyNonmodal}), is then given by:
\bea
E_w(t) & = & w_m^2 \frac{1}{2} \int \limits_0^\infty \frac{1}{\beta^2} | D c|^2 + | c|^2 \, dz, \\
E_\zeta(t) &=& \frac{1}{2} \frac{\Rey_\delta^2}{4} \int \limits_0^\infty \frac{1}{\beta^2} \left( \zeta_m^2 a^2 + 2 \zeta_m w_m ab + w_m^2 b^2 \right)\,dz.
\eea
We can thus write:
\bea
E_w(t_0) &=& w_m^2 A_0, \\
E_w(t) &=& w_m^2 A_1, \\
E_\zeta(t_0) &=& \Rey_\delta^2 \zeta_m^2 B_0, \\
E_\zeta(t) &=& \Rey_\delta^2 \left( \zeta_m^2 B_1 + 2 \zeta_m w_m B_2 + w_m^2 B_3 \right),
\eea
where $ A_0 $, $ A_1 $, $ B_0 $, $ B_1 $, $ B_2 $ and $ B_3 $ are independent
of $ \Rey_\delta $. 
The normalization constraint for the initial condition reads:
\be
E_w(t_0) + E_\zeta(t_0) = w_m^2 A_0 + \Rey_\delta^2 \zeta_m^2 B_0 = 1,
\ee
From which we find:
\be
w_m^2 = \frac{1}{A_0} \left( 1 - \Rey_\delta^2 B_0 \zeta_m^2 \right)
\ee
As the right hand side needs to be positive for all $ \Rey_\delta $, this
motivates
the following ansatz for $ \zeta_m $ 
in the limit of $ \Rey_\delta \rightarrow \infty $:
\be
\zeta_m = \frac{d}{\Rey_\delta^\theta},
\ee
where $ \theta \ge 1 $ and $ d $ some constant. 
For the energy at time $ t $, we can write:
\bea
E(t) &=& w_m^2 A_1 + \Rey_\delta^2 \left( \zeta_m^2 B_1 + 2 \zeta_m w_m B_2 + w_m^2 B_3 \right) \\
&=& \frac {1}{{ A_0}}
\Big( 2d \,{{\Rey_\delta}}^{-\theta+2} \sqrt{ A_0}\,{
  B_2}\,\sqrt {
  { {
      \left( {{\Rey_\delta}}^{2\,\theta} - {B_0}\,{{ \Rey_\delta}}^{2}{d}^{2} \right) {{\Rey_\delta}}^{-2\,\theta}}}
} \\
& & \quad \quad +{d}^{
2} \left( {A_0}\,{B_1}-{A_1}\,{B_0} \right) {{\Rey_\delta}}^{2
-2\,\theta}-{{\Rey_\delta}}^{-2\,\theta+4}{B_0}\,{B_3}\,{d}^{2}+{
B_3}\,{{\Rey_\delta}}^{2}+{A_1} \Big). \nonumber
\eea
As the energy is maximum for the optimal perturbation, we must have
\be
\frac{\partial E}{\partial \theta} = 0.
\ee
Solving this equation for $ \theta $ gives us four solutions
\be
\theta_{1,2,3,4} = 1/2\,{\frac {1}{\ln  \left( {\Rey_\delta} \right) } \left( -\ln  \left( 2
  \right) +2\,\ln  \left( \pm {\frac {d}{{B_2}}\sqrt { \frac{F_{\pm}}{A_0} }
  } \right)  \right) },
\ee
where
\bea
F_{\pm} &=&  \pm \sqrt { D } +
          \left( {{B_3}}^{2}{{\Rey_\delta}}^{4}+2\,{A_1}\,{B_3}\,{{\Rey_\delta
          }}^{2}+{{A_1}}^{2} \right) {{B_0}}^{2} \nonumber \\
          & & +
          \left(  \left( -2\,{B_1}\,{B_3}+4\,{{B_2}}^{2} \right) {{\Rey_\delta}}^{2}-2\,{A_1}\,{
            B_1} \right) {A_0}\,{B_0}+{{A_0}}^{2}{{B_1}}^{2}\\
D &=& \left(  \left( -{B_3}\,{{\Rey_\delta}}^{2}-{A_1} \right) {B_0}+{
  A_0}\,{B_1} \right)^{2} \\
& &      \left(  \left( {B_3}\,{{\Rey_\delta}}^{2
            }+{A_1} \right) ^{2}{{B_0}}^{2}-2\,{A_0}\, \left(  \left( {
B_1}\,{B_3}-2\,{{B_2}}^{2} \right) {{\Rey_\delta}}^{2}+{A_1}\,
{B_1} \right) {B_0}+{{A_0}}^{2}{{B_1}}^{2} \right)  \nonumber 
\eea
Taking the limit $ \Rey_\delta \rightarrow \infty $, we obtain:
\be
 \lim_{\Rey_\delta \rightarrow \infty} \theta_i = 2 \quad \mbox{for} \quad i = 1,2,3,4.
\ee
From this it follows, that for $ \Rey_\delta >> 1 $, we have
approximately
\be
\tilde{\zeta}(z,t_0)  \propto \frac{1}{\Rey_\delta^2},
\ee
from which relation (\ref{fig:scalingStreaksInfty}) can directly be obtained. 

\end{appendix}

\bibliographystyle{jfmabbrv}
%\bibliography{references}

\begin{thebibliography}{45}
\expandafter\ifx\csname natexlab\endcsname\relax\def\natexlab#1{#1}\fi

\bibitem[Andersson {\em et~al.\/}(2001)Andersson, Brandt, Bottaro \&
  Henningson]{AnderssonBrandtBottaroHenningson2001}
{\sc Andersson, P. , Brandt, L. , Bottaro, A.  \& Henningson, D.~S. } 2001 On
  the breakdown of boundary layer streaks. {\em Journal of Fluid Mechanics\/}
  {\bf 428}, 29--60.

\bibitem[Benjamin(1966)]{Benjamin1966}
{\sc Benjamin, T.~B. } 1966 Internal waves of finite amplitude and permanent
  form. {\em Journal of Fluid Mechanics\/} {\bf 25}, 241--270.

\bibitem[Bertolotti {\em et~al.\/}(1992)Bertolotti, Herbert \&
  Spalart]{BertolottiHerbertSpalart1992}
{\sc Bertolotti, F. , Herbert, T.  \& Spalart, P. } 1992 Linear and nonlinear
  stability of the {B}lasius boundary layer. {\em Journal of Fluid Mechanics\/}
  {\bf 242}, 441--474.

\bibitem[Biau(2016)]{Biau2016}
{\sc Biau, D. } 2016 Transient growth of perturbations in stokes oscillatory
  flows. {\em Journal of Fluid Mechanics\/} {\bf 794}, 10.

\bibitem[Blondeaux {\em et~al.\/}(2012)Blondeaux, Pralits \&
  Vittori]{BlondeauxPralitsVittori2012}
{\sc Blondeaux, P. , Pralits, J.  \& Vittori, G. } 2012 Transition to
  turbulence at the bottom of a solitary wave. {\em Journal of Fluid
  Mechanics\/} {\bf 709}, 396--407.

\bibitem[Brandt {\em et~al.\/}(2004)Brandt, Schlatter \&
  Henningson]{BrandtSchlatterHenningson2004}
{\sc Brandt, L. , Schlatter, P.  \& Henningson, D.~S. } 2004 Transition in
  boundary layers subject to free-stream turbulence. {\em Journal of Fluid
  Mechanics\/} {\bf 517}, 167--198.

\bibitem[Butler \& Farrell(1992)]{ButlerFarrell1992}
{\sc Butler, K.~M.  \& Farrell, B.~F. } 1992 Three-dimensional optimal
  perturbations in viscous shear flow. {\em Physics of Fluids A\/} {\bf 4},
  1637--1650.

\bibitem[Carr \& Davies(2006)]{CarrDavies2006}
{\sc Carr, M.  \& Davies, P.~A. } 2006 The motion of an internal solitary wave
  of depression over a fixed bottom boundary in a shallow, two-layer fluid.
  {\em Physics of Fluids\/} {\bf 18}, 016601--10.

\bibitem[Carr \& Davies(2010)]{CarrDavies2010}
{\sc Carr, M.  \& Davies, P.~A. } 2010 Boundary layer flow beneath an internal
  solitary wave of elevation. {\em Physics of Fluids\/} {\bf 22}, 026601--1--8.

\bibitem[Corbett \& Bottaro(2000)]{CorbettBottaro2000}
{\sc Corbett, P.  \& Bottaro, A. } 2000 Optimal perturbations for boundary
  layers subject to stream-wise pressure gradient. {\em Physics of Fluids\/}
  {\bf 12}~(1), 120--130.

\bibitem[Corbett \& Bottaro(2001)]{CorbettBottaro2001}
{\sc Corbett, P.  \& Bottaro, A. } 2001 Optimal linear growth in swept boudary
  layers. {\em Journal of Fluid Mechanics\/} {\bf 435}, 1--23.

\bibitem[Davis \& von Kerczek(1973)]{DavisKerczek1973}
{\sc Davis, S.~H.  \& von Kerczek, C. } 1973 A reformulation of energy
  stability theory. {\em Archive for Rational Mechanics and Analysis\/} pp.
  112--117.

\bibitem[Ellingsen \& Palm(1975)]{EllingsenPalm1975}
{\sc Ellingsen, T.  \& Palm, E. } 1975 Hydrodynamic stability. {\em Physics of
  Fluids\/} {\bf 18}, 487.

\bibitem[Fenton(1972)]{Fenton1972}
{\sc Fenton, J. } 1972 A ninth-order solution for the solitary wave. {\em
  Journal of Fluid Mechanics\/} {\bf 53}, 257--271.

\bibitem[Frigo \& Johnson(2005)]{fftw}
{\sc Frigo, M.  \& Johnson, S.~G. } 2005 The design and implementation of
  {FFTW3}. In {\em Proceedings of the IEEE\/}, , vol.~93, pp. 216--231.

\bibitem[Galassi {\em et~al.\/}(2009)Galassi, Davies, Theiler, Gough, Jungman,
  Alken, Booth \& Rossi]{gsl}
{\sc Galassi, M. , Davies, J. , Theiler, B. , Gough, B. , Jungman, G. , Alken,
  P. , Booth, M.  \& Rossi, F. } 2009 {\em {GNU} Scientific Library Reference
  Manual\/}. Network Theory Ltd.

\bibitem[Gaster(2016)]{Gaster2016}
{\sc Gaster, M. } 2016 Boundary layer transition initiated by a random
  excitation. In {\em Book of Abstracts $ 24^{th} $ International Congress of
  Theoretical and Applied Mechanics\/}.

\bibitem[Grimshaw(1971)]{Grimshaw1971}
{\sc Grimshaw, R. } 1971 The solitary wave in water of variable depth. part 2.
  {\em Journal of Fluid Mechanics\/} {\bf 46}, 611--622.

\bibitem[Gustavsson(1991)]{Gustavsson1991}
{\sc Gustavsson, L.~H. } 1991 Energy growth of three-dimensional disturbances
  in plane {P}oiseuille flow. {\em Journal of Fluid Mechanics\/} {\bf 224},
  241--260.

\bibitem[Herbert(1988)]{Herbert1988}
{\sc Herbert, T. } 1988 Secondary instability of boundary layers. {\em Annual
  Review of Fluid Mechanics\/} {\bf 20}, 487--526.

\bibitem[Jimenez(2013)]{Jimenez2013}
{\sc Jimenez, J. } 2013 How linear is wall-bounded turbulence? {\em Physics of
  Fluids\/} {\bf 25}, 110814--1--19.

\bibitem[Joseph(1966)]{Joseph1966}
{\sc Joseph, D.~D. } 1966 Nonlinear stability of the boussinesq equations by
  the method of energy. {\em Archive for Rational Mechanics and Analysis\/}
  {\bf 22}, 163.

\bibitem[von Kerczek \& Davis(1974)]{KerczekDavis1974}
{\sc von Kerczek, C.  \& Davis, S.~H. } 1974 Linear stability theory of
  oscillatory stokes layers. {\em Journal of Fluid Mechanics\/} {\bf 62},
  753--773.

\bibitem[Levin \& Henningson(2003)]{LevinHenningson2003}
{\sc Levin, O.  \& Henningson, D.~S. } 2003 Exponential vs algebra growth and
  transition prediction in boundary layer flow. {\em Flow, Turbulence and
  Combustion\/} {\bf 70}, 183--210.

\bibitem[Liu \& Orfila(2004)]{LiuOrfila2004}
{\sc Liu, P. L.-F.  \& Orfila, A. } 2004 Viscous effects on transient long-wave
  propagation. {\em Journal of Fluid Mechanics\/} {\bf 520}, 83--92.

\bibitem[Liu {\em et~al.\/}(2007)Liu, Park \& Cowen]{LiuParkCowen2007}
{\sc Liu, P. L.-F. , Park, Y.~S.  \& Cowen, E.~A. } 2007 Boundary layer flow
  and bed shear stress under a solitary wave. {\em Journal of Fluid
  Mechanics\/} {\bf 574}, 449--463.

\bibitem[Luchini \& Bottaro(2014)]{LuchiniBottaro2014}
{\sc Luchini, P.  \& Bottaro, A. } 2014 Adjoint equations in stability
  analysis. {\em Annual Review of Fluid Mechanics\/} {\bf 46}, 493--517.

\bibitem[Luo \& Wu(2010)]{LuoWu2010}
{\sc Luo, J.  \& Wu, X. } 2010 On the linear instability of a finite stokes
  layer: Instantaneous versus floquet modes. {\em Physics of Fluids\/} {\bf
  22}, 1--13.

\bibitem[Miles(1980)]{Miles1980}
{\sc Miles, J.~W. } 1980 Solitary waves. {\em Annual Review of Fluid
  Mechanics\/} {\bf 12}, 11--43.

\bibitem[Ozdemir {\em et~al.\/}(2013)Ozdemir, Hsu \&
  Balachandar]{OzdemirHsuBalachandar2013}
{\sc Ozdemir, C.~E. , Hsu, T.-J.  \& Balachandar, S. } 2013 Direct numerical
  simulations of instability and boundary layer turbulence under a solitay
  wave. {\em Journal of Fluid Mechanics\/} {\bf 731}, 545--578.

\bibitem[Park {\em et~al.\/}(2014)Park, Verschaeve, Pedersen \&
  Liu]{ParkVerschaevePedersenLiu2014}
{\sc Park, Y.~S. , Verschaeve, J. C.~G. , Pedersen, G.~K.  \& Liu, P. L.-F. }
  2014 Corrigendum and addendum for boundary layer flow and bed shear stress
  under a solitary wave. {\em Journal of Fluid Mechanics\/} {\bf 753},
  554--559.

\bibitem[Sadek {\em et~al.\/}(2015)Sadek, Parras, Diamessis \&
  Liu]{SadekParrasDiamessisLiu2015}
{\sc Sadek, M.~M. , Parras, L. , Diamessis, P.~J.  \& Liu, P. L.-F. } 2015
  Two-dimensional instability of the bottom boundary layer under a solitary
  wave. {\em Physics of Fluids\/} {\bf 27}, 044101--1--25.

\bibitem[Sanderson \& Curtin(2016)]{armadillo}
{\sc Sanderson, C.  \& Curtin, R. } 2016 Armadillo: a template-based {C}++
  library for linear algebra. {\em Journal of Open Source Software\/} {\bf 1},
  26.

\bibitem[Schmid(2007)]{Schmid2007}
{\sc Schmid, P.~J. } 2007 Nonmodal stability theory. {\em Annual Review of
  Fluid Mechanics\/} {\bf 39}, 129--162.

\bibitem[Schmid \& Henningson(2001)]{SchmidHenningson2001}
{\sc Schmid, P.~J.  \& Henningson, D.~S. } 2001 {\em Stability and Transition
  in Shear Flows\/}. New York: Springer-Verlag.

\bibitem[Shaikh \& Gaster(1994)]{ShaikhGaster1994}
{\sc Shaikh, F.~N.  \& Gaster, M. } 1994 The non-linear evolution of modulated
  waves in a boundary layer. {\em Journal of Engineering Mathematics\/} {\bf
  28}, 55--71.

\bibitem[Shen(1994)]{Shen1994}
{\sc Shen, J. } 1994 Efficient spectral-galerkin method i. direct solvers for
  the second and fourth order equations using legendre polynomials. {\em Siam
  Journal of Scientific Coputing\/} {\bf 15}, 1489--1505.

\bibitem[Shen(1995)]{Shen1995}
{\sc Shen, J. } 1995 Efficient spectral-galerkin method ii. direct solvers of
  second fourth order equations by using chebyshev polynomials. {\em SIAM
  Journal of Scientific Computing\/} {\bf 16}~(1), 74--87.

\bibitem[Shuto(1976)]{Shuto1976}
{\sc Shuto, N. } 1976 Transformation of nonlinear long waves. In {\em
  Proceedings of 15th Conference on Coastal Enginearing\/}.

\bibitem[Sumer {\em et~al.\/}(2010)Sumer, Jensen, S{\o}rensen, Freds{\o}e, Liu
  \& Carstensen]{SumerJensenSorensenFredsoeLiuCarstensen2010}
{\sc Sumer, B.~M. , Jensen, P.~M. , S{\o}rensen, L.~B. , Freds{\o}e, J. , Liu,
  P. L.-F.  \& Carstensen, S. } 2010 Coherent structures in wave boundary
  layers. part 2. solitary motion. {\em Journal of Fluid Mechanics\/} {\bf
  646}, 207--231.

\bibitem[Tanaka {\em et~al.\/}(2011)Tanaka, Winarta, Suntoyo \&
  Yamaji]{TanakaWinartaSuntoyoYamaji2011}
{\sc Tanaka, H. , Winarta, B. , Suntoyo \& Yamaji, H. } 2011 Validation of a
  new generation system for bottom boundary layer beneath solitary wave. {\em
  Coastal Engineering\/} {\bf 59}, 46--56.

\bibitem[Trefethen {\em et~al.\/}(1993)Trefethen, Trefethen, Reddy \&
  Driscoll]{TrefethenTrefethenReddyDriscoll1993}
{\sc Trefethen, L.~N. , Trefethen, A.~E. , Reddy, S.~C.  \& Driscoll, T.~A. }
  1993 Hydrodynamic stability witwith eigenvalues. {\em Science\/} {\bf 261},
  578--584.

\bibitem[Verschaeve \& Pedersen(2014)]{VerschaevePedersen2014}
{\sc Verschaeve, J. C.~G.  \& Pedersen, G.~K. } 2014 Linear stability of
  boundary layers under solitary waves. {\em Journal of Fluid Mechanics\/} {\bf
  761}, 62--104.

\bibitem[Vittori \& Blondeaux(2008)]{VittoriBlondeaux2008}
{\sc Vittori, G.  \& Blondeaux, P. } 2008 Turbulent boundary layer under a
  solitary wave. {\em Journal of Fluid Mechanics\/} {\bf 615}, 433--443.

\bibitem[Vittori \& Blondeaux(2011)]{VittoriBlondeaux2011}
{\sc Vittori, G.  \& Blondeaux, P. } 2011 Characteristics of the boundary layer
  at the bottom of a solitary wave. {\em Coastal Engineering\/} {\bf 58},
  206--213.

\end{thebibliography}

\end{document}